\documentclass[preprint,aps,nofootinbib]{revtex4}

\usepackage{dcolumn}% Align table columns on decimal point
\usepackage{bm}% bold math
\usepackage{epsfig}
\usepackage{graphicx,subfigure}% Include figure files

\usepackage{amsmath}
\usepackage{amssymb}
\usepackage{color}
\usepackage[usenames,dvipsnames]{xcolor}
\usepackage{hyperref}

\def\ga{\mathrel{\mathpalette\fun >}}
\def\fun#1#2{\lower3.6pt\vbox{\baselineskip0pt\lineskip.9pt
  \ialign{$\mathsurround=0pt#1\hfil##\hfil$\crcr#2\crcr\sim\crcr}}}

\def\simlt{\stackrel{<}{{}_\sim}}
\def\simgt{\stackrel{>}{{}_\sim}}

\input epsf

\newcommand{\be}{\begin{equation}}
\newcommand{\ee}{\end{equation}}
\newcommand{\bea}{\begin{eqnarray}}
\newcommand{\eea}{\end{eqnarray}}

\begin{document}

\begin{flushright}
%\vspace*{-1.5cm}
%\vspace{-0.2cm}ANL-HEP-PR-13-50\\
%\vspace{-0.2cm}EFI-13-27\\
%\vspace{-0.2cm}FERMILAB-PUB-13-455-T\\
%\vspace{-0.2cm}MCTP-13-31\\
%\vspace{-0.2cm}NSF-KITP-13-XXX
\end{flushright}

%\vspace*{0.1cm}

\title{Impersonating the Standard Model Higgs Boson: \\
Alignment without Decoupling}

%\vspace*{0.2cm}

\author{
%\vspace{0.2cm} 
\mbox{\bf Marcela Carena$^{\,a,b,c}$, Ian Low$^{\,d,e,f}$, Nausheen R. Shah$\,^g$, and Carlos E.~M.~Wagner$^{\,b,c,e}$}
 }
\affiliation{
%\vspace*{.2cm}
$^a$  \mbox{Fermi National Accelerator Laboratory, P.O. Box 500, Batavia, IL 60510}\\
$^b$  \mbox{Enrico Fermi Institute, University of Chicago, Chicago, IL 60637}\\
$^c$  \mbox{Kavli Institute for Cosmological Physics, University of Chicago, Chicago, IL 60637}\\
$^d$\mbox{Kavli Institute for Theoretical Physics, University of California, Santa Barbara, CA 93106}\\
$^e$ \mbox{High Energy Physics Division, Argonne National Laboratory, Argonne, IL 60439}\\
$^f$ \mbox{Department of Physics and Astronomy, Northwestern University, Evanston, IL 60208} \\
$^g$\mbox{Michigan Center for Theoretical Physics, Department of Physics,} \\
\mbox{University of Michigan, Ann Arbor, MI 48109}\\
}

\begin{abstract}
In models with an extended Higgs sector there exists an alignment limit, in which the lightest CP-even Higgs boson mimics the Standard Model Higgs. The alignment limit is commonly associated with the decoupling limit, where all non-standard scalars are significantly heavier than the $Z$ boson. However, alignment  can occur irrespective of the mass scale of the rest of the Higgs sector. In this work we discuss the general conditions that lead to ``alignment without decoupling", therefore allowing for the existence of additional non-standard Higgs bosons at the weak scale. The values of  $\tan\beta$ for which  this happens are derived in terms of the effective Higgs quartic couplings in general two-Higgs-doublet models as well as 
in supersymmetric theories, including  the MSSM and the NMSSM. Moreover, we study the information encoded in  the variations of the SM Higgs-fermion couplings  to explore regions  in the $m_A - \tan\beta$ parameter space.
\end{abstract}
\thispagestyle{empty}

\maketitle

%%%%%%%%%%%%%%%%%%%%%%%%%%%%%%%%%%%%%%%
%%%%%%%%%%%%%%%%%%%%%%%%%%%%%%%%%%%%%%%
 \section{Introduction}
 %%%%%%%%%%%%%%%%%%%%%%%%%%%%%%%%%%%%%%%
%%%%%%%%%%%%%%%%%%%%%%%%%%%%%%%%%%%%%%%

The Standard Model (SM) with one Higgs doublet is the simplest realization of electroweak symmetry breaking and provides a very good description of all data collected so far at hadron and lepton colliders. This includes measurements associated with the recently discovered 125 GeV Higgs boson at the CERN LHC \cite{:2012gk,:2012gu}.  In this model, the Higgs field receives a vacuum expectation value (VEV), $v\approx 246$ GeV, which breaks the electroweak gauge symmetry and gives masses to the fundamental fermions and gauge bosons. The couplings of these particles to the  Higgs boson are fixed by their masses and $v$. On the other hand, Higgs self interactions are controlled by the quartic coupling in the Higgs potential, which in turn is given by the Higgs mass and $v$. Therefore, interactions of the SM Higgs boson with fermions, gauge bosons and with itself are completely determined.

Extensions of the SM commonly  lead to modifications of the Higgs couplings, especially if there exist new particles interacting with the Higgs or if there is an extended Higgs sector. The size of these modifications may be naively estimated in  the decoupling limit to be:
\be
\label{order}
  {\cal{O}} \left(\frac{v^2}{m_{\rm new}^2}\right)  \ \approx  \  {\cal{O}} ( 5 \% ) \ \times \left(\frac{1 {\rm\  TeV}}{m_{\rm new}}\right)^2\ ,
\ee
where $m_{\rm new}$ is the scale of new particles.  Therefore, for new particles 
below the TeV scale, changes in the Higgs couplings from the SM expectations are quite small. Such an estimate supports the fact that due to the large uncertainties in present measurements,  no significant deviations from the SM Higgs properties should be identifiable in present data, if all new particles are at or above the TeV scale.  At the same time, it stresses the need  for precision Higgs measurements to uncover possible signs of new physics.

Conversely,  Eq.~(\ref{order}) implies that,  if in the future, refined  measurements of the properties of the 125 GeV Higgs boson continue to be consistent with those of a SM Higgs boson,  no new light particles interacting with the SM-like Higgs are to be expected. However, this estimate is only valid in the so called decoupling limit, where all non-standard Higgs bosons are significantly heavier than the  $Z$ gauge boson. On the other hand, current searches for these particles do not exclude the possibility of additional Higgs bosons 
in the hundred to several hundred GeV mass range. Given that  initial data appears to disfavor large deviations with respect to the SM Higgs description~\cite{Low:2012rj}, it is of special interest to consider models of extended Higgs sectors containing  a CP-even Higgs that has properties mimicking quite precisely the SM ones, even if the non-standard Higgs bosons are light. 

 A well-known example is that of  general two-Higgs-doublet models (2HDMs)~\cite{Branco:2011iw,Craig:2012vn}, in which the heavy CP-even Higgs could be the SM-like Higgs boson.  However, in this case the 2HDM parameter space becomes  very restrictive, with  masses of the non-standard scalars of the order of the $W$ and $Z$ boson masses, and is severely constrained by data \cite{Christensen:2012ei}. On the other hand, the possibility of the lightest CP-even Higgs mimicking the SM Higgs, referred to as ``alignment" in Ref.~\cite{Craig:2013hca}, is much less constrained and  usually associated with the decoupling limit. The less known and more interesting case of alignment  without recourse to decoupling deserves further study. 
 
 A few examples  of ``alignment without  decoupling''  have been considered in the literature.  
The first one was presented over a decade ago by Gunion and Haber in Ref.~\cite{Gunion:2002zf}. Their main focus was to emphasize the SM-like behavior of  the lightest CP-even Higgs of a 2HDM in the decoupling limit. However, they also demonstrated that it can behave like a  SM Higgs  without decoupling the non-SM-like scalars.  Much more recently, a similar situation was discussed in Ref.~\cite{Delgado:2013zfa}, where an extension of the Minimal Supersymmetric Standard Model (MSSM) with a triplet scalar was studied. It was found that after integrating out the triplet scalar, a SM-like Higgs boson and  additional light scalars 
 are left in the spectrum for  low values of $\tan\beta \alt 10$. Another recent study, Ref.~\cite{Craig:2013hca},  presented  a scanning over the parameter space of general 2HDMs. Solutions were found fulfilling  alignment  without decoupling and the phenomenological implications were investigated.

It is obvious that the possibility of alignment without  decoupling would have far-reaching implications  for physics beyond the SM searches. However, its existence has remained obscure and has sometimes been attributed to accidental cancellations in the scalar potential. A simple way to understand how one of the CP-even Higgs bosons in a 2HDM mimics  the SM Higgs is to realize that the alignment limit  occurs whenever the mass eigenbasis in the CP-even sector aligns with the  basis in which the electroweak gauge bosons receive all of their masses from only one of the Higgs doublets~\footnote{This would imply that the other, non-standard CP-even Higgs has no tree-level couplings to the gauge bosons. However, there are still couplings to SM fermions in general. Therefore the non-standard Higgs boson is not inert.}. From this perspective, it is clear that the alignment limit does not require the non-standard Higgs bosons to be heavy. After  presenting the general conditions for the alignment limit  in 2HDMs, we  analyze in detail  the possible implications for well motivated models containing two Higgs doublets. In particular, we  consider  the MSSM as well as its generalization to the next-to-minimal supersymmetric standard model (NMSSM), where an extra singlet is added. Along the way, we analyze the extent to which  precision measurements of Higgs-fermion couplings could be useful in probing regions of parameters that are difficult to access through  direct non-standard Higgs boson searches.

This article is organized as follows. In the next section we define the notation and briefly review the scalar potential and the Higgs couplings in general, renormalizable 2HDMs. In Section \ref{sect:3} we derive the alignment condition in the decoupling regime in terms of the eigenvectors of the CP-even Higgs mass matrix, which provides a simple analytical understanding of alignment. We then write down the general conditions for alignment without decoupling.  In Section \ref{sect:4} we study the alignment limit in general 2HDMs and provide new perspectives on previous works. Detailed studies on the parameter space of the MSSM and beyond are presented in Section \ref{sect:5}, which is followed by the conclusion in Section \ref{sect:6}.

%%%%%%%%%%%%%%%%%%%%%%%%%%%%%%%%%%%%%%%
%%%%%%%%%%%%%%%%%%%%%%%%%%%%%%%%%%%%%%%
 \section{Overview of 2HDM}
 %%%%%%%%%%%%%%%%%%%%%%%%%%%%%%%%%%%%%%%
%%%%%%%%%%%%%%%%%%%%%%%%%%%%%%%%%%%%%%%

%%%%%%%%%%%%%%%%%%%%%%%%%%%%%%%%%%%%%%%
 \subsection{Scalar Potential}
 %%%%%%%%%%%%%%%%%%%%%%%%%%%%%%%%%%%%%%%

 We follow the notation in Ref.~\cite{Haber:1993an} for the scalar potential of the most general two-Higgs-doublet extension of the SM:
\bea
\label{eq:generalpotential}
V &=& m_{11}^2 \Phi_1^\dagger \Phi_1+m_{22}^2 \Phi_2^\dagger \Phi_2-m_{12}^2 (\Phi_1^\dagger \Phi_2 +{\rm h.c.}) +\frac12 \lambda_1 ( \Phi_1^\dagger \Phi_1)^2+\frac12 \lambda_2 ( \Phi_2^\dagger \Phi_2)^2 \nonumber \\
&& +\lambda_3 ( \Phi_1^\dagger \Phi_1)( \Phi_2^\dagger \Phi_2)+\lambda_4 ( \Phi_1^\dagger \Phi_2)( \Phi_2^\dagger \Phi_1) \nonumber \\
&&+ \left\{ \frac12 \lambda_5 ( \Phi_1^\dagger \Phi_2)^2 + [ \lambda_6 (\Phi^\dagger_1\Phi_1)+ \lambda_7 (\Phi^\dagger_2\Phi_2)]\Phi_1^\dagger\Phi_2 + {\rm h.c.} \right\}\ ,
\eea
where
\be
\Phi_i = \left( \begin{array}{c}
       \phi_i^+ \\
       \frac1{\sqrt{2}}(\phi_i^0+i a_i^0)
       \end{array} \right) \ .
 \ee      
We will assume CP conservation and that the minimum of the potential is at
\be
\langle \Phi_i \rangle = \frac1{\sqrt{2}}  \left( \begin{array}{c}
        0\\
        v_i
         \end{array} \right) \ ,
\ee         
where 
\be
v\equiv \sqrt{v_1^2+v_2^2} \approx 246 \ {\rm GeV} \ , \quad t_\beta \equiv \tan\beta = \frac{v_2}{v_1} \ .
\ee
We choose $0\le \beta \le \pi/2$ so that $t_\beta \ge 0$ and write $v_1=v \cos\beta\equiv v c_\beta$ and $v_2=v \sin\beta\equiv v s_\beta$. The five mass eigenstates are: two CP-even scalars, $H$ and $h$, with $m_h \le m_H$, one CP-odd scalar, $A$, and a charged pair, $H^\pm$. The mass parameters, $m_{11}^2$ and $m_{22}^2$, can be eliminated by imposing the minimization condition \cite{Haber:1993an}:
\bea
&&m_{11}^2-t_\beta m_{12}^2 + \frac12 v^2 c_\beta^2(\lambda_1+3\lambda_6t_\beta+\tilde{\lambda}_3 t_\beta^2 +\lambda_7 t_\beta^3) =0\ , \\
&&m_{22}^2-t_\beta^{-1} m_{12}^2 + \frac12 v^2 s_\beta^2(\lambda_2+3\lambda_7t_\beta^{-1}+\tilde{\lambda}_3 t_\beta^{-2} +\lambda_6 t_\beta^{-3}) =0\ ,
\eea
where $\tilde{\lambda}_3 =\lambda_3+\lambda_4+\lambda_5$. It then follows that \cite{Haber:1993an}
\be
m_A^2 =\frac{2m_{12}^2}{s_{2\beta}} - \frac12 v^2(2\lambda_5+\lambda_6 t_\beta^{-1} +\lambda_7t_\beta) \ ,
\ee
and the mass-squared matrix for the CP-even scalars can be expressed as
\be
\label{eq:cpemass}
{\cal M}^2  = \left(\begin{array}{cc}
                                                    {\cal M}_{11}^2 &  {\cal M}_{12}^2 \\
                                                     {\cal M}_{12}^2 &  {\cal M}_{22}^2 
                                                     \end{array}\right)
                                                     \equiv
                                                     m_A^2 \left(\begin{array}{cc}
                                                    s_\beta^2 & -s_\beta c_\beta \\
                                                     -s_\beta c_\beta & c_\beta^2 
                                                     \end{array}\right) 
                                 + v^2    \left(\begin{array}{cc}
                                                            L_{11} & L_{12} \\
                                                            L_{12} & L_{22}
                                                            \end{array}\right) \ ,
\ee
where                                                            
\bea                                 
    L_{11} &=& \lambda_1 c_\beta^2 +2\lambda_6 s_\beta c_\beta+\lambda_5 s_\beta^2 \ , \label{eqL11}\\
    L_{12}&=& (\lambda_3+\lambda_4)s_\beta c_\beta +\lambda_6 c_\beta^2 +\lambda_7 s_\beta^2\ ,\label{eqL12}
 \\
     L_{22}&=&  \lambda_2 s_\beta^2 +2\lambda_7 s_\beta c_\beta+\lambda_5 c_\beta^2 \ .
     \label{eqL22}
 \eea
The mixing angle, $\alpha$, in the CP-even sector is defined as  
\be\label{ralpha}
\left(\begin{array}{c}
                    H\\
                    h
        \end{array}\right) =
              \left(\begin{array}{cc}
                    c_\alpha & s_\alpha \\
                    -s_\alpha & c_\alpha
        \end{array}\right) 
        \left(\begin{array}{c}
                    \phi_1^0\\
                    \phi_2^0
        \end{array}\right)   \equiv R(\alpha)      \left(\begin{array}{c}
                    \phi_1^0\\
                    \phi_2^0
        \end{array}\right)\ ,                                
\ee
where $s_\alpha\equiv \sin\alpha$ and $c_\alpha\equiv \cos\alpha$. This leads to
\be
\label{eq:master}
R^T(\alpha) 
              \left(\begin{array}{cc}
                    m_H^2 & 0 \\
                    0 &  m_h^2
        \end{array}\right) R(\alpha) = \left(\begin{array}{cc}
                                                    {\cal M}_{11}^2 &  {\cal M}_{12}^2  \\
                                                     {\cal M}_{12}^2  &  {\cal M}_{22}^2  
                                                     \end{array}\right) \ .
\ee
From the $(1,2)$ component in the above equation we see
\be 
(m_H^2-m_h^2) s_\alpha c_\alpha = {\cal M}_{12}^2 \ ,
\ee
which implies $s_\alpha c_\alpha$ has the same sign as ${\cal M}_{12}^2$. There are two possible sign choices:
\bea
\label{eq:signI}
{\rm (I)} & & -\frac{\pi}2 \le \alpha \le \frac{\pi}2: \quad c_\alpha \ge 0 \ \ {\rm and} \ \ {\rm Sign}(s_\alpha)= {\rm Sign}({\cal M}_{12}^2) \ , \\
\label{eq:signII}
{\rm (II)}& & \phantom{-}\ 0 \le \alpha \,\le \pi: \quad  s_\alpha \ge 0 \ \ {\rm and}\ \ {\rm Sign}(c_\alpha)={\rm Sign}({\cal M}_{12}^2) \ .
\eea
We will discuss the implications of these two sign choices in some detail below.

The eigenvector associated with the eigenvalue $m_{h}^2$ corresponds to the second row in $R(\alpha)$, Eq.~(\ref{ralpha}), and satisfies
 \be
 \label{eq:eigeneq}
  \left(\begin{array}{cc}
                                                    {\cal M}_{11}^2 &  {\cal M}_{12}^2  \\
                                                     {\cal M}_{12}^2  &  {\cal M}_{22}^2  
                                                     \end{array}\right)
 \ \left( \begin{array}{c}
        -s_\alpha \\
            c_\alpha
             \end{array} \right)
  = m_h^2\ \left( \begin{array}{c}
                                                                                                                           -s_\alpha \\
                                                                                                                           c_\alpha
                                                                                                                           \end{array} \right)\ ,
\ee
giving rise to two equivalent representations for  $t_\alpha\equiv \tan\alpha$:
\be
\label{eq:talpha}
t_\alpha = \frac{{\cal M}_{12}^2}{{\cal M}_{11}^2-m_h^2} =\frac{{\cal M}_{22}^2-m_h^2} {{\cal M}_{12}^2}\ .
\ee
The equivalence of the two representations is guaranteed by the characteristic equation, ${\rm Det}({\cal M}^2 -m_h^2\ { I})=0$, where $I$ is the $2\times 2$ identity matrix. Moreover, since
\be
\label{eq:massorder}
m_h^2 \le {\cal M}_{ii}^2 \le m_H^2 \ , \qquad {\rm for} \ \ \  i=1,2 \ ,
\ee
due to the ``level repulsion" of eigenvalues of symmetric matrices, in both representations  ${\rm Sign}(t_\alpha) = {\rm Sign}({\cal{M}}^2_{12})$, consistent with the sign choices specified above.

Eq.~(\ref{eq:talpha}) allows us to solve for the mixing angle, $\alpha$, in terms of $\{{\cal M}_{11}^2, {\cal M}_{12}^2, m_h^2\}$ or $\{{\cal M}_{22}^2, {\cal M}_{12}^2, m_h^2\}$, depending on one's preference. For example, in the sign choice (I) we have the following two representations:
 \bea
 \label{eq:salpha1}
  s_\alpha& =& \frac{{\cal M}_{12}^2}{\sqrt{({\cal M}_{12}^2)^2 +({\cal M}_{11}^2-m_h^2)^2}}\ , \quad \quad
 m_H^2= \frac{{\cal M}_{11}^2({\cal M}^2_{11}-m_h^2)+ ({\cal M}^2_{12})^2}{{\cal M}^2_{11}-m_h^2} , \\
 \label{eq:salpha2}
  s_\alpha &=& {\rm Sign}({\cal M}_{12}^2)\frac{{\cal M}_{22}^2 - m_h^2}{\sqrt{({\cal M}_{12}^2)^2 +({\cal M}_{22}^2-m_h^2)^2}}\ , \quad
 m_H^2= \frac{{\cal M}_{22}^2({\cal M}^2_{22}-m_h^2)+ ({\cal M}^2_{12})^2}{{\cal M}^2_{22}-m_h^2} ,
 \eea
 where the expression for $m_H^2$ follows from solving for the corresponding eigenvalue equation for $m_H^2$. 
 
One can verify that Eqs.~(\ref{eq:salpha1}) and (\ref{eq:salpha2}) lead to the expected limiting behavior  when ${\cal M}_{12}^2\to 0$. For example, for Eq.~(\ref{eq:salpha1}), if ${\cal M}_{11}^2 > {\cal M}_{22}^2$, the smaller mass eigenvalue, $m_h^2$, is given by ${\cal M}_{22}^2$. Then in Eq.~(\ref{eq:salpha1}) we have $s_\alpha\to 0$ and $m_H^2\to {\cal M}_{11}^2$.  As expected the lightest CP-even Higgs is mostly $\Phi_2$ in this case. On the other hand, if ${\cal M}_{11}^2 < {\cal M}_{22}^2$ then $h$ is mostly $\Phi_1$ and $s_\alpha\to 1$, since
 \be
 ({\cal M}_{11}^2-m_h^2)= \frac{ ({\cal M}^2_{12})^2}{|{\cal M}_{11}^2-{\cal M}_{22}^2|} + {\cal O}\left(({\cal M}^2_{12})^4\right)\ ,
 \ee
 which also implies $m_H^2\to {\cal M}_{22}^2$ in this case. The behavior of Eq.~(\ref{eq:salpha2}) can be verified in a similar fashion.

%%%%%%%%%%%%%%%%%%%%%%%%%%%%%%%%%%%%%%%
\subsection{Higgs Couplings}
\label{sect:3A}
%%%%%%%%%%%%%%%%%%%%%%%%%%%%%%%%%%%%%%%

The Higgs boson couplings to gauge bosons in 2HDMs follow from gauge invariance and have the same parametric dependence on the CP-even mixing angle, $\alpha$, and the angle $\beta$ in any 2HDM, namely,
\be
g_{hVV} =  s_{\beta-\alpha}\, {g}_{V} \ , \qquad g_{HVV} = c_{\beta-\alpha}\, g_{V}  \ ,
\label{hVV}
\ee
where $g_{V}= 2i m_V^2/v$ is the SM value for $V=W,Z$ bosons. 

The fermion couplings, on the other hand, take different forms in different 2HDMs. However, it is common to require the absence of tree-level flavor-changing neutral currents (FCNC) by imposing the Glashow-Weinberg condition \cite{Glashow:1976nt}. This condition requires fermions with the same quantum numbers to couple to a single Higgs doublet and leads to four different types of 2HDMs \cite{Branco:2011iw}.\footnote{Typically the Glashow-Weinberg condition requires a discrete symmetry: $\Phi_1 \to -\Phi_1$, which demands $\lambda_6=\lambda_7=0$ in the general scalar potential given in Eq.~(\ref{eq:generalpotential}).} Amongst them the most popular ones are: the type I model, where all SM fermions couple to one doublet, and the type II model, where the  up-type fermions couple to one doublet and  down-type fermions  couple to the other. In one of the other two models, up-type quarks and leptons couple to the same doublet, while down-type quarks couple to the other. The remaining one  has all the quarks coupled to one Higgs doublet while the leptons couple to the other one.  In what follows we base the discussion on the type II model, although  our analysis can be easily adapted to all four types of 2HDMs. 

In type II models, where at tree-level $\Phi_1$ and $\Phi_2$ only couple to down-type and up-type fermions, respectively, the tree-level Higgs couplings to fermions are
\bea
\label{eq:ghdd}
g_{hdd} &=& -\frac{s_{\alpha}}{c_\beta} \, g_{f} = (s_{\beta-\alpha} - t_\beta \,c_{\beta-\alpha}) \, g_f\ , \quad  g_{huu} = \frac{c_{\alpha}}{s_\beta} \, g_{f}=(s_{\beta-\alpha} + t_\beta^{-1}\, c_{\beta-\alpha}) \, g_{f} \ ,\\
\label{eq:gHdd}
g_{Hdd} &=& \phantom{-}\frac{c_{\alpha}}{c_\beta} \, g_{f} =(c_{\beta-\alpha} + t_\beta\, s_{\beta-\alpha}) \, g_{f} \ , \quad  g_{Huu} = \frac{s_{\alpha}}{s_\beta} \, g_{f}= (c_{\beta-\alpha} - t_\beta^{-1}\, s_{\beta-\alpha}) \, g_{f}  \ ,
\label{hff}
\eea
where $g_f=im_f/v$ is the  coupling of the Higgs to the corresponding fermions  in the SM. 

We are interested in the alignment limit, where the lightest CP-even Higgs mimics the SM one. We will begin by solving for the conditions for which the Higgs couplings to fermions have the same magnitude as in the SM: $|g_{huu}/g_f|=|g_{hdd}/g_f|=1$. There are four possibilities, which can be divided in two cases: 
\begin{eqnarray}
\mbox{i}) &&\quad g_{hdd}=g_{huu}= \pm g_f , \nonumber\\
\mbox{ii})&&\quad g_{hdd}=-g_{huu}= \pm g_f. \nonumber
\end{eqnarray}
Demanding case i) leads to 
\be
s_\alpha = \mp c_\beta \ , \qquad c_\alpha = \pm s_\beta \ ,
\ee
which then implies
\be
\label{eq:alignc}
c_{\beta-\alpha} = 0 \qquad {\rm and} \qquad s_{\beta-\alpha} = \pm 1\ .
\ee
Couplings of the CP-even Higgs bosons now become
\be
 g_{hVV}\to \pm g_V  \ , \  g_{hff} \to \pm g_{f}  \ , \ g_{HVV} \to 0 \ , \ g_{Hdd} \to \pm t_\beta \,  g_f \ ,\  g_{Huu} \to \mp t_\beta^{-1} \, g_f \ ,\\
\ee
where the upper and lower signs correspond to $s_{\beta-\alpha}=1$ and $-1$, respectively.  This is the alignment limit. The heavy CP-even Higgs couplings to SM gauge bosons vanish in this limit since  it does not acquire a VEV. In other words, the alignment limit is the limit where the mass eigenbasis in the CP-even sector coincides with the basis where the gauge bosons receive all of their masses from one of the doublets.  As such, the non-SM-like CP-even Higgs does not couple to the gauge bosons at the tree-level. However, in this basis $H$ still has non-vanishing couplings to SM fermions. This feature remains true in all four types of 2HDMs, as can be seen, for example, by inspecting Table 2 in Ref.~\cite{Craig:2013hca}. It is important to observe that $s_{\beta-\alpha}=\pm 1$ results in an overall sign difference in the couplings of the SM-like Higgs and, hence,   has no physical consequences. 

On the other hand,  fulfillment of case ii) requires 
\be
\label{requirement}
s_\alpha = \mp c_\beta \ , \qquad c_\alpha = \mp s_\beta \ ,
\ee
which gives
\be
c_{\beta-\alpha} =  \mp s_{2\beta} \ , \qquad s_{\beta-\alpha} = \pm c_{2\beta} \ .
\ee
We see that the $hVV$ coupling does not tend to the SM value in this case and alignment is not reached. However, in the limit $t_\beta \gg 1$, 
\be
s_{2\beta} =  \frac{2t_\beta}{1+t_\beta^2} \approx   \frac{2}{t_\beta} \ , \quad
c_{2\beta} = \frac{1-t_\beta^2}{1+t_\beta^2}  \approx -1\ ,
\ee
we observe that the CP-even Higgs couplings become, to linear order in $t_\beta^{-1}$,
\bea
g_{hVV}&=&\mp g_V\ , \quad   g_{hdd}=\pm g_f\ , \quad g_{huu}= \mp g_{f}\ ,  \\
 g_{HVV} &=&\mp 2 t_\beta^{-1} g_V\ ,\quad  g_{Hdd} = \mp t_\beta g_f\ , \quad g_{Huu} = \mp t_\beta^{-1} g_f\ .
\eea
Hence, if Eq.~(\ref{requirement}) is required, one obtains that the lightest CP-even Higgs couplings to down-type fermions have the opposite sign as compared to its couplings to  both the vector bosons and up-type fermions, although all couplings have the same strength as in the SM. If, instead of the large $t_\beta$ limit, one takes $t_\beta \ll 1$, then it is straightforward to check that $g_{huu}$ has the opposite sign to $g_{hdd}$ and $g_{hVV}$. It is worth noting that, in type II 2HDMs, $t_\beta \ll 1$ leads to an unacceptably large top Yukawa coupling and should be avoided. However, the scenario of ``wrong-sign'' down-type fermion couplings of the SM-like Higgs in the large $t_\beta$ limit is clearly of phenomenological importance. A detailed study of this scenario is beyond the scope of the present work.

Similar arguments can be made in the case in which it is the heavy Higgs that behaves as the SM Higgs. For this to occur, 
\be
s_{\beta-\alpha} = 0
\ee
and therefore $c_{\beta-\alpha} = \pm 1$.  In the following, we shall concentrate on the most likely case  in which the lightest CP-even Higgs satisfies the alignment condition. The heavy Higgs case can be treated in an analogous way.

We also comment on the $Hhh$ coupling since it may have a significant impact on strategies in direct searches \cite{Craig:2013hca}. The coupling of the heavy Higgs to the lightest Higgs is given by
\begin{eqnarray} 
\label{gHhh}
g_{Hhh} & = & \frac{v}{4} \left[ -12 \ \lambda_1 c_\beta  c_\alpha s_\alpha^2 - 12 \ \lambda_2  s_\beta s_\alpha c_\alpha^2 
 +    \tilde{\lambda}_3 (-4 c_{\alpha - \beta}  + 6 s_{2\alpha} s_{\alpha+\beta})   \right.
\nonumber\\
&  &\left.\phantom{\frac{v}4} + 3 \lambda_6 (-4 s_\alpha^2 s_{\alpha+\beta} + 8 s_\alpha c_\alpha^2 c_\beta) 
  +   3 \lambda_7 (8 s^2_\alpha c_\alpha s_\beta - 4 c_\alpha^2 s_{\alpha +\beta}) \right]\ .
\end{eqnarray} 
One can rewrite Eq.~(\ref{gHhh}) as
\begin{eqnarray}
\label{eq:gHhh1}
g_{Hhh} & = & - 3 v s_\beta c_\beta^3 \left\{ \left[ s_{\alpha \beta} c_{\alpha \beta} \left( \lambda_1 s_{\alpha \beta} + \tilde{\lambda}_3 t_\beta^2 c_{\alpha \beta}
+ \lambda_6 t_\beta ( 2 c_{\alpha \beta} + s_{\alpha \beta} )\right) + \lambda_7 t_\beta^3 c_{\alpha \beta}^3 \right] \right.
\nonumber\\
& - & \left.  \left[ s_{\alpha \beta} c_{\alpha \beta} \left( \lambda_2 t_\beta^2 c_{\alpha \beta} + \tilde{\lambda}_3  s_{\alpha \beta}
+ \lambda_7 t_\beta ( 2 s_{\alpha \beta} + c_{\alpha \beta} )\right) + \lambda_6 t_\beta^{-1} s_{\alpha \beta}^3 \right] \right\} - \tilde{\lambda}_3 c_{\alpha-\beta} ,
\end{eqnarray}
where $s_{\alpha \beta} \equiv (-s_\alpha/c_\beta)$ and $c_{\alpha \beta} \equiv (c_\alpha/s_\beta)$ tend to 1 in the alignment limit.  We shall demonstrate in the next section that the alignment conditions in general 2HDMs imply that the $Hhh$ coupling vanishes.

%%%%%%%%%%%%%%%%%%%%%%%%%%%%%%%%%%%%%%%
%%%%%%%%%%%%%%%%%%%%%%%%%%%%%%%%%%%%%%%
\section{Alignment without decoupling}
\label{sect:3}
%%%%%%%%%%%%%%%%%%%%%%%%%%%%%%%%%%%%%%%
%%%%%%%%%%%%%%%%%%%%%%%%%%%%%%%%%%%%%%%

%%%%%%%%%%%%%%%%%%%%%%%%%%%%%%%%%%%%%%%
\subsection{Derivation of the Conditions for Alignment}  
%%%%%%%%%%%%%%%%%%%%%%%%%%%%%%%%%%%%%%%

One of the main results of this work is to find  the generic conditions for obtaining alignment without decoupling. The decoupling limit, where the low-energy spectrum contains only the SM and no new light scalars, is only a subset of the more general alignment limit  in Eq.~(\ref{eq:alignc}). In particular, quite generically, there exist regions of parameter space where one attains the alignment limit with new light scalars not far above $m_h=125$ GeV.  

It is instructive to first derive the alignment limit in the usual decoupling regime but in a slightly different manner.  Consider the eigenvalue equation of the CP-even Higgs mass matrix, Eq.~(\ref{eq:eigeneq}), which,  using  Eq.~(\ref{eq:cpemass}), becomes
\be
\label{eq:mastereq}
   \left(\begin{array}{cc}
                                                    s_\beta^2 & -s_\beta c_\beta \\
                                                     -s_\beta c_\beta & c_\beta^2 
                                                     \end{array}\right)  \  \left( \begin{array}{c}
        -s_\alpha \\
            c_\alpha
             \end{array} \right)
  =  -\frac{ v^2}{m_A^2}    \left(\begin{array}{cc}
                                                            L_{11} & L_{12} \\
                                                            L_{12} & L_{22}
                                                            \end{array}\right) \ 
                                                             \  \left( \begin{array}{c}
        -s_\alpha \\
            c_\alpha
             \end{array} \right)+ \frac{m_h^2}{m_A^2}\ \left( \begin{array}{c}   
                                                                                                                           -s_\alpha \\
                                                                                                                           c_\alpha
                                                                                                                           \end{array} \right)
       \ .
\ee 
Decoupling  is defined by taking all non-SM-like scalar masses to be much heavier than  the SM-like Higgs mass, $m_A^2 \gg v^2, m_h^2$. Then we see that at leading order in $v^2/m_A^2$ and $m_h^2/m_A^2$, the right-hand side of Eq.~(\ref{eq:mastereq}) can be ignored, and the eigenvalue equation reduces to 
\be
\label{eq:alignlimit}
 \left(\begin{array}{cc}
                                                    s_\beta^2 & -s_\beta c_\beta \\
                                                     -s_\beta c_\beta & c_\beta^2 
                                                     \end{array}\right)  \  \left( \begin{array}{c}
        -s_\alpha \\
            c_\alpha
             \end{array} \right)
  \approx 0 \ ,
  \ee
leading to the well-known decoupling limit \cite{Gunion:2002zf}: $c_{\beta - \alpha} = 0$. This is also exactly the alignment limit. 

Here we make the key observation that while decoupling achieves alignment by neglecting the right-hand side of Eq.~(\ref{eq:mastereq}),  alignment can also be obtained if the right-hand side of Eq.~(\ref{eq:mastereq}) vanishes {\em identically}, independent of $m_A$:
\be
\label{eq:alignmentmatrix}
 v^2    \left(\begin{array}{cc}
                                                            L_{11} & L_{12} \\
                                                            L_{12} & L_{22}
                                                            \end{array}\right) \ 
                                                               \left( \begin{array}{c}
        -s_\alpha \\
            c_\alpha
             \end{array} \right)
  = m_h^2\ \left( \begin{array}{c}   
                                                                                                                           -s_\alpha \\
                                                                                                                           c_\alpha
                                                                                                                           \end{array} \right)\ .
\ee       
More explicitly, since $s_\alpha = -c_\beta$ in the alignment limit, we can re-write the above matrix equation as two algebraic equations:~\footnote{The same conditions can also be derived using results presented in Ref.~\cite{Gunion:2002zf}.}
 \bea
 \label{SMconditionI}
{\rm(C1)}&:& m_h^2 = v^2 L_{11} + t_\beta v^2 L_{12} = v^2\left( \lambda_1 c_\beta^2+ 3\lambda_6 s_\beta c_\beta +\tilde{\lambda}_3 s_\beta^2 + \lambda_7 t_\beta s_\beta^2\right) \ , \\
 \label{SMconditionII}
{\rm (C2)}&:& m_h^2 =  v^2 L_{22} + \frac1{t_\beta} v^2 L_{12} = v^2\left( \lambda_2 s_\beta^2+ 3\lambda_7 s_\beta c_\beta +\tilde{\lambda}_3 c_\beta^2 + \lambda_6 t_\beta^{-1} c_\beta^2\right)\ .
 \eea
 Recall that  $\tilde{\lambda}_3=(\lambda_3+\lambda_4+\lambda_5)$. In the above $m_h$  is the SM-like Higgs mass, measured to be about 125 GeV,  and $L_{ij}$ is known once a model is specified. Notice that (C1) depends on all the quartic couplings in the scalar potential except $\lambda_2$, while (C2) depends on all the quartics but $\lambda_1$.  If there exists a $t_\beta$ satisfying the above equations, then the alignment limit would occur for arbitrary values of $m_A$ and {\it does not} require non-SM-like scalars to be heavy!

 Henceforth we will consider the coupled equations given in Eqs.~(\ref{SMconditionI}) and (\ref{SMconditionII}) as required conditions for alignment. When the model parameters satisfy them,  the lightest CP-even Higgs boson behaves exactly like a SM Higgs boson even if the non-SM-like scalars are light. A detailed analysis of the physical solutions will be presented in the next Section.

%%%%%%%%%%%%%%%%%%%%%%%%%%%%%%%%%%%%%%%
\subsection{Departure from Alignment}
\label{sect:3c}
%%%%%%%%%%%%%%%%%%%%%%%%%%%%%%%%%%%%%%%

Phenomenologically it seems  likely that alignment will only be realized approximately, rather than exactly. Therefore it is important to consider small departures from the alignment limit, which we do in this subsection. 

Since the alignment limit is characterized by $c_{\beta-\alpha}=0$, it is customary to parametrize the departure from alignment by considering a Taylor-expansions in $c_{\beta-\alpha}$ \cite{Gunion:2002zf, Craig:2013hca}, which defines the deviation of the  $g_{hVV}$ couplings from the SM values. However, this parametrization has the drawback that deviations in the Higgs coupling to down-type fermions are really controlled by $t_\beta \, c_{\beta-\alpha}$, which could be ${\cal O}(1)$ when $t_\beta$ is large. Therefore, we choose to parametrize the departure from the alignment limit by a parameter $\eta$ which is related to $c_{\beta - \alpha}$ by 
\be
c_{\beta-\alpha} = t_\beta^{-1} \eta \ , \qquad s_{\beta-\alpha} = \sqrt{1-t_\beta^{-2} \eta^2} \ .
\ee 
Then at leading order in $\eta$, the Higgs couplings become \bea
g_{hVV} &\approx& \left(1-\frac12 t_\beta^{-2} \eta^2 \right) g_V \ , \quad\quad 
   g_{HVV} \approx t_\beta^{-1}\eta \ g_V \ , \\
\label{eq:ghddepsilon}
g_{hdd} &\approx&  (1 -  \eta) \, g_f \ ,\quad \quad\quad \quad \quad \;\; \;
g_{Hdd} \approx t_\beta (1+t_\beta^{-2} \eta) g_{f}   \ , \\
  g_{huu}&\approx& (1+t_\beta^{-2} \eta) \, g_f  \ ,\quad \quad \quad \quad \;\;
  g_{Huu} \approx -t_\beta^{-1}(1-\eta) g_f \ .
\eea
We see $\eta$ characterizes the departure from the alignment limit of not only $g_{hdd}$ but also $g_{Huu}$. On the other hand, the deviation in the $g_{huu}$ and $g_{Hdd}$ are given by $t_\beta^{-2}\eta$, which is doubly suppressed in the large $t_\beta$ regime. Moreover, terms neglected above are of order $\eta^2$ and are never multiplied by positive powers of $t_\beta$, which could invalidate the expansion in $\eta$ when $t_\beta$ is large.

There are some interesting features regarding the pattern of deviations. First, whether the coupling to fermions is suppressed or enhanced relative to the SM values, is determined by the sign of $\eta$: $g_{hdd}$ and $g_{Huu}$ are suppressed (enhanced) for positive (negative) $\eta$, while the trend in $g_{huu}$ and $g_{Hdd}$ is the opposite. In addition, as $\eta\to$  0, the approach to the SM values is the fastest in $g_{hVV}$ and the slowest in $g_{hdd}$. This is especially true in the large $t_\beta$ regime, which motivates focusing on precise measurements of $g_{hdd}$ in type II 2HDMs.

Our parametrization of $c_{\beta-\alpha}=t_\beta^{-1}\eta$ can also be obtained by modifying Eq.~(\ref{eq:alignlimit}), which defines the alignment limit, as follows:
\be
\label{eq:nearalignlimit}
 \left(\begin{array}{cc}
                                                    s_\beta^2 & -s_\beta c_\beta \\
                                                     -s_\beta c_\beta & c_\beta^2 
                                                     \end{array}\right)  \  \left( \begin{array}{c}
        -s_\alpha \\
            c_\alpha
             \end{array} \right)
  = t_\beta^{-1} \eta \left( \begin{array}{c}
        -s_\beta \\
           c_\beta
             \end{array} \right) \ .
  \ee
The eignevalue equation for $m_h$ in Eq.~(\ref{eq:alignmentmatrix}) is modified accordingly,
\be
 v^2    \left(\begin{array}{cc}
                                                            L_{11} & L_{12} \\
                                                            L_{12} & L_{22}
                                                            \end{array}\right) \ 
                                                               \left( \begin{array}{c}
        -s_\alpha \\
            c_\alpha
             \end{array} \right)
  = m_h^2\ \left( \begin{array}{c}   
                                                                                                                           -s_\alpha \\
                                                                                                                           c_\alpha
                                                                                                                           \end{array} \right)
-
 m_A^2\,t_\beta^{-1} \eta\ \left( \begin{array}{c}   
                                                                                                                           -s_\beta \\
                                                                                                                           c_\beta
                                                                                                                           \end{array} \right)
\ .
\ee
From the above, taking $\eta \ll 1$ and expanding to first order in $\eta$, we obtain the ``near-alignment conditions'',
 \bea
\label{eq:nearalign1}
{\rm(C1')}&:& m_h^2 = v^2 L_{11} + t_\beta v^2 L_{12}+\eta \left[ t_\beta (1+t_\beta^{-2}) v^2 L_{12}-m_A^2 \right] \ , \\
\label{eq:nearalign2}
{\rm (C2')}&:& m_h^2 =  v^2 L_{22} +{t_\beta}^{-1} v^2 L_{12}-\eta \left[ t_\beta^{-1} (1+t_\beta^{-2}) v^2 L_{12}-m_A^2 \right] \ .
 \eea
We will return to study these two conditions  in the next section, after first analyzing solutions for alignment without decoupling in general 2HDMs. 
 
%%%%%%%%%%%%%%%%%%%%%%%%%%%%%%%%%%%%%%%
%%%%%%%%%%%%%%%%%%%%%%%%%%%%%%%%%%%%%%% 
 \section{Alignment in General 2HDM}
 \label{sect:4}
 %%%%%%%%%%%%%%%%%%%%%%%%%%%%%%%%%%%%%%%
%%%%%%%%%%%%%%%%%%%%%%%%%%%%%%%%%%%%%%%

 In what follows we solve for the alignment conditions (C1) and (C2), assuming all the scalar couplings are independent of $t_\beta$. This is not true in general, as radiative corrections to the scalar potential often introduce a $t_\beta$ dependence in the quartic couplings that are not present at the tree-level. However, this assumption allows us to analyze the solutions analytically and obtain the necessary intuition to understand more complicated situations.

 When all the quartics are independent of $t_\beta$, the conditions (C1) and (C2) may be re-written as cubic equations in $t_\beta$, with coefficients that depend on $m_h$ and  the quartic couplings in the scalar potential,
 \bea
 \label{eq:alignmentexpI}
{\rm(C1)}&:& (m_h^2 - \lambda_1 v^2 ) + (m_h^2 - \tilde{\lambda}_3 v^2) t_\beta^2 = v^2(3\lambda_6 t_\beta + \lambda_7 t_\beta^3) \ ,\\
 \label{eq:alignmentexpII}
{\rm(C2)}&:& (m_h^2 - \lambda_2 v^2 ) + (m_h^2 - \tilde{\lambda}_3 v^2) t_\beta^{-2} = v^2(3\lambda_7 t_\beta^{-1} + \lambda_6 t_\beta^{-3}  )   \ .
 \eea
Alignment without decoupling occurs only if there is (at least) a common physical solution for $t_\beta$ between the two cubic equations.\footnote{Since $t_\beta >0$ in our convention, a physical solution means a real positive root of the cubic equation.} From this perspective it may appear that alignment without decoupling is a rare and fine-tuned phenomenon. However, as we will show below, there are  situations where a common physical solution would exist between (C1) and (C2) without fine-tuning. 

Regarding the coupling of the heaviest CP-even Higgs to the lightest one, 
it is now easy to see from Eqs.~(\ref{eq:alignmentexpI}) and (\ref{eq:alignmentexpII}) that each term inside the square brackets in Eq.~(\ref{eq:gHhh1}) tends to  $m_h^2 (1 + t_\beta^2)/v^2$ in the alignment limit, and hence, as stated in Ref.~\cite{Craig:2013hca}, $g_{Hhh}$ vanishes.

%%%%%%%%%%%%%%%%%%%%%%%%%%%%%%%%%%%%%%%
\subsection{Alignment for Vanishing Values of $\lambda_{6,7}$}
%%%%%%%%%%%%%%%%%%%%%%%%%%%%%%%%%%%%%%%

It is useful to consider solutions to the alignment conditions (C1) and (C2) when $\lambda_6=\lambda_7=0$ and $\lambda_1 = \lambda_2$, which can be enforced by the symmetries $\Phi_1 \to -\Phi_2$ and $\Phi_1 \to \Phi_2$. Then (C1) and (C2) collapse into  quadratic equations: 
 \bea
 \label{eq:quadI}
{\rm(C1)}&\to& (m_h^2 - \lambda_1 v^2 ) + (m_h^2 - \tilde{\lambda}_3 v^2) t_\beta^2 = 0\ ,\\
 \label{eq:quadII}
{\rm(C2)}&\to&  (m_h^2 - \tilde{\lambda}_3 v^2) +(m_h^2 - \lambda_1 v^2 ) t_\beta^{2}  = 0 \ .
 \eea
We see that a solution exists for $t_\beta=1$ whenever 
\be
\label{eq:finetune0}
 \lambda_{\rm SM} = \frac{\lambda_1 + \tilde{\lambda}_3}{2} \, ,
\ee
where  we have expressed the SM-like Higgs mass as  
\be
 m_h^2 = \lambda_{\rm SM} v^2 \ .
 \ee
 From Eq.~(\ref{eq:finetune0}) we see  that the above solution, $t_\beta=1$, is obviously special, since it demands $\lambda_{\rm SM}$ to be the average  of $\lambda_1$ and $\tilde{\lambda}_3$.

We next relax the $\lambda_1 = \lambda_2$ condition while still keeping $\lambda_6=\lambda_7=0$. Recall that the Glashow-Weinberg condition \cite{Glashow:1976nt} on the absence of tree-level FCNC requires a discrete symmetry, $\Phi_1 \to -\Phi_1$, which enforces $\lambda_6=\lambda_7=0$ at the tree-level. The two quadratic equations have a common root if and only if the  determinant of the Coefficient Matrix of the two quadratic equations vanishes,
\be
\label{eq:finetune1}
{\rm Det} \left( \begin{array}{cc}
                         m_h^2 -\tilde{\lambda}_3 v^2 \ \ & m_h^2 -\lambda_1 v^2 \\
                         m_h^2 -\lambda_2 v^2 \ \ & m_h^2 -\tilde{\lambda}_3 v^2  
                         \end{array} \right) 
                         = 
                         (m_h^2 -\tilde{\lambda}_3 v^2)^2 - (m_h^2 -\lambda_1 v^2) (m_h^2 -\lambda_2 v^2) = 0 \ .
\ee
Then the positive root can be expressed in terms of $(\lambda_1, \tilde{\lambda}_3)$,
\be
\label{eq:l670}
 t_\beta^{(0)} =  \sqrt{\frac{{\lambda_1 -\lambda_{\rm SM}}}{\lambda_{\rm SM}-{\tilde{\lambda}_3 }}}\ .
 \ee
 We see from Eqs.~(\ref{eq:finetune1}) and (\ref{eq:l670}), that a real value of $t_\beta^{(0)}$ can exist only if the set of parameters $\{\lambda_{\rm SM}, \lambda_1, \lambda_2, \tilde{\lambda}_3\}$ has one of the two  orderings 
 \be
 \label{eq:special1}
 \lambda_1, \lambda_2  \ \ge \ \lambda_{\rm SM}\ \ge\ \tilde{\lambda}_3  \ ,
 \ee
 or
 \be
 \label{eq:special2}
 \lambda_1, \lambda_2  \ \le \ \lambda_{\rm SM}\ \le\ \tilde{\lambda}_3 \ .
 \ee
A solution for $t_\beta^{(0)}$ can be found using the following procedure: once one of the conditions in Eqs.~(\ref{eq:special1}) or (\ref{eq:special2}) is satisfied, Eq.~(\ref{eq:l670}) leads to the alignment solution  $t_\beta^{(0)}$ for a given $(\lambda_1, \tilde{\lambda}_3)$. However,  Eq.~(\ref{eq:finetune1}) must also be satisfied, which is then used to solve for the desired $\lambda_2$ so that $t_\beta^{(0)}$ is  a root of  (C2) as well.  More specifically, the relations
\be
\label{eq:tb024}
\lambda_2 - \lambda_{\rm SM} \  = \ \frac{\lambda_{\rm SM} - \tilde{\lambda}_3}{\left(t_\beta^{(0)}\right)^2} \ =  \ \frac{\lambda_1 - \lambda_{\rm SM}}{\left(t_\beta^{(0)}\right)^4}
\ee
must be fulfilled.  Therefore,  the alignment solution demands a specific relationship between the quartic couplings of the 2HDM.  In addition, it is clear from Eqs.~(\ref{eq:l670}) and~(\ref{eq:tb024}) that if all the quartic couplings are ${\cal O}(1)$, $t_\beta^{(0)}\sim {\cal O}(1)$ as well, unless $\tilde{\lambda}_3$  and $\lambda_2$ are  very close to $\lambda_{\rm SM}$, or $\lambda_1$ is taken to be much larger than $\lambda_{\rm SM}$. For examples, $t_\beta^{(0)}\sim 5$ could be achieved for $(\lambda_1, \tilde{\lambda}_3, \lambda_2) \sim (1., 0.23,0.261)$, or for  $(\lambda_1, \tilde{\lambda}_3, \lambda_2) \sim (5., 0.07,0.263)$.  Our discussion so far applies to alignment limit scenarios studied, for instance,  in Refs.~\cite{Delgado:2013zfa,Craig:2013hca}, both of which set  $\lambda_6=\lambda_7=0$.

%%%%%%%%%%%%%%%%%%%%%%%%%%%%%%%%%%%%%%%
\subsection{Alignment for Non-Zero $\lambda_{6,7}$}
%%%%%%%%%%%%%%%%%%%%%%%%%%%%%%%%%%%%%%%

The symmetry $\Phi_1 \to -\Phi_1$ leading to $\lambda_6 = \lambda_7 = 0$ is broken softly by $m_{12}$. Thus a phenomenologically interesting scenario is to consider small but non-zero $\lambda_{6,7}$. Therefore, in this subsection we study solutions to the alignment conditions (C1) and (C2) under the assumptions
\be
\lambda_{6,7} \ \ll \ 1\ .
\ee

Although general solutions of cubic algebraic equations exist, much insight can be gained  by first solving for the cubic roots of (C1) as a  perturbation to the quadratic solution $t_\beta^{(0)} $,
\bea
\label{eq:lno01}
  t_\beta^{(\pm)}&=& t_\beta^{(0)} \pm  \frac32 \frac{\lambda_6}{\lambda_{\rm SM}-\tilde{\lambda}_3}  \pm  \frac{\lambda_7(\lambda_1 -\lambda_{\rm SM})}{(\lambda_{\rm SM}-\tilde{\lambda}_3)^2}  + {\cal O}(\lambda_6^2,\lambda_7^2)\ .
\eea
The solutions $t_\beta^{(\pm)}$ lie in the same branch as $t_\beta^{(0)}$, to which they reduce in the limit $\lambda_{6,7}\to 0$. In addition, both solutions are again
 %of 
 ${\cal O}(1)$ given our assumptions. More importantly, similar to $t_\beta^{(0)}$, specific fine-tuned relations between the quartic couplings are required to ensure $t_\beta^{(\pm)}$ are also cubic roots of (C2). 

However, a new solution also appears,
\bea
 \label{eq:lno02}
  {t}_\beta^{(1)} &=& \frac{\lambda_{\rm SM}-\tilde{\lambda}_3}{\lambda_7}-  \frac{3\lambda_6}{\lambda_{\rm SM}-\tilde{\lambda}_3}-  \frac{\lambda_7(\lambda_1 -\lambda_{\rm SM})}{(\lambda_{\rm SM}-\tilde{\lambda}_3)^2} + {\cal O}(\lambda_6^2,\lambda_7^2)\ .
 \eea
The solution $t_\beta^{(1)}$ belongs to a new branch that disappears when $\lambda_7\to 0$ and exists provided the condition
\be
{\rm Sign}(\lambda_{\rm SM} - \tilde{\lambda}_3)  = {\rm Sign}( \lambda_7)
\label{eq:largetan}
\ee 
is satisfied.  For $|\lambda_{\rm SM} - \tilde{\lambda}_3| \gg |\lambda_7|$, as is natural due to the assumption $|\lambda_7| \ll 1$,  we are led to $t_\beta^{(1)} \gg 1$. As an example, for $(\lambda_1, \tilde{\lambda}_3,\lambda_6,\lambda_7)=(0.5, -0.1,0.01,0.01)$, one obtains  ${t}_\beta^{(1)}\sim 35$ by solving for the cubic root of (C1) exactly.  Lower values of $t_\beta^{(1)}= {\cal{O}}(10)$ may be obtained for somewhat larger values of $\lambda_7$ and/or larger values of $\tilde{\lambda}_3$. 

The $t_\beta^{(1)}$ solution is an example of alignment without decoupling that does not require fine-tuning. This is because the condition (C2), in the limits $\lambda_6, \lambda_7 \ll 1$ and $t_\beta \gg 1$, becomes insensitive to all quartic couplings but $\lambda_2$:
\be
\label{eq:l2mlsm}
 m_h^2 - \lambda_2 v^2  =  {\cal O}\left(\frac1{t_\beta^2}, \frac{\lambda_7}{t_\beta}, \frac{\lambda_6}{t_\beta^3}\right) \ .
 \ee
Unlike the fine-tuned relation in Eq.~(\ref{eq:finetune1}), in this case $\lambda_2$ is determined by the input parameter $m_h$, or equivalently  $\lambda_{\rm SM}$, and is insensitive to other quartic couplings in the scalar potential. Therefore, provided the condition given in Eq.~(\ref{eq:largetan}) is fulfilled, the value of the quartic couplings, $\tilde{\lambda}_3$ and $\lambda_7$, are still free parameters and thus can be varied,  leading to different values of $t_\beta$  for which alignment occurs. 

For the purpose of demonstration, let us again use the example below Eq.~(\ref{eq:largetan}), $(\lambda_1, \tilde{\lambda}_3,\lambda_6,\lambda_7)=(0.5, -0.1,0.01,0.01)$.  The condition that ${t}_\beta^{(1)}\sim 35$ is also a root of (C2) requires
\be
\lambda_2 \approx 0.26 + \left(\frac{\tilde{\lambda}_3}{-0.1}\right) \times 8\times 10^{-5} -\left( \frac{\lambda_7}{0.01}\right)\times 8 \times 10^{-4} - \left(\frac{\lambda_6}{0.01}\right)\times 8 \times 10^{-7} \ .
\ee
From this we see that the required value of  $\lambda_2$ is very insensitive to the values of the other quartic couplings in the potential, and is determined only by $m_h$. 

The solution $t_\beta^{(1)}$ is perhaps the most interesting among the three branches of solutions because its existence does not require specific relationships amongst the quartic couplings, and to our knowledge has never been studied in the literature. The crucial observation to arrive at this scenario of alignment without decoupling is to turn on small but non-vanishing $\lambda_7$, which arises automatically in 2HDMs without tree-level FCNC. In this case, we see alignment without decoupling is not only a generic feature of the model, but also a ``natural" phenomenon, and can occur at $t_\beta = {\cal{O}}(10)$ for which direct searches for non-standard Higgs bosons become difficult.  In fact, in the next section we will see that this solution can be realized in one of the most popular models for beyond the SM physics, the MSSM.

%%%%%%%%%%%%%%%%%%%%%%%%%%%%%%%%%%%%%%%
\subsection{Departure from Alignment}
%%%%%%%%%%%%%%%%%%%%%%%%%%%%%%%%%%%%%%%

So far we have analyzed solutions for the alignment conditions (C1) and (C2) in general 2HDMs. However, it is likely that the alignment limit, if realized in Nature at all, is only approximate and the value of $t_\beta$ does not need to coincide with the value at the exact alignment limit. It is therefore important to study the approach to alignment and understand patterns of deviations in the Higgs couplings in the ``near-alignment limit,'' which was introduced in Section \ref{sect:3c}.

Although we derived the near-alignment conditions (C1$^\prime$) and (C2$^\prime$) in Eqs.~(\ref{eq:nearalign1}) and (\ref{eq:nearalign2}) using the eigenvalue equations, it is convenient to consider the (near-)alignment limit from a slightly different perspective. Adopting the sign choice (I) in Eq.~(\ref{eq:signI}) and using the expression for the mixing angle, $\alpha$, in Eq.~(\ref{eq:salpha1}), we can re-write the $g_{hdd}$ and $g_{huu}$ couplings as follows
\bea
g_{hdd}&=&-\frac{s_\alpha}{c_\beta}\, g_f  =  \frac{\cal A}{\sqrt{{\cal A}^2 c_\beta^2 + {\cal B}^2 s_\beta^2}}\, g_f \ ,\\
\label{sinalphabeta}
g_{huu}&=&\phantom{-}\frac{c_\alpha}{s_\beta}\, g_f  =  \frac{{\cal B}}{\sqrt{{\cal A}^2 c_\beta^2 + {\cal B}^2 s_\beta^2}} \, g_f \ .
\eea
where
\begin{eqnarray}
 {\cal A} & = & -\frac{{\cal{M}}_{12}^2}{c_\beta} = \left(m_A^2- (\lambda_3 + \lambda_4) v^2 \right) s_\beta - \lambda_7 v^2  s_\beta t_\beta - \lambda_6 v^2 c_\beta  \ , \\
 {\cal B} & = & \frac{{\cal{M}}_{11}^2 - m_h^2}{s_\beta} = \left( m_A^2 + \lambda_5 v^2 \right) s_\beta + \lambda_1 v^2 \frac{c_\beta}{t_\beta} + 2 \lambda_6 v^2 c_\beta - \frac{m_h^2}{s_\beta}\ .
\end{eqnarray}
Again it is instructive to consider first taking the pseudo-scalar mass to be heavy: $m_A\to \infty$. In this limit we have ${\cal A} \to m_A^2 s_\alpha$ and ${\cal B} \to m_A^2 s_\alpha$, leading to $-s_\alpha/c_\beta \to 1$ and $c_\alpha/s_\beta \to 1$. We recover the familiar alignment-via-decoupling limit. On the other hand, alignment without decoupling could occur by setting directly
\be
\label{eq:AB}
{\cal A} = {\cal B}\, ,
\ee
where, explicitly,  
\be 
{\cal B}-{\cal A} = \frac1{s_\beta}\left(- m_h^2 + \tilde{\lambda}_3 v^2 s_\beta^2 + \lambda_7 v^2 s_\beta^2 t_\beta + 3 \lambda_6 v^2 s_\beta c_\beta + \lambda_1 v^2 c_\beta^2\right) =0\ ,
\ee
is nothing but the alignment condition (C1) in Eq.~(\ref{SMconditionI}). The alignment condition (C2) would be obtained if the representation in Eq.~(\ref{eq:salpha2}) is used instead, leading to ${\cal A}=-{\rm Sign}({\cal M}_{12}^2)({\cal M}_{22}^2 - m_h^2)/c_\beta$ and ${\cal B}= |{\cal M}_{12}^2|/s_\beta$. Further, $m_h$ is the mass of the lightest CP-even Higgs boson and ${\cal{M}}_{ii}^2 - m_h^2>0, \;i=\{ 1,2\}$ by Eq.~(\ref{eq:massorder}). Therefore  Eq.~(\ref{eq:AB}) implies
\be
{\cal A} \ge 0 \qquad {\rm and} \qquad {\cal B} \ge 0 \ 
\ee
at the alignment limit.

Now in the near-alignment limit, where the alignment is only approximate, one can  derive
\bea
g_{hdd}&=&  \frac{{\cal A}}{{\cal B}\sqrt{1-(1- {\cal A}^2/{\cal B}^2)c_\beta^2}} \, g_f 
\label{eq:ghdd11} \\
 &=& \left[ 1 - s_\beta^2 \left(1 - \frac{{\cal A}}{\cal B}\right)+ {\cal O}\left((1-{\cal A}/{\cal B})^2\right)\right]\, g_f \ ,
\label{sinalphabetaexp}
\eea
which, when comparing with Eq.~(\ref{eq:ghddepsilon}), implies
\be
\eta = s_\beta^2 \left(1- \frac{{\cal A}}{{\cal B}}\right) = s_\beta^2 \frac{{\cal B}-{\cal A}}{{\cal B}} \ .
\label{B-A}
\ee
Therefore, the $g_{hdd}$ coupling is enhanced (suppressed) if ${\cal B}-{\cal A}<0$ ($>0$). It is easy to verify that the above equation is identical to the near-alignment condition (C1$^\prime$) in Eq.~(\ref{eq:nearalign1}). The condition (C2$^\prime$) could again be obtained using Eq.~(\ref{eq:salpha2}).

%==============================================================================
\begin{figure}
\begin{center}
\includegraphics[width=0.44\textwidth]{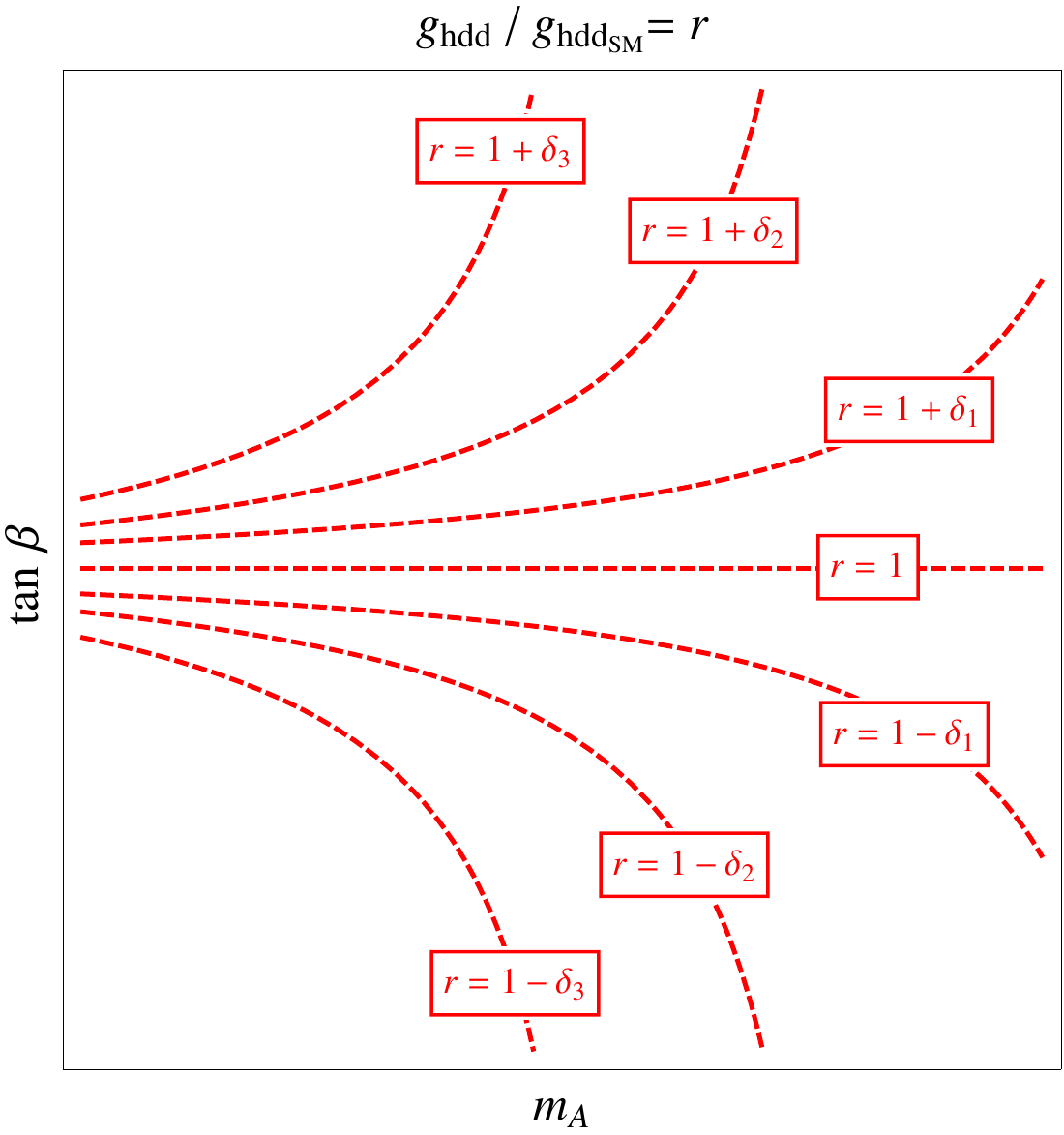}  \\
\end{center}
\caption{General behavior of contours with constant $g_{hdd}/g_{hdd_{\rm SM}} = r$ in the $m_A - \tan\beta$ plane. $r=1$ corresponds to the alignment limit. At constant $t_\beta$, moving toward smaller $m_A$ results in larger deviations from $r=1$. In the plot $\delta_i$ for $i=1,2,3$ can be 
either positive or negative and $|\delta_3| > |\delta_2| > |\delta_1|$.  }
\label{fig:generic}
\end{figure}
%==============================================================================

It is useful to analyze Eq.~(\ref{sinalphabetaexp}) in different instances. For example, when $\lambda_6 = \lambda_7 = 0$, one obtains
\be
g_{hdd} \simeq \left[ 1 + s_\beta \frac{ \left( \lambda_{\rm SM}  - \tilde{\lambda}_3 s_\beta^2 - \lambda_1 c_\beta^2 \right) v^2}{{\cal B}}\right] \, g_f \ .
\label{ghddlowtb}
\ee
Hence, for $\tilde{\lambda}_3 > \lambda_{\rm SM} > \lambda_1$, a suppression of $g_{hdd}$ will take place for values of $t_\beta$ larger than the ones necessary to achieve the alignment limit.  On the contrary, for $\lambda_1 > \lambda_{\rm SM} > \tilde{\lambda}_3$, larger values of $t_\beta$ will lead to an enhancement of $g_{hdd}$.

On the other hand, for $\lambda_7 \neq 0$ and large values of $t_\beta$, one obtains
\be
g_{hdd} \simeq \left[1 + s_\beta \frac{ \left( \lambda_{\rm SM} - \tilde{\lambda}_3 - \lambda_7 t_\beta \right) v^2}{{\cal B}}\right] \, g_f \ ,
\label{ghddlargetb}
\ee
which shows that for $\lambda_{\rm SM} > \tilde{\lambda}_3$ and $\lambda_7$ positive, $g_{hdd}$ is suppressed at values of $t_\beta$ larger than those necessary to obtain the alignment limit, and vice versa. 

One can in fact push the preceding analysis further by deriving the condition giving rise to a particular deviation from alignment. More specifically, the algebraic equation dictating the contour $g_{hdd}/g_f = r$, where $r \neq 1$, can be obtained by using Eq.~(\ref{eq:ghdd11}):
\be
m_A^2 = \frac{1}{R(\beta)-1}\frac{{\cal A}-{\cal B}}{s_\beta }  +  \frac{m_h^2}{s_\beta^2} -  v^2 \lambda_5 -  \lambda_1 v^2 t_\beta^{-2} - 2 \lambda_6 v^2 t_\beta^{-1} \ ,
\label{boundmA}
\ee
where
\bea
R(\beta) &=& \frac{t_\beta \, r}{\sqrt{1+t_\beta^2 - r^2}} \ .
\eea
When $r$ is close to unity, the above equation becomes
\bea
R(\beta) &\approx& 1+ \frac{r-1}{s_\beta^2}\ .
\eea
Several comments are in order. First, for $r \approx 1 -\eta$ with $\eta \ll 1$, $R(\beta)\approx 1+ \eta/s_\beta^2$. Second, once all the scalar quartic couplings are known, which in general could also depend on $t_\beta$,  Eq.~(\ref{boundmA}) gives the contour corresponding to $g_{hdd}/g_f = r$ in the $m_A - \tan\beta$ plane. Third, if we consider a slice of constant $t_\beta$ away from the alignment limit then larger values of $m_A$ correspond to values of $R(\beta)$,  and hence $r$, closer to 1. Therefore, large deviations from $r=1$ lie in regions with small $m_A$ and $t_\beta$ far from the alignment limit. These considerations allow for an understanding of the general behavior of contours with constant $r$ in the $m_A - \tan\beta$ plane, which is shown in Fig.~\ref{fig:generic}. To a large extent, the various examples we will consider later simply correspond to zooming in on Fig.~\ref{fig:generic}  in different regions of parameters of interest in representative scenarios like the MSSM and the NMSSM.   However, as we shall explain in the next section,  radiative corrections induce a departure of the MSSM Higgs sector from the type II 2HDM behavior.  This does not change the qualitative behavior shown in Fig.~\ref{fig:generic}, but leads to  a modification of the contours of constant $r$ at large values of $t_\beta$.

%%%%%%%%%%%%%%%%%%%%%%%%%%%%%%%%%%%%%%%
%%%%%%%%%%%%%%%%%%%%%%%%%%%%%%%%%%%%%%%
\section{Alignment in Supersymmetry}
\label{sect:5}
%%%%%%%%%%%%%%%%%%%%%%%%%%%%%%%%%%%%%%%
%%%%%%%%%%%%%%%%%%%%%%%%%%%%%%%%%%%%%%%

In this section we first give a detailed overview of the Higgs mass dependance on the general 2HDM quartics and the constraints this implies for the MSSM parameters, given $m_h \approx 125$ GeV. We then present detailed analyses of alignment without decoupling in the MSSM and in the NMSSM.

%%%%%%%%%%%%%%%%%%%%%%%%%%%%%%%%%%%%%%%
\subsection{MSSM Higgs Mass and Quartic Couplings}
%%%%%%%%%%%%%%%%%%%%%%%%%%%%%%%%%%%%%%%

 The tree-level Higgs sector in the MSSM belongs to the so-called type II 2HDM, where one doublet couples to the up-type fermions, denoted by $H_u$, and the other doublet couples to the down-type fermions, denoted by $H_d$. Both the tree-level and higher-order contributions to the CP-even mass matrix are well-known. At tree-level we have
\be
{\cal M}_{\rm MSSM,tree}^2  = \left(\begin{array}{cc}
                                                   m_A^2 s_\beta^2 + m_Z^2 c_\beta^2 &  -(m_A^2+m_Z^2) s_\beta c_\beta \\
                                                    -(m_A^2+m_Z^2) s_\beta c_\beta & m_A^2 c_\beta^2 + m_Z^2 s_\beta^2
                                                     \end{array}\right)\ .
\ee
Typically one is interested in the region where $m_A \ga m_Z$ and $t_\beta \agt 1$.\footnote{Values of $t_\beta \lesssim 1$  lead to such large values of the top-quark Yukawa coupling that the perturbative consistency of the theory is lost well below the grand unification scale, $M_G \simeq 2 \times 10^{16}$~GeV. In this work we shall  assume that $t_\beta$ is moderate or large so that the perturbativity of the top Yukawa  is not a concern.} Then  ${\cal M}_{11}^2 > {\cal M}_{22}^2$ in ${\cal M}_{\rm MSSM,tree}^2$  and it is conventional to use the sign choice (I) in Eq.~(\ref{eq:signI}), $-\pi/2 \le \alpha \le \pi/2$. In addition, ${\cal M}_{12}^2 <0$ at  tree-level and one is further restricted to $-\pi/2 \le \alpha \le 0$. However, beyond tree-level one could have $\alpha > 0$ in the MSSM.

Only four of the quartic couplings are non-zero at tree-level in the MSSM, 
\bea
\lambda_1&=&\lambda_2 = \frac14 (g_1^2+g_2^2) = \frac{m_Z^2}{v^2} \ , \\
\lambda_3&=&\frac14 (g_1^2-g_2^2) = -\frac{m_Z^2}{v^2} +\frac12 g_2^2 \ , \\
\lambda_4 &=& -\frac12 g_2^2  \ , \\
\lambda_5&=&\lambda_6=\lambda_7=0\ . 
\eea
 Therefore we see that the Higgs sector of the MSSM at tree-level is an example of 2HDMs in which the condition $\lambda_6 = \lambda_7 = 0$ and $\lambda_1 = \lambda_2$ is fulfilled.  Moreover,  $ \tilde{\lambda}_3<0$ and $\lambda_{1,2} = -\tilde{\lambda}_3 < \lambda_{\rm SM}$, so the alignment conditions, Eqs.~(\ref{eq:special1}) and (\ref{eq:special2}) cannot be fulfilled.  As a result, alignment without decoupling never happens at the tree-level  in the MSSM.

At the loop level, however, $\lambda_{1-4}$ are modified and, furthermore, the remaining three couplings $\lambda_{5-7}$ acquire non-zero values. These radiative corrections to the quartic couplings depend relevantly on the values of $t_\beta$.

At moderate or large values of $t_\beta\equiv v_u/v_d$, $H_u$ acquires a VEV, $v_u \alt v$, while $v_d \ll v$. Therefore, the SM-like Higgs is approximately identified with the real component of $H_u^0$, and its squared mass is approximately given by the (2,2) component of the CP-even Higgs mass matrix. More precisely, multiplying both sides of Eq.~(\ref{eq:eigeneq}) from the left by the row vector $(-s_\alpha, c_\alpha)$ and using the alignment relation, $(\beta -\alpha) =  \pi/2$, we obtain
\begin{eqnarray}
m_h^2   & =  & {\cal{M}}^2_{22} s_\beta^2 + 2 {\cal{M}}^2_{12} s_\beta c_\beta  + {\cal{M}}_{11} c_\beta^2
\nonumber\\
& = & v^2 \left( \lambda_2\, s_\beta^4 +  4 \lambda_7 s_\beta^3 c_\beta + 2  \tilde{\lambda}_3\ s_\beta^2 c_\beta^2 + 4\lambda_6 s_\beta c_\beta^3 +   \lambda_1 \, c_\beta^4 \right)\;,  
\nonumber\\
               & = & m_Z^2 c_{2\beta}^2 + v^2 \left(\Delta \lambda_2\, s_\beta^4 +  4 \lambda_7 s_\beta^3 c_\beta + 2 \Delta \tilde{\lambda}_3\ s_\beta^2 c_\beta^2 + 4\lambda_6 s_\beta c_\beta^3 +  \Delta \lambda_1 \, c_\beta^4 \right)\;,  
               \label{eq:mhl2}  
\end{eqnarray}
where the $\Delta \lambda_i$'s denote a change of the corresponding quartic coupling, $\lambda_i$, due to radiative corrections. Hence we see that the tree-level value of the  SM-like Higgs mass in the MSSM is bounded above by $m_Z$,
\begin{equation}
\left( m_h^2 \right)^{\rm tree} \le m_Z^2 c_{2\beta}^2\ .
\label{eq:mhtree}
\end{equation}
Since we are focusing on $t_\beta \agt 1$,  this upper bound is maximized for large values of $t_\beta$. It is well known that in the MSSM loop corrections to the quartic couplings are necessary to raise the SM-like Higgs mass from values below $m_Z$ to values consistent with the LHC measured value, $m_h \simeq 125$~GeV.  Note that since the upper bound on the tree-level $m_h$ is minimized for $t_\beta = 1$, the  radiative corrections  required to raise the Higgs mass from its small tree-level value must be very large for $t_\beta \sim1$. Such radiative corrections may only be obtained for very heavy scalar top-quarks, with masses far above the TeV scale.

At moderate or large values of $t_\beta$, $s_\beta \simeq 1$ and the Higgs mass is mostly governed by $\lambda_2$, as can be seen from Eq.~(\ref{eq:mhl2}), although other terms may become relevant for smaller values of $t_\beta$. The most important contributions to the quartic couplings come from the stop sector. When the two stop masses are close to each other, $(m_{\tilde{t}_2}^2 - m_{\tilde{t}_1}^2 )<  0.5 (m_{\tilde{t}_1}^2 + m_{\tilde{t}_2}^2)$,  one can approximate the Higgs mass  including the most relevant  two loop corrections, namely~\cite{Okada:1990vk,mhiggsRG1}
 \begin{eqnarray}
m_h^2& \simeq & m_Z^2 c_{2\beta}^2+\frac{3}{8 \pi^2} h_t^4 \ v^2 \left[ \frac{1}{2}\tilde{X}_t + t
+\frac{1}{16\pi^2}\left(\frac{3 h_t^2}{2}-32\pi\alpha_s
\right)\left(t\, \tilde{X}_t +t^2\right) \right]\,, \label{mhsm}
\end{eqnarray}
where $h_t$ is the top Yukawa coupling, $M_{\rm SUSY}^2 = (m^2_{\tilde{t}_1} + m^2_{\tilde{t}_2})/2$  and $t=\log { M_{\rm SUSY}^2}/{m_t^2}$. The parameter $\tilde{X_t}$ is defined as
\begin{eqnarray}
\label{stopmix}
\tilde{X}_{t} & =& \frac{2 \tilde{A}_t^2}{M_{\rm SUSY}^2}
                  \left(1 - \frac1{12}\frac{\tilde{A}_t^2}{ M_{\rm SUSY}^2} \right)\;,
\nonumber \\
\tilde{A}_t & = & A_t-\mu\cot\beta\; ,
\end{eqnarray}
where $A_t$ is the trilinear Higgs-stop coupling, $\mu$ is the Higgsino mass parameter and the running couplings in the $\overline{\rm MS}$ scheme must be evaluated at the top quark mass scale.  

In addition to the stop sector,  $m_h^2$ also receives negative radiative corrections proportional to the fourth power of the bottom and/or tau Yukawa couplings. However, these corrections become relevant only at very large values of $t_\beta$, where the bottom and/or tau Yukawa couplings become comparable to the top Yukawa (see, for example, Ref.~\cite{Carena:2011aa})
\begin{equation}
\Delta m_h^2 \simeq
 - \frac{v^2 h_b^4 \mu^4}{32 \pi^2 M_{\rm SUSY}^4} - \frac{v^2h_\tau^4 \mu^4}{96 \pi^2 M_{\tilde{\tau}}^4} \; .
\end{equation}
where $\mu$ is the higgsino mass parameter.
In the above $M_{\tilde{\tau}}^2$ is the average  stau mass-squared and $h_{b} (h_\tau)$ is the Yukawa coupling for the bottom quark ($\tau$ lepton). We have also assumed  the sbottom masses to be of the same order as the stop masses and,  for simplicity, neglected higher loop corrections.

%%%%%%%%%%%%%%%%%%%%%%%%%%%%%%%%
\subsection{Couplings of the Down -Type Fermions to the Higgs in the MSSM}
%%%%%%%%%%%%%%%%%%%%%%%%%%%%%

It is also important to recall that the MSSM Higgs sector is a type II 2HDM only at the tree-level. Beyond tree-level, however, supersymmetry breaking effects induce $H_u$ couplings to $down$-type fermions, denoted by $\Delta h_d$. These loop-induced couplings modify the relation between the down-type fermion Yukawa couplings and their running masses, namely
\begin{equation}
h_{b/\tau} \simeq \frac{\sqrt{2}~m_{b/\tau} }{v c_\beta(1 +  \epsilon_{b/\tau}t_\beta)}\;,
\label{hbtau}
\end{equation}
where $\epsilon_{b/\tau} = (\Delta h_{b/\tau}/h_{b/\tau})$ are  the one-loop corrections whose dominant contribution depends on the sign of $(\mu M_{3})$ and $(\mu M_{2})$, respectively\cite{deltamb2,deltamb}. Positive values of $(\mu M_3)$ induce positive contributions to $\epsilon_b$ which reduces the bottom Yukawa coupling, $h_b$. Smaller values of $h_b$,  in turn, reduce the negative sbottom effect on the Higgs mass and increase the values of $t_\beta$ for which the theory remains perturbative up to the GUT scale. Negative values of $(\mu M_3)$, instead, give the opposite trend. On the contrary, positive values of $(\mu M_2)$ tend to induce negative values for $\epsilon_{\tau}$, increasing the $\tau$-Yukawa coupling and hence the impact of the stau sector.  

These effects give small negative corrections to the Higgs mass, $\mathcal{O}(\sim$ few GeV), however, they could be quite relevant to the couplings at hand. Including these loop effects, the couplings of the lightest Higgs to the bottom quarks and the tau leptons are~\cite{Carena:2002es},
\be
\label{ghdd}
g_{hd{d}} 
 = -\frac{m_d  \ s_\alpha }{v \ c_\beta (1 + \Delta_d)} \left(1  - \frac{\Delta_d}
{ t_\beta \ t_\alpha} \right), 
\ee
where $\Delta_d \equiv \epsilon_d   t_\beta$ and $d =b$ or $\tau$ respectively. Although the down-type couplings, $g_{hd{d}}$, depend in a relevant way on $\epsilon_d$, in the alignment limit we have $t_\alpha t_\beta \to -1$. Thus the down couplings approach the SM values independently of  $\epsilon_d$.

%%%%%%%%%%%%%%%%%%%%%%%%%%%%%%%%
\subsection{Alignment: $\mu \ll M_{\rm SUSY}$ and  Moderate $\tan\beta$}
%%%%%%%%%%%%%%%%%%%%%%%%%%%%%

When  $|\mu| \ll M_{\rm SUSY}$ and $1 \ll t_\beta\ll m_t/m_b$,  then both the $\mu$ induced and $t_\beta$ enhanced corrections associated with the bottom and tau-Yukawa couplings are negligible ~\cite{mhiggsRG1}. Therefore, the only relevant radiative corrections affecting the Higgs sector are those coming from  the top-stop sector, affecting $\Delta \lambda_2$ and leading to Eq.~(\ref{mhsm}).  In this case  $\lambda_6$ and $\lambda_7$ remain very small and the conditions for alignment to occur are still  determined to a good approximation by Eqs.~(\ref{eq:special1}) and (\ref{eq:special2}).  However, neither of these two alignment conditions are  fulfilled in this corner of the  MSSM, which has the following relation
\be
\lambda_{\rm SM} > \lambda_{1} > \tilde{\lambda}_3\ .
\ee
We can reach the same conclusion by using Eq.~(\ref{eq:salpha1}) for $s_\alpha$ in this regime, 
\be
s_{\alpha}  =    \frac{ -(m_A^2 + m_Z^2) s_\beta c_\beta}{ \sqrt{(m_A^2 + m_Z^2)^2 s_\beta^2 c_\beta^2 +  \left(m_A^2 s_\beta^2 +m_Z^2 c_\beta^2 - m_h^2 \right)^2}}  \;,
\label{salphatree}
\ee
which, for $m_A \simgt 2 m_h$ and moderate $t_\beta$ implies
\be
- \frac{s_{\alpha}}{c_\beta}   \simeq   \frac{m_A^2 + m_Z^2}{m_A^2 - m_h^2} \ .
\label{salphabeta}
\ee
This clearly demonstrates that in this case the deviation of $(-s_\alpha/c_\beta)$ from 1 depends only on $m_A$ and is independent of $t_\beta$. In other words, alignment is only achieved in the  decoupling limit, $m_A^2 \gg m_Z^2, m_h^2$.  

%==============================================================================
\begin{figure}
\begin{center}
\includegraphics[width=0.48\textwidth]{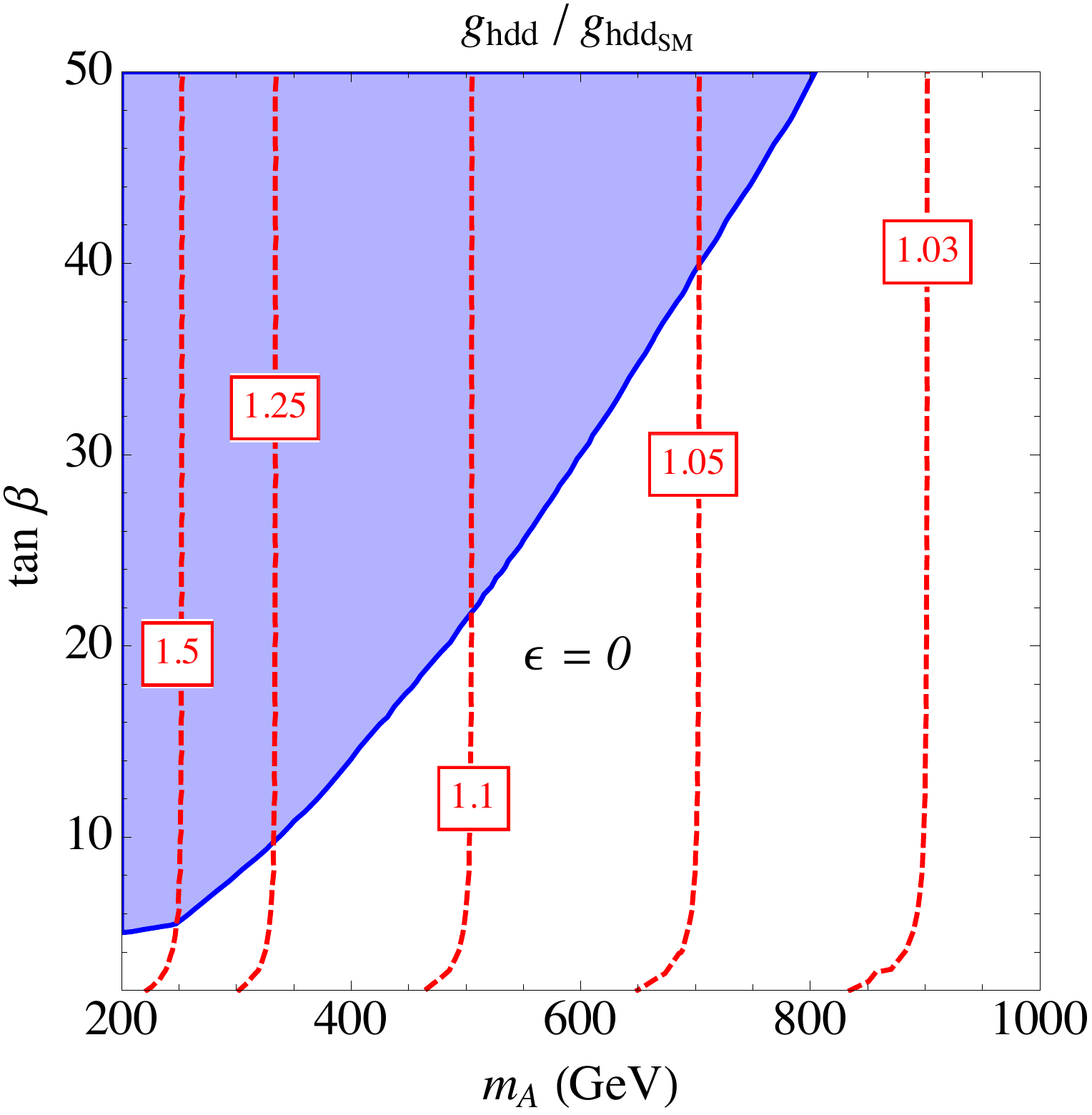}  \\
\end{center}
\caption{Ratio of the value of the down-type fermion couplings to Higgs bosons to their SM values in the case of low $\mu$ ($L_{1j}\sim0$),  as obtained from  Eq.~(\ref{salphatree}), and $\epsilon_d \simeq 0$. }
\label{Treesalphabeta}
\end{figure}
%==============================================================================

This also agrees with our expressions regarding the approach to the alignment limit via decoupling, Eq.~(\ref{B-A}). In this regime $\lambda_{5,6,7}$ are very small implying
\be 
{\cal B} \simeq  m_A^2 - m_h^2, \;\;\;\;\;\;\;\;\;  {\rm and}   \;\;\;\;\;\;\;\;  {\cal B-A}  \simeq -(m_Z^2 + m_h^2) \  .
\ee
In Fig.~\ref{Treesalphabeta} we display the value of $-s_\alpha/c_\beta$  in the $m_A - \tan_\beta$ plane, for low values of $\mu$, for which the radiative corrections to the matrix element $L_{11}$ and $L_{12}$ are small, Eq.~(\ref{salphatree}).  As expected from our discussion above, the down-type fermion couplings to the Higgs become independent of $t_\beta$ in large regions of parameter space, and the ratio to their SM values is well described by Eq.~(\ref{salphabeta}).  We see that under these conditions, a measurement of the down-type fermion coupling to the Higgs  that deviates from the SM value  by 3\% or less  will allow us to infer that the  non-standard Higgs boson masses are at or above the TeV range. The strong enhancements of the down-type couplings at low values of $m_A$ would also imply a correlated enhancement of the Higgs width, leading to a suppression of the decay branching ratio of the Higgs into photons and weak gauge bosons.  The above results imply that such suppression is only weakly dependent on $t_\beta$, which explains the numerical results of Ref.~\cite{Patrick} where the relevant Higgs decays are studied.

%%%%%%%%%%%%%%%%%%%%%%%%%%%%%
\subsection{Alignment: $\mu \sim {\cal{O}}(M_{\rm SUSY})$ and Large  $\tan\beta$}
%%%%%%%%%%%%%%%%%%%%%%%%%%%%%

The situation is quite different for $\mu \sim {\cal{O}}(M_{\rm SUSY})$, since in this case there may be important radiative corrections to quartic couplings, leading to non-vanishing values for $\lambda_{5,6,7}$.  Instead of just taking the alignment conditions in terms of the quartic couplings, we shall rewrite them in terms of the radiative corrections to the matrix elements  ${\cal M}_{11}$ and ${\cal M}_{12}$, since it allows us to get a clear idea of where the important contributions are coming from. We shall then make contact with previous expressions. Quite generally, from Eq.~(\ref{eq:salpha1}), 
\begin{eqnarray}
s_{\alpha} & = &   \frac{- (m_A^2 + m_Z^2) s_\beta c_\beta + v^2(s_\beta^2\, \Delta L_{12} +s_\beta c_\beta\, \Delta \tilde{L}_{12} )}{ \sqrt{\left[(m_A^2 + m_Z^2) s_\beta c_\beta - v^2 (s_\beta^2\, \Delta L_{12} +s_\beta c_\beta\, \Delta \tilde{L}_{12} )\right]^2 +  \left(m_A^2 s_\beta^2 +m_Z^2 c_\beta^2 - m_h^2 + v^2\, \Delta L_{11} s_\beta^2 \right)^2}}  \nonumber\\
\end{eqnarray}
where, as before, the $\Delta L_{ij}$ denote variation due to radiative corrections. We have further separated out the corrections to the $L_{12}$ component into $\Delta L_{12}$ and $\Delta \tilde{L}_{12}$, which contribute with different $t_\beta$ factors. In terms of the quartics these loop corrections are
\begin{equation}
\Delta L_{12} \simeq \lambda_7, \;\;\;\;\;\;\;\;\;\; \Delta \tilde{L}_{12} \simeq \Delta \left( \lambda_3 + \lambda_4 \right), \;\;\;\;\;\;\;\;\;\; \Delta L_{11} \simeq \lambda_5,
\;\;\;\;\;\;\;\;\;\; \Delta L_{22} \simeq \lambda_2 .
\end{equation}
In the above, we have only kept terms which are relevant for moderate or large values of $t_\beta$, and we have included $\Delta L_{22}$ for future use. In particular, since $c_\beta^2 \ll 1$, we have dropped the $\lambda_1$ term which is proportional to $c_\beta^2$ and the $\lambda_6 c_\beta$ term, since  $\lambda_6$ is already a small quantity, generated by radiative corrections. We have kept the  $\Delta\tilde{L}_{12}$ contribution to the matrix element $L_{12}$ since it  has the same $t_\beta$ dependance as the tree-level contribution, and for sizable $\mu$ but small $\mu A_f$, with $f =\{ t,b, \tau\}$, may also be competitive to the radiatively generated $\lambda_7$ contribution. 

Since $t_\beta \gg 1$, and hence $s_\beta \simeq 1$, the condition in Eqs.~(\ref{SMconditionI}) and (\ref{SMconditionII}) now read
\bea
m_h^2& =& -m_Z^2  + v^2  \left(\Delta L_{11} + \Delta \tilde{L}_{12}+ t_\beta  \Delta L_{12} \right)\ , \\
m_h^2& =& m_Z^2  + v^2  \left(\Delta L_{22}  +  c_\beta^2 \Delta \tilde{L}_{12}+ c_\beta  \Delta L_{12} \right)\ . 
\eea
Observe that for moderate or large values of $t_\beta$ the second expression above just shows that the Higgs mass is strongly governed by $\lambda_2$, while the first expression shows that  one reaches the alignment limit for values of $t_\beta$ given by 
\begin{equation}
t_\beta \simeq \frac{m_h^2 + m_Z^2 - v^2 (\Delta L_{11}+\Delta \tilde{L}_{12})}{v^2 \Delta L_{12}} = \frac{m_h^2 - v^2 \tilde{\lambda}_3}{v^2 \lambda_7} \,
\end{equation}
in agreement with  Eq.~(\ref{eq:largetan}) derived in the previous section.

%%%%%%%%%%%%%%%%%%%%%%%%%%%%%
\subsubsection{Radiative Loop Corrections}
%%%%%%%%%%%%%%%%%%%%%%%%%%%%%

The radiative corrections to the quartic couplings $\tilde{\lambda}_3$ and $\lambda_7$, for small differences between the values of the two stops, sbottoms and stau masses,  have been computed previously in Refs.~\cite{Haber:1993an},~\cite{mhiggsRG1},~\cite{Carena:2011aa}. Using these expressions one obtains 
\bea
\label{Loop12}
 \Delta L_{12}  &\simeq& \frac{1}{32 \pi^2}  \left[ h_t^4 \frac{\mu A_t}{M_{\rm SUSY}^2}  \left(  \frac{A_t^2}{M_{\rm SUSY}^2} - 6 \right)
+ h_b^4  \frac{\mu^3 A_b}{M_{\rm SUSY}^4}  + \frac{h_\tau^4}{3} \frac{\mu^3 A_{\tau}}{M_{\tilde{\tau}}^4} \right], \\
\left( \Delta\tilde{L}_{12} + \Delta L_{11} \right) & \simeq&  \frac{3 \ \mu^2 }{16 \pi^2 M_{\rm SUSY}^2} \left[ h_t^4  \left(1-\frac{A_t^2}{2 M_{\rm SUSY}^2} \right) +h_b^4  \left(1-\frac{A_b^2}{2 M_{\rm SUSY}^2} \right) \right.\nonumber \\
&&\quad \left. \phantom{\frac23} +h_\tau^4 \frac{M_{\rm SUSY}^2}{3 M_{\tilde{\tau}}^2}\left(1-\frac{A_\tau^2}{2 M_{\tilde{\tau}}^2}\right)
\right]\ , 
\label{Loop122}
\eea
where, for simplicity, we have ignored  two-loop corrections.

Observe that for moderate values of $|A_t| < \sqrt{6} M_{\rm SUSY}$, the top contributions to $\Delta L_{12}$, ($\lambda_7$), become positive for negative values of $A_t$ and negative for positive ones. On the other hand, for $|A_t| > \sqrt{6} M_{\rm SUSY}$, the sign of $\Delta L_{12}$ is given by the sign of $A_t$. Interestingly enough, the radiative corrections to $\lambda_2$ (and therefore to $m_h$) are maximized at $|A_t| \simeq \sqrt{6}M_{\rm SUSY}$, leading to a Higgs mass of order 130~GeV for stop masses of the order of 1~TeV  (see, for example, Refs.~\cite{Okada:1990vk,mhiggsRG1}).  Therefore, one can get consistency with the measured Higgs mass for values of $|A_t|$ larger or smaller than $\sqrt{6} M_{\rm SUSY}$. The sbottom and stau contributions to $\lambda_7$ become relevant only at large values of $t_\beta$ and are positive for $(\mu A_{b,\tau}) > 0$. Regarding, $(\Delta\tilde{L}_{12} + \Delta L_{11})$, $\Delta \tilde{\lambda}_3$, again we see that the sbottom and stau contributions will only become relevant for large values of $t_\beta$. However, the sign of all of the corrections, including the stops, does not depend on the sign of $A_f$, but rather would be positive (negative) if $|A_f| <  (>) \sqrt{2} M_{\rm SUSY}$. Further, noting the different coefficients in Eqs.~(\ref{Loop12}) and (\ref{Loop122}), observe that values of $| \Delta\tilde{L}_{12} + \Delta L_{11} |$ may be pushed to  larger values compared to values of $|\Delta L_{12}|$.

Keeping these considerations in mind, in our numerical work we will take representative values of these loop corrections to be $32 \pi^2 \Delta L_{12}= \{-1,5\}$ and  $32 \pi^2 (\Delta\tilde{L}_{12} + \Delta L_{11} )=\pm 25$. Such values can be naturally obtained for non-extreme values of $(\mu/M_{\rm SUSY})$ and $(A_f/M_{\rm SUSY})$ and lead to either no alignment or alignment at $t_\beta\sim 20$, respectively. 

%%%%%%%%%%%%%%%%%%%%%%%%%%%%%
\subsubsection{Values of $\tan\beta$ at Alignment}
%%%%%%%%%%%%%%%%%%%%%%%%%%%%%

One can write the large $t_\beta$ alignment condition in the MSSM as
\be
t_\beta \simeq  \frac{120-32   \pi^2 \left(\Delta L_{11} + \Delta \tilde{L}_{12} \right)
}{ 32 \pi^2 \Delta L_{12}}
\label{eq:tanbetaL12}
\ee
where we have made use of the fact that all contributions to $\lambda_7$ in Eq.~(\ref{Loop12}) are proportional to $1/(32 \pi^2) \sim {\cal{O}}(1/300)$ and rescaled both the denominator and numerator by a factor of $32\pi^2$.  The numerator of Eq.~(\ref{eq:tanbetaL12}) tends to remain positive and large after the inclusion of the radiative corrections.
%, 
Therefore, in order to obtain sensible values of $t_\beta$ consistent with a perturbative description of the theory, $0 < t_\beta \lesssim 100$, it is necessary that $32 \pi^2 \Delta L_{12}$ be positive and  larger than one.  This can only be achieved for large values of $|\mu|$ and of some of the trilinear couplings, $A_f$. 

One important implication that can be inferred from Eqs.~(\ref{Loop12}) and (\ref{Loop122}) is that the values of $\Delta L_{12}$ depend in a relevant way on the values of the bottom and tau Yukawa couplings. Since these couplings grow with $t_\beta$, it is clear that $\Delta L_{12}$ ($\lambda_7$) is not independent of $t_\beta$.  This leads to new solutions for the alignment condition at very large values of $t_\beta$, that would not exist if $\lambda_7$ were independent of $t_\beta$.

%==============================================================================
\begin{figure}
\begin{center}
\begin{tabular}{c c}
(i) & (ii)\\
\includegraphics[width=0.48\textwidth]{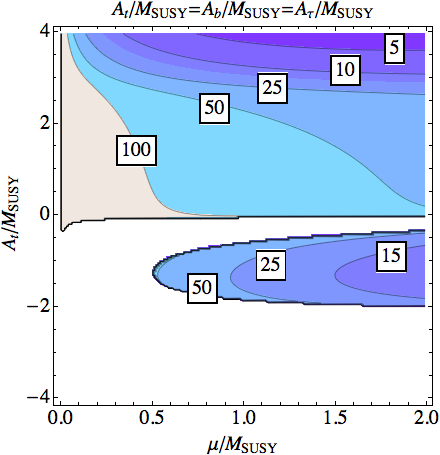}  &
\includegraphics[width=0.48\textwidth]{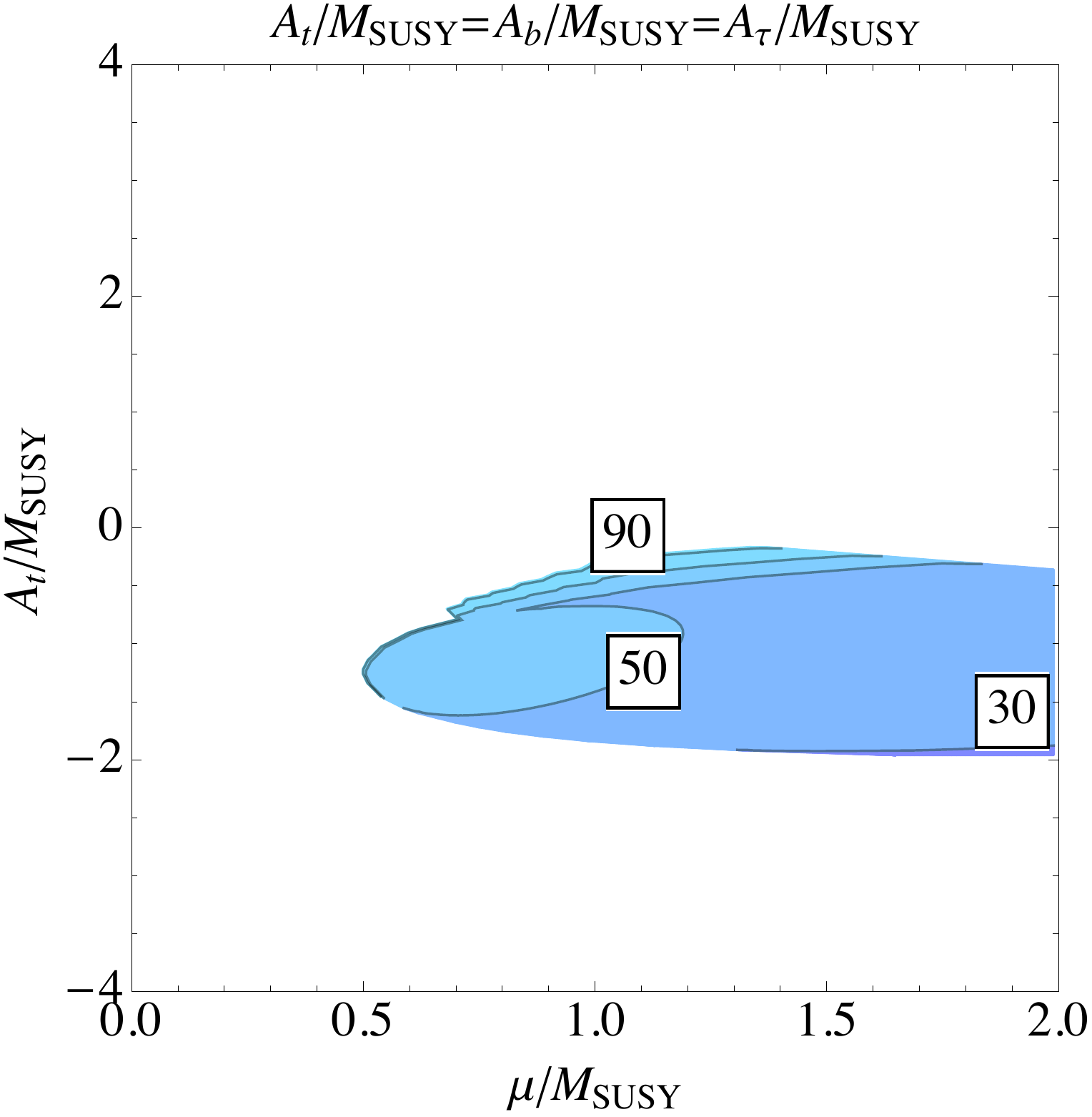}\\
\end{tabular}
\end{center}
\caption{Values of $t_\beta$, at which alignment without decoupling occurs, using 
Eqs.~(\ref{Loop12})-(\ref{Loop122}) . We assume common masses, $M_{\rm{SUSY}}$ and include the contribution from the bottom and tau Yukawas,  fixing $A_b=A_\tau=A_t$. The branch of solutions displayed in (i) (left panel) would exist even if one neglected the bottom and tau Yukawa couplings. Those in (ii) (right panel) appear due to the extra $t_\beta$ dependence associated with the down-type fermion Yukawa couplings.}
\label{tanbeta}
\end{figure}
%==============================================================================
%
%
%==============================================================================
\begin{figure}
\begin{center}
\begin{tabular}{c c}
(i) & (ii)\\
\includegraphics[width=0.48\textwidth]{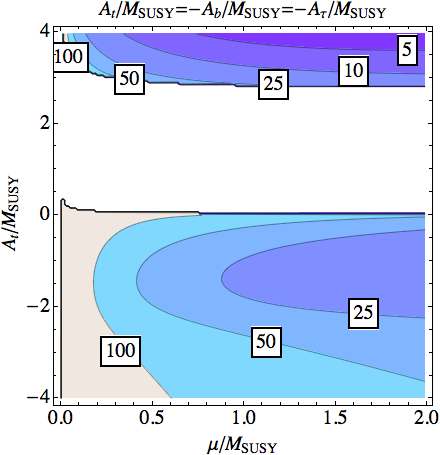} &
\includegraphics[width=0.48\textwidth]{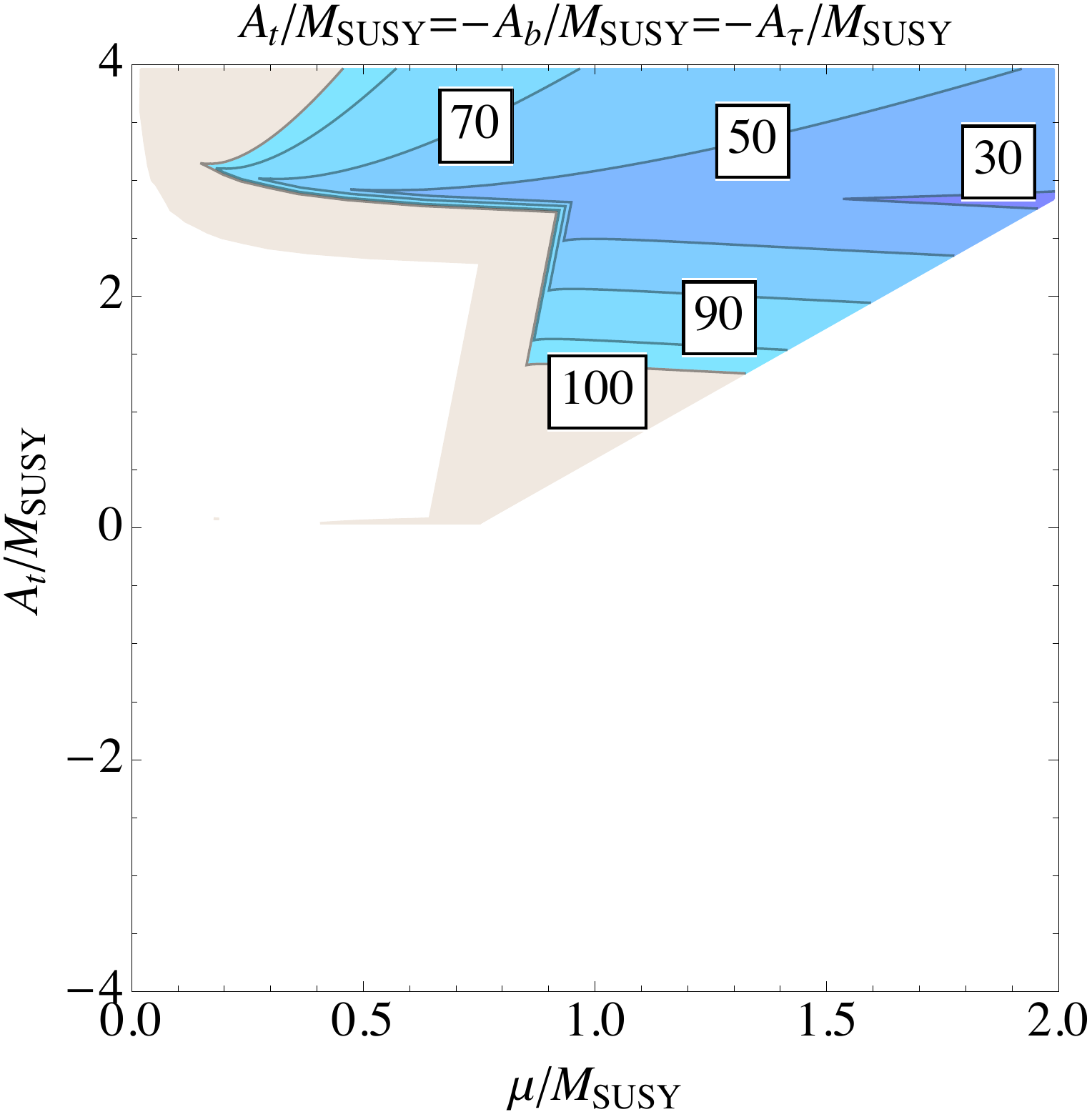} 
\end{tabular}
\end{center}
\caption{Same as Fig.~\ref{tanbeta} but now showing alignment values of $t_\beta$ under the assumptions $A_b=A_\tau=-A_t$. }
\label{tanbeta2}
\end{figure}
%==============================================================================
%
%==============================================================================
\begin{figure}
\begin{center}
\begin{tabular}{c c}
(i) & (ii)\\
\includegraphics[width=0.48\textwidth]{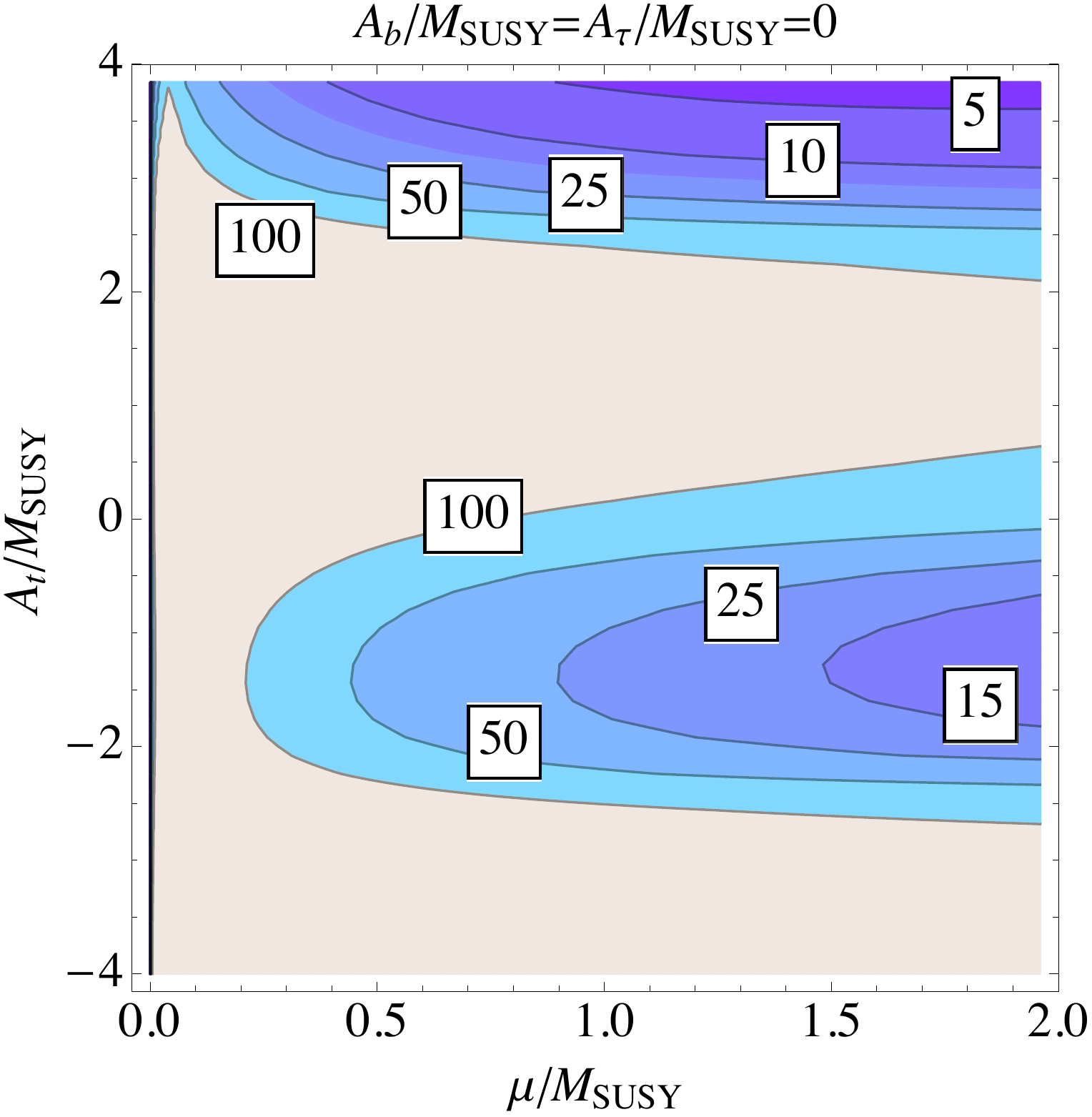} &
\includegraphics[width=0.48\textwidth]{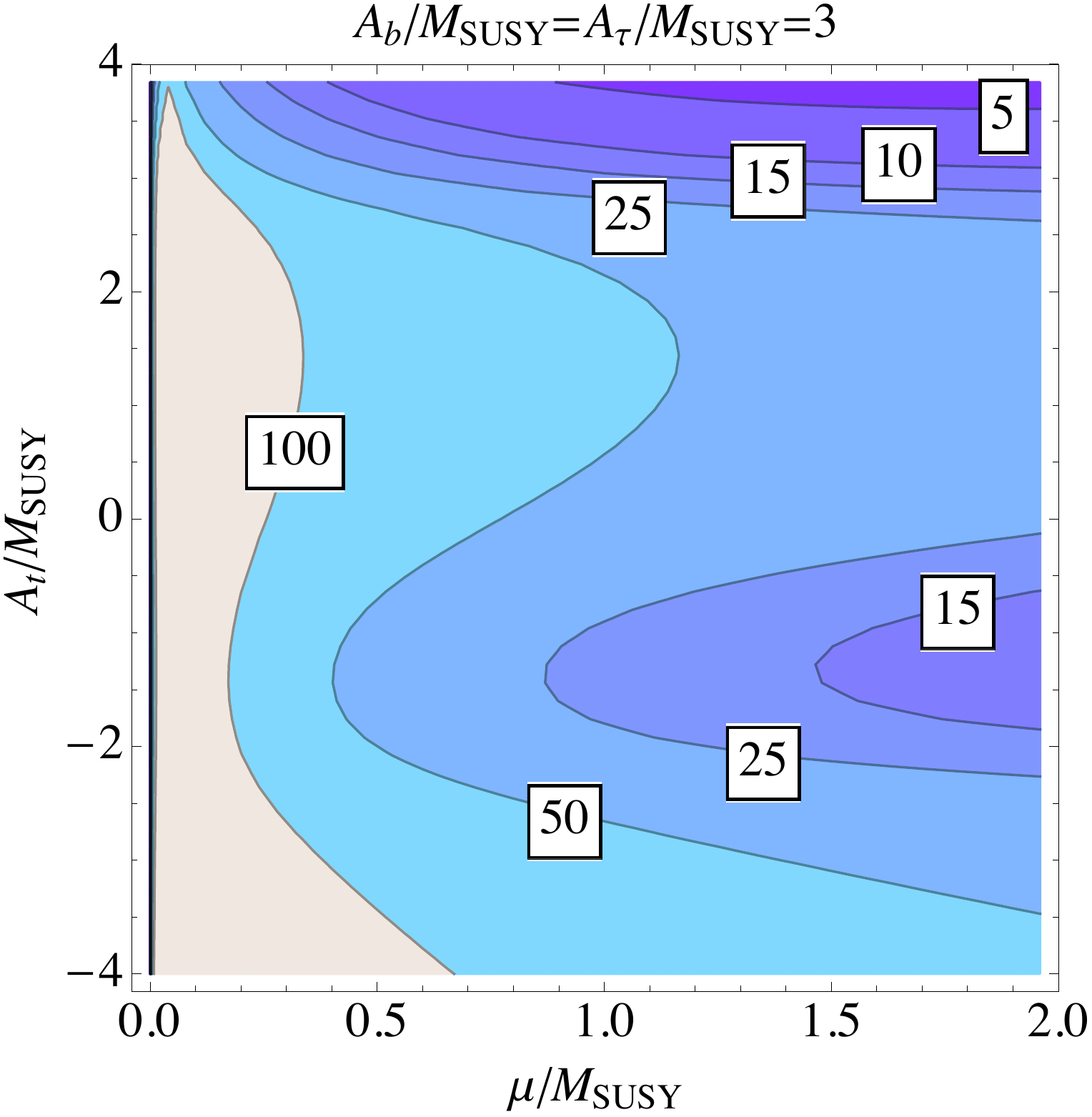} 
\end{tabular}
\end{center}
\caption{Same as Fig.~\ref{tanbeta} but now showing alignment values of $t_\beta$ under the assumptions: (i) $A_b=A_\tau= 0$ and (ii) $A_b/M_{\rm{SUSY}}=A_\tau/M_{\rm{SUSY}}=3$.  Only one set of solutions appear in these cases.  }
\label{tanbeta3}
\end{figure}
%==============================================================================
%

Fig.~\ref{tanbeta} shows contour plots of the values of $t_\beta$ where alignment is achieved 
for different values for the ratio $(\mu/M_{\rm SUSY})$ and for positive/negative values of $(A_t/M_{\rm SUSY})$, keeping the  $t_\beta$ dependance in the Yukawas explicit. The value of $t_\beta$ is obtained by solving the corresponding algebraic equation exactly, without the approximations done in Eq.~(\ref{eq:tanbetaL12}). In Fig.~\ref{tanbeta} we chose equal values of the $(A_f/M_{\rm SUSY})$ parameters. We find two different roots for the algebraic equation and display them in Figs.~\ref{tanbeta}(i) and \ref{tanbeta}(ii).  The second root, displayed in Fig.~\ref{tanbeta}(ii) appears due to the $t_\beta$ dependence of the bottom and tau Yukawa couplings.

In Fig.~\ref{tanbeta2} we show similar results, but for $A_t = - A_b = -A_\tau$.  Moderate values of $t_\beta$ may be obtained for either large values of $(A_t/M_{\rm SUSY})$ or for negative values of this parameter. The second root, displayed in Fig.~\ref{tanbeta2}(ii), moves now to positive values of $A_t$.  

Finally, in Fig.~\ref{tanbeta3} we show the effect of setting the $A_b$ and $A_\tau$ parameters to zero or to large values. Only one root appears in both cases, and the difference between the results in the left and right panel of Fig.~\ref{tanbeta3} is only visible for very large values of $t_\beta$,  which is when the bottom and tau Yukawa couplings become relevant. 

Looking at the results presented in Figs.~\ref{tanbeta}-\ref{tanbeta3}, we see that in general, one obtains a wide range in both the values of $t_\beta$ at alignment and the  associated parameter space where this would occur. However note that values of $t_\beta \lesssim 10$ at alignment are only obtained for very large values of $(A_f/ M_{\rm SUSY})$, which could be significantly constrained from both the Higgs mass and the stability of the vacuum. A detailed analysis of the phenomenological implications for such a scenario is beyond the scope of this work.  

%%%%%%%%%%%%%%%%%%%%%%%%%%%%%
\subsubsection{Departure from Alignment}
%%%%%%%%%%%%%%%%%%%%%%%%%%%%%

As explained previously section, it is important to study the departure from the alignment condition. The effect of this will be most readily visible in the couplings of the lightest SM-like Higgs boson to down-type fermions. Re-writing Eq.~(\ref{ghdd}), close to alignment for small values of $\epsilon_d$, this coupling reads, approximately
\be
\frac{g_{hdd}}{g_f}  \simeq  -\frac{s_\alpha}{c_\beta} + \epsilon_d t_\beta \left( 1 + \frac{s_\alpha}{c_\beta} \right). 
\ee
This means that positive (negative) values of $\epsilon_d$ tend to suppress (enhance) the coupling departures from the SM values, a tendency that is enforced for larger values of $t_\beta$.  On the other hand, at values of $t_\beta$ larger than the ones leading to alignment, positive values of  $\Delta L_{12}$ tend to suppress the departure of $g_{hdd}$  from SM values. Hence, the effect of negative values of $\epsilon_d$ may be partially compensated (enhanced) for positive (negative) values of $\Delta L_{12}$, and vice versa.

%
%==============================================================================
\begin{figure}
\begin{center}
\begin{tabular}{c c}
(i) & (ii)\\
\includegraphics[width=0.48\textwidth]{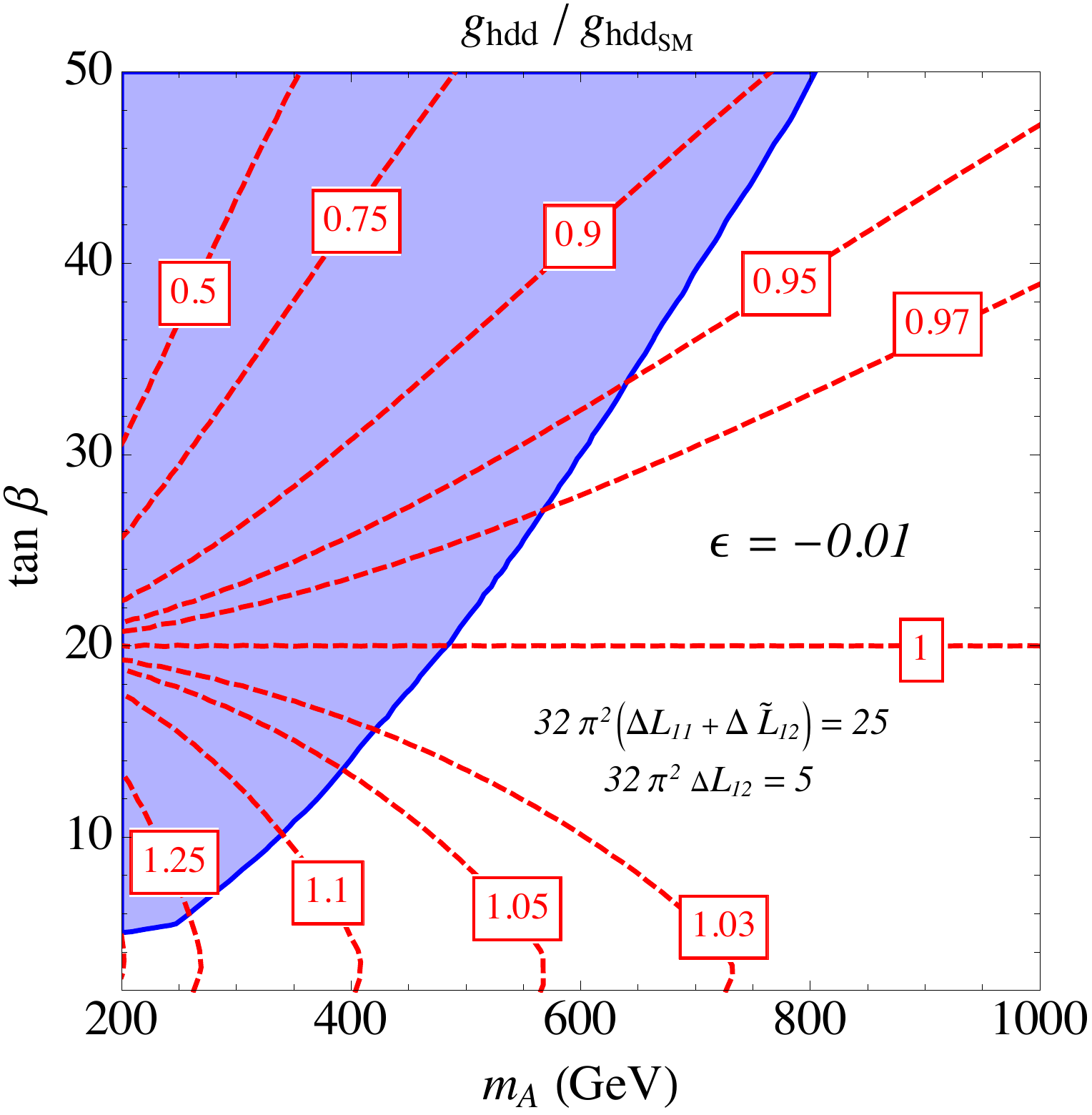}  &
\includegraphics[width=0.48\textwidth]{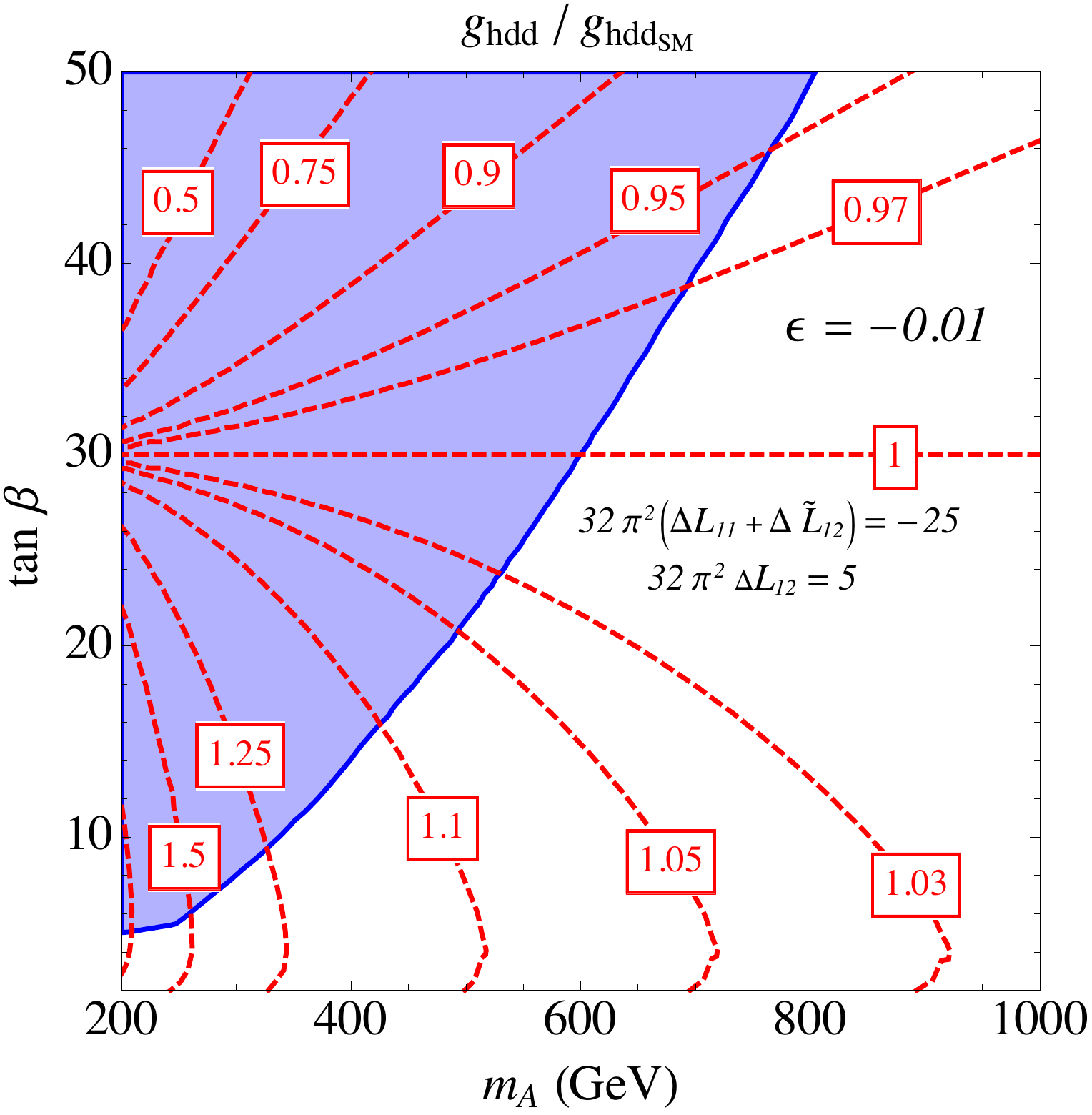}  \\
(iii) & (iv)\\
\includegraphics[width=0.48\textwidth]{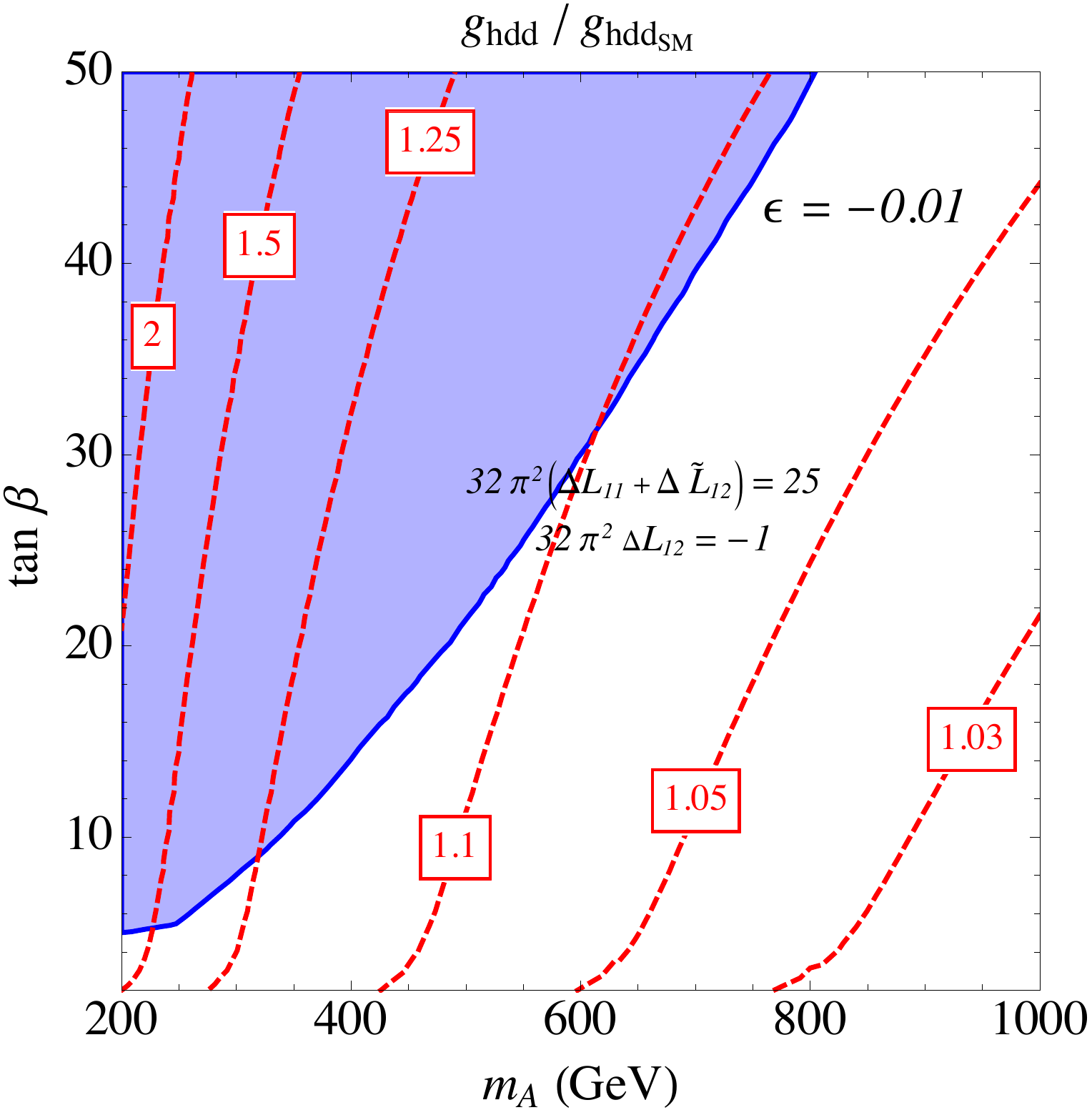}  &
\includegraphics[width=0.48\textwidth]{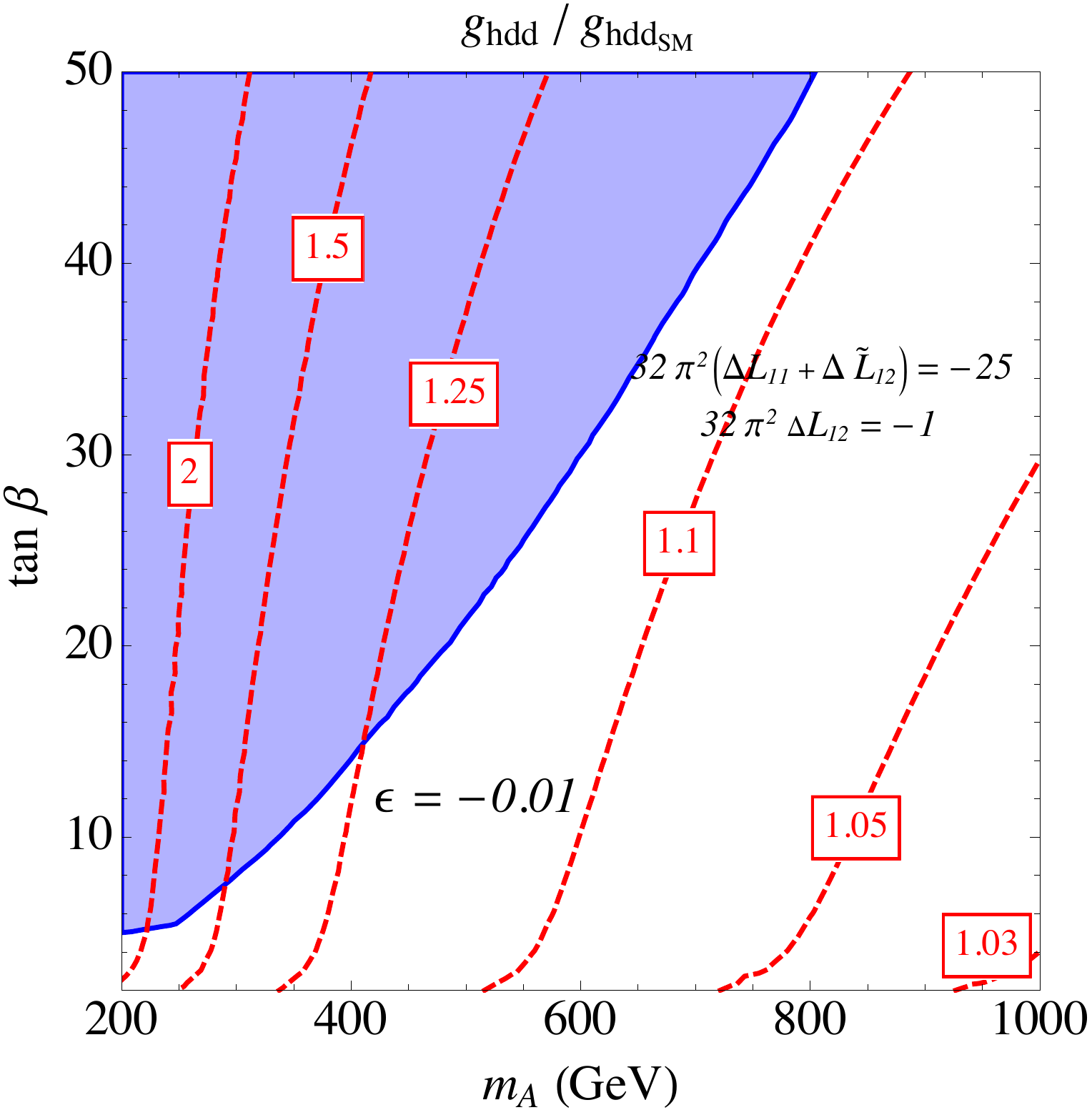}  
\end{tabular}
\end{center}
\caption{Blue shaded region denotes current LHC limits. The ratio of the Higgs coupling to down-type quarks to the SM limit is shown by the red dashed contours for $\epsilon_d =-0.01$. The top two panels have $(32 /\pi^2) \Delta L_{12}=5$ and the lower ones $(32/\pi^2)\Delta L_{12}=-1$. The left panels are for $(32 /\pi^2)\Delta L_{11}=25$ and right for $(32 /\pi^2)\Delta L_{11}=-25$. $\Delta\tilde{L}_{12}=0$ in these figures since its contribution in the mixing angle is suppressed by $c_\beta$ and is effectively negligible.}
\label{ghdd1}
\end{figure}
%==============================================================================
%
%
%==============================================================================
\begin{figure}
\begin{center}
\begin{tabular}{c c}
(i) & (ii)\\
\includegraphics[width=0.48\textwidth]{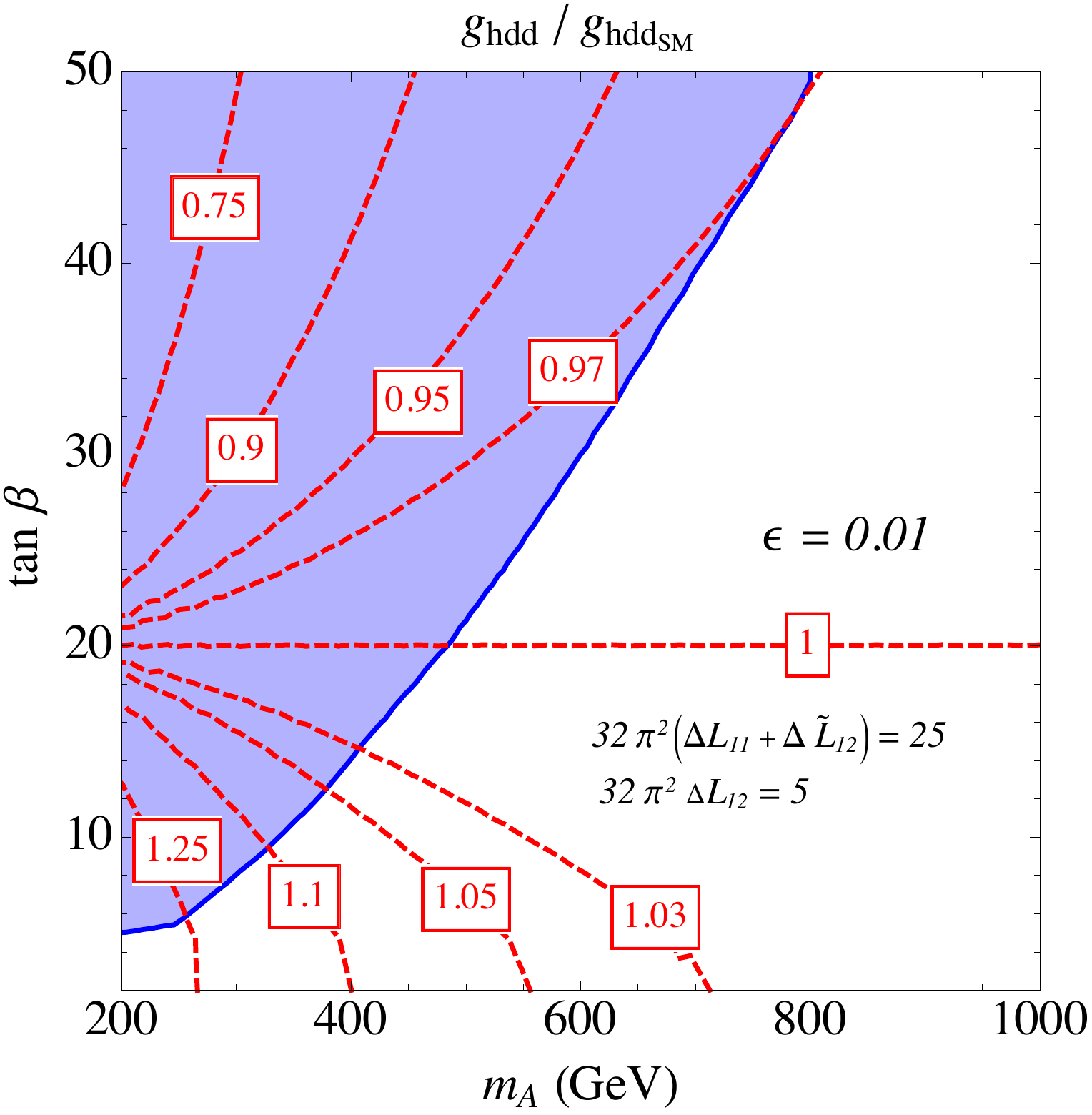}  &
\includegraphics[width=0.48\textwidth]{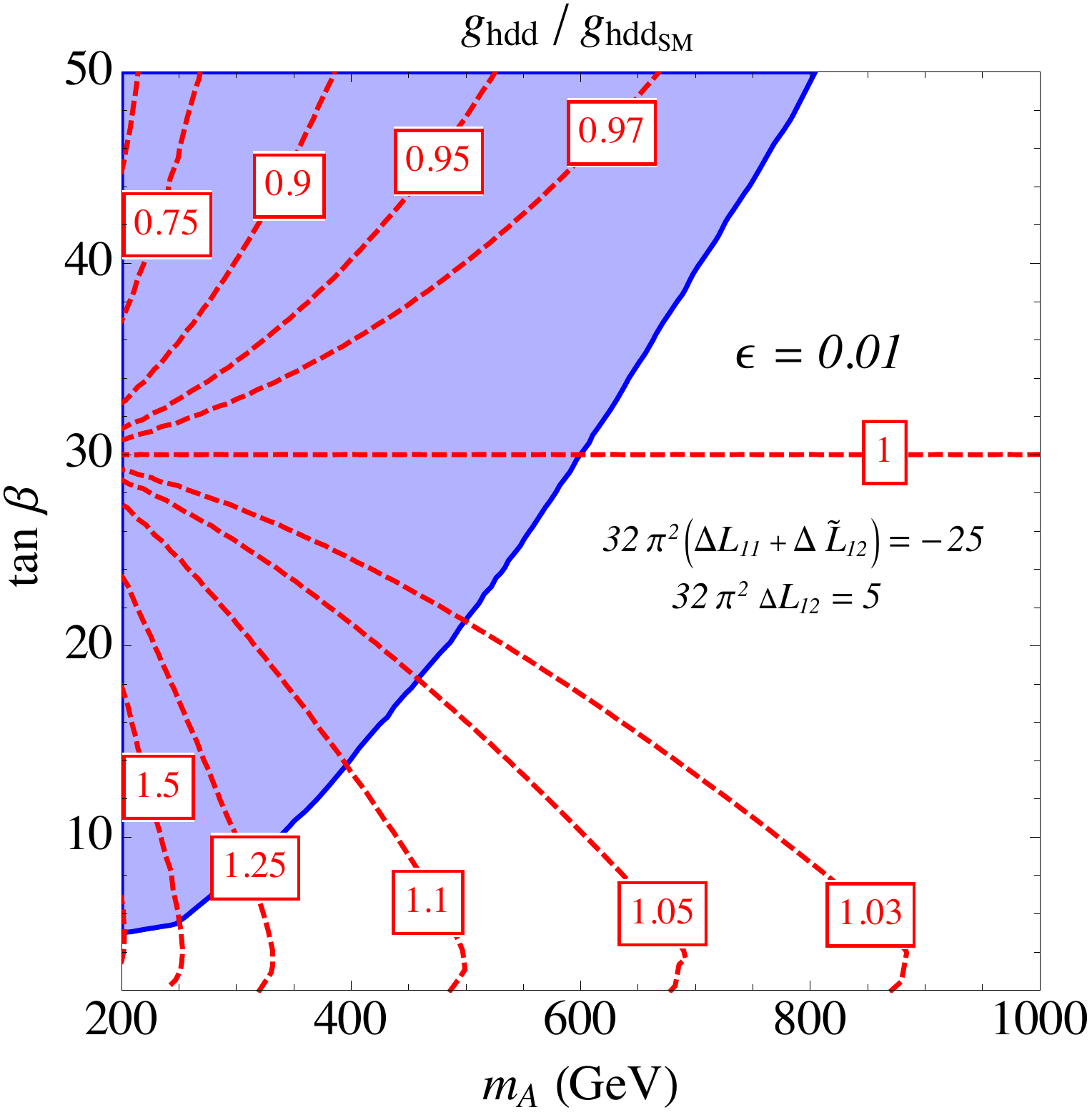}  \\
(iii) & (iv)\\
\includegraphics[width=0.48\textwidth]{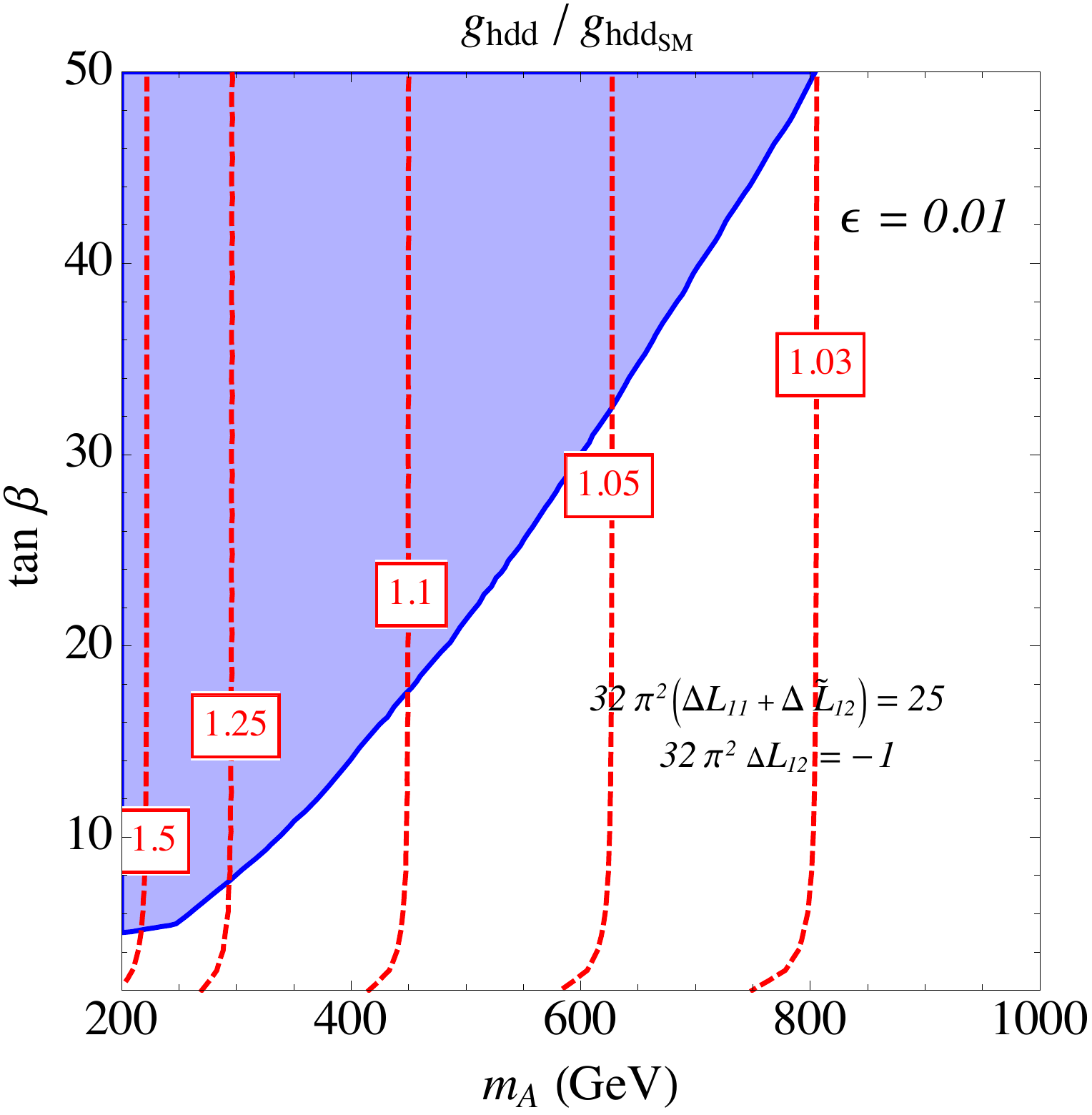}  &
\includegraphics[width=0.48\textwidth]{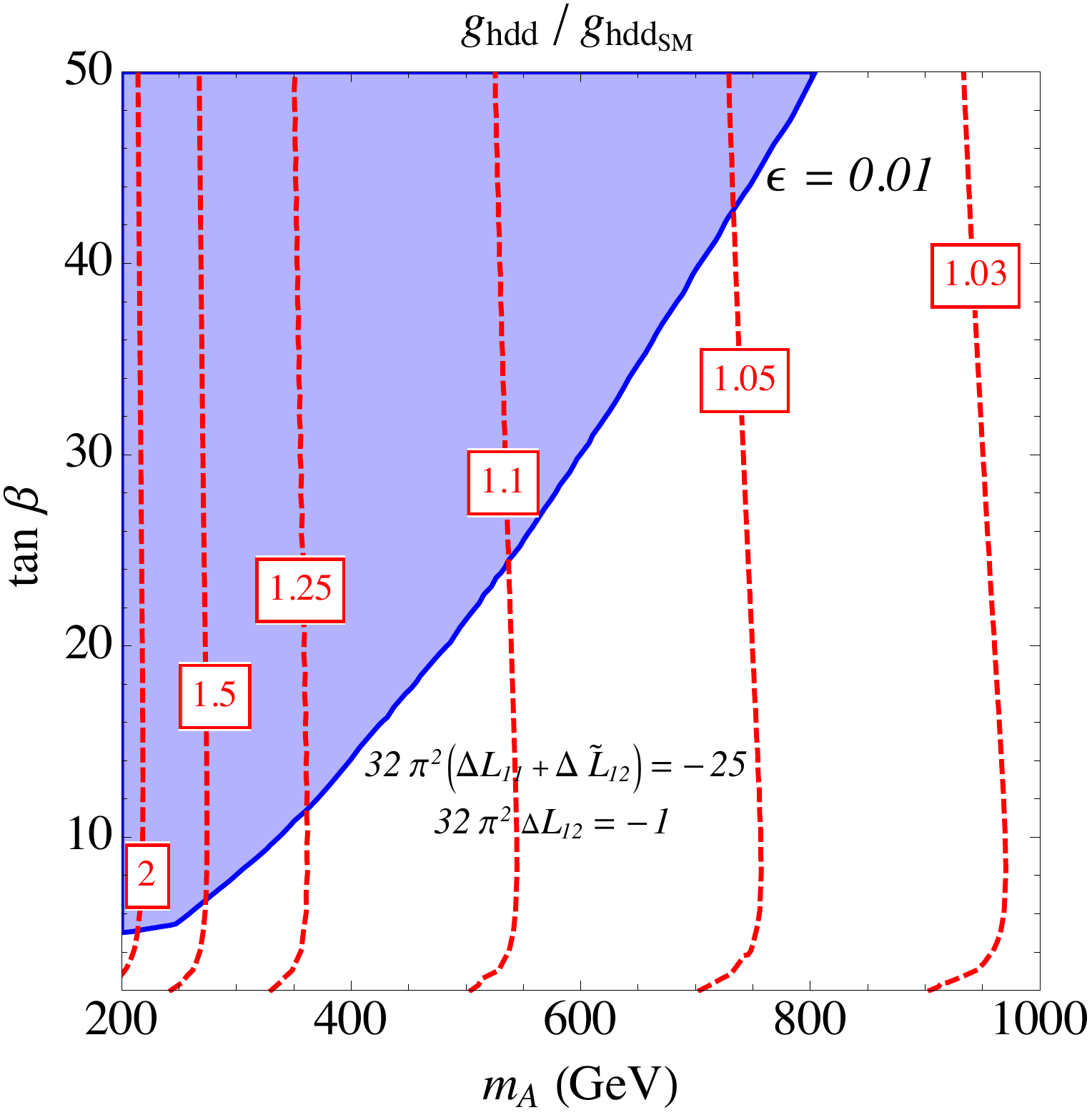}  
\end{tabular}
\end{center}
\caption{Blue shaded region denotes current LHC limits. The ratio of the Higgs coupling to down-type quarks to the SM limit is shown by the red dashed contours for $\epsilon_d =0.01$. The top two panels have $(32 /\pi^2) \Delta L_{12}=5$ and the lower ones $(32/\pi^2)\Delta L_{12}=-1$. The left panels are for $(32 /\pi^2)\Delta L_{11}=25$ and right for $(32 /\pi^2)\Delta L_{11}=-25$.  $\Delta\tilde{L}_{12}$ is again chosen to be 0. }
\label{ghdd2}
\end{figure}
%==============================================================================

In Figs.~\ref{ghdd1} and \ref{ghdd2} we display deviations with respect to the SM of the Higgs-to-down-type-fermion couplings in the $m_A-\tan\beta$ plane
 for fixed representative values of $32 \pi^2 \Delta L_{ij}$ and for two different signs of the one-loop contribution to the Yukawa couplings,  $\epsilon_d = -0.01$ and $\epsilon_d = 0.01$, which could be considered to be associated with the $\tau$ and bottom couplings, respectively (see Eq.~(\ref{hbtau})). We are neglecting the $t_\beta$ dependence of the quartic couplings which eliminates additional  alignment solutions that may appear at very large values of $t_\beta$, but makes the interpretation of the results more transparent.  Moreover, we show regions that are excluded by current direct searches at the LHC~\cite{LHCtautauMSSM}~\footnote{Larger values of $\mu$ than those assumed in the $m_{h}^{max}$ scenario would lead to slighter stronger bounds~\cite{Carena:2013qia}.}. 

For negative values of $32 \pi^2 \Delta L_{12}$, as shown in the lower panels,  there is no alignment solution and the deviation of the couplings from the SM values depend mostly on $m_A$ and not on $t_\beta$.  The value of the down-type fermion couplings to the Higgs is always enhanced with respect to the SM value, as happens whenever $t_\beta$ is below the one associated with the alignment solution. On the other hand, in the presence of alignment, as is the case in the upper panels of Figs.~\ref{ghdd1} and \ref{ghdd2}, suppression of the Higgs couplings to down fermions may be obtained for $t_\beta$ larger than the ones leading to alignment (see Eq.~(\ref{ghddlargetb})). Additionally, even when the $t_\beta$ at alignment  is large, there can be significant variations in the $g_{hdd}$ couplings at much smaller values of $t_\beta$, as can be seen from the top panels in both Figs.~\ref{ghdd1} and \ref{ghdd2}.

In the absence of any $\epsilon_d$ corrections, the deviations in $g_{hdd}$ couplings are flavor universal. The impact of non-zero $\epsilon_d$ can be seen by comparing Figs.~\ref{ghdd1} and \ref{ghdd2}. These corrections are enhanced by $t_\beta$  in $g_{hdd}$. Any deviation in the ratio of $g_{hbb}/g_{h\tau\tau}$ from its SM value, $m_b/m_\tau$, should predominantly come from the $\epsilon_d$ dependance of these couplings. Further, independently of the value of $\Delta L_{12}$, the largest deviation in $g_{hbb}/g_{h\tau\tau}$ occurs at low values of $m_A$ and larger $t_\beta$, which are  constrained by direct searches for non-standard Higgs bosons. 

Figs.~\ref{ghdd1} and \ref{ghdd2} also illustrate the so-called wedge region which is difficult to access using direct searches. It is commonly assumed that measurements of the Higgs couplings to down-type fermions could effectively constrain this wedge region.  However, as can be seen from these figures and from  Eq.~(\ref{ghddlargetb}), these constraints depend strongly on the precise value of $t_\beta$ leading to alignment, and become weaker when this value becomes smaller.

%%%%%%%%%%%%%%%%%%%%%%%%%%%%%
\subsection{Beyond the MSSM}
%%%%%%%%%%%%%%%%%%%%%%%%%%%%%

In the previous section it was shown that  in the region of MSSM parameter space where $\mu \ll M_{SUSY}$ and $t_\beta$ is moderate, alignment without decoupling never occurs because the quartic coupling $\tilde{\lambda}_3$ is too small and the alignment conditions in Eqs.~(\ref{eq:special1}) and (\ref{eq:special2}) are never fulfilled. However, one could increase the value of $\tilde{\lambda}_3$ by augmenting the MSSM with either a singlet scalar, as in the case of NMSSM \cite{Ellwanger:2009dp,Espinosa:1991gr}, or a triplet scalar ~\cite{Delgado:2013zfa}. In these models, a gauge singlet, $S$, in the case of NMSSM, and  a triplet scalar, $\Sigma$, in the case of triplet augmented MSSM, are added to the superpotential with the following cubic couplings, among others,
\be\label{dellam3}
\Delta_S {\cal W} = \lambda S H_u H_d \ , \qquad \Delta_T {\cal W} = \lambda  H_u \Sigma H_d   \ .
\ee
Since the Higgs with a mass of 125 GeV appears to be mostly a doublet scalar by all accounts \cite{Low:2012rj}, it is reasonable to consider a limit where the singlet and triplet scalars are much heavier than the doublet scalars so that the singlet/doublet or triplet/doublet mixing is small. As such, the singlet and triplet scalars can be integrated out of the low energy spectrum. However, if the superpartners of singlet and triplet scalars are also integrated out at the same time, one would just regain the MSSM at low energies and no new insights could be obtained relative to what was already discussed previously. Therefore we shall consider a limit where only the scalar components of the singlet and triplet superfield are integrated out \cite{Lu:2013cta}. In this scenario  new contributions to $\tilde{\lambda}_3$ in the scalar potential are generated, which are not present in the MSSM. Then a solution for alignment without decoupling could be found in the branch of $t_\beta \sim {\cal O}(1)$.

More explicitly, after integrating out the singlet or triplet fields, one finds a correction to the quartic coupling $\tilde{{\lambda}}_3$  given by,
\be
\delta {\tilde{\lambda}}_3 = c \ \lambda^2 \;,
\ee
where $c=1$ or $c=1/2$ for the NMSSM \cite{Nomura:2005rk} or the triplet-augmented MSSM \cite{Delgado:2013zfa}, respectively.

Alignment in this case would occur if the values of $\lambda$ are such that the relations in Eqs.~(\ref{eq:finetune1}) and~(\ref{eq:special2}) are satisfied.
Since $\lambda_1\simeq m_Z^2/v^2 \simeq 0.137$ and thus $(\lambda_{\rm SM} - \lambda_1) \simeq 0.123$, using Eq.~(\ref{eq:tb024}) we find
\be
\label{eq:lambda3}
\tilde{\lambda}_3 - \lambda_{\rm SM} \simeq   \frac{0.123}{t_\beta^2}  \ ,\qquad {\lambda}_{\rm SM} - \lambda_2  \simeq  \frac{0.123}{t_\beta^4}\ .
\ee
Then, in order to obtain a solution for $t_\beta \simgt 2$, not only does one have to obtain large radiative corrections to $\lambda_2$ to raise its value to be very close to  $\lambda_{\rm SM}$, but also  $\tilde{\lambda}_3$ has to be adjusted so that Eq.~(\ref{eq:lambda3}) is satisfied simultaneously. Using the tree level value of $\tilde{\lambda}_3^{\rm MSSM} \simeq -0.137$, we see  the new coupling in the superpotential must be
\be
\label{eq:lambda}
\lambda^2 = \frac{1}{c} \left( 0.397 + \frac{0.123}{t_\beta^2}\right).
\ee
In the triplet-augmented MSSM ($c = 1/2$),  for $t_\beta \simeq 2.7$, one gets $\lambda \simeq 0.9$, which is in excellent agreement with Eq.~(2.21) in Ref.~\cite{Delgado:2013zfa}.  Larger values of $t_\beta$ can be obtained by simultaneously adjusting $\lambda$ and $\lambda_2$ so that both conditions in Eq.~(\ref{eq:lambda3}) are satisfied.

The NMSSM has been widely studied in the literature, since for small values of $t_\beta$ large corrections to the SM-like Higgs mass may be obtained due to the non-decoupling $F$-terms described above.  This tree-level correction reduces the need for large radiative corrections from the stop sector, ameliorating the fine-tuning problem. For this to happen without spoiling the perturbative consistency of the theory up to the GUT scale, one must have $\lambda \simlt 0.7$ \cite{Ellwanger:2009dp}. On the other hand, since the tree-level Higgs mass is given by
\be
\left( m_h^2 \right)^{\rm tree}  = M_Z^2 c_{2 \beta}^2  + 2 \lambda^2 v^2 s_{\beta}^2 c_{\beta}^2\; ,
\ee
a large  tree-level contribution to the Higgs mass, beyond the one obtained in the MSSM, may only be obtained if  $\lambda > \sqrt{2} M_Z/v \simeq 0.5$, and a sizable reduction of 
fine tuning demands even larger values of $\lambda$ and $t_\beta = {\cal{O}}(1)$.  Therefore,  there is a small range of values of $\lambda$ and $t_\beta$ for which the fine-tuning is reduced without spoiling the perturbative consistency of the theory.

%==============================================================================
\begin{figure}[h!]
\begin{center}
\begin{tabular}{c c}
(i) & (ii)\\
&\\
\includegraphics[width=0.48\textwidth]{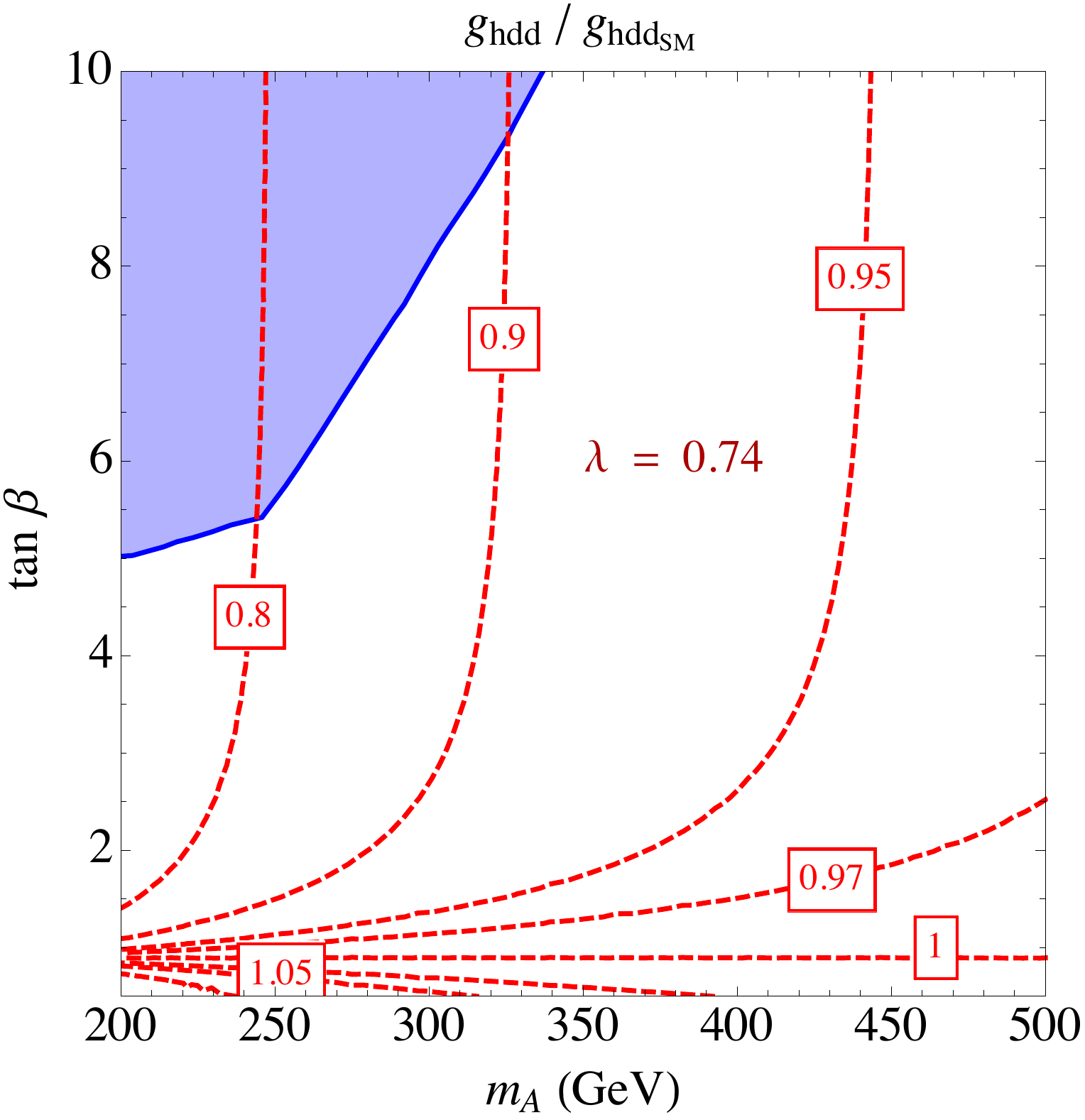}  &
\includegraphics[width=0.48\textwidth]{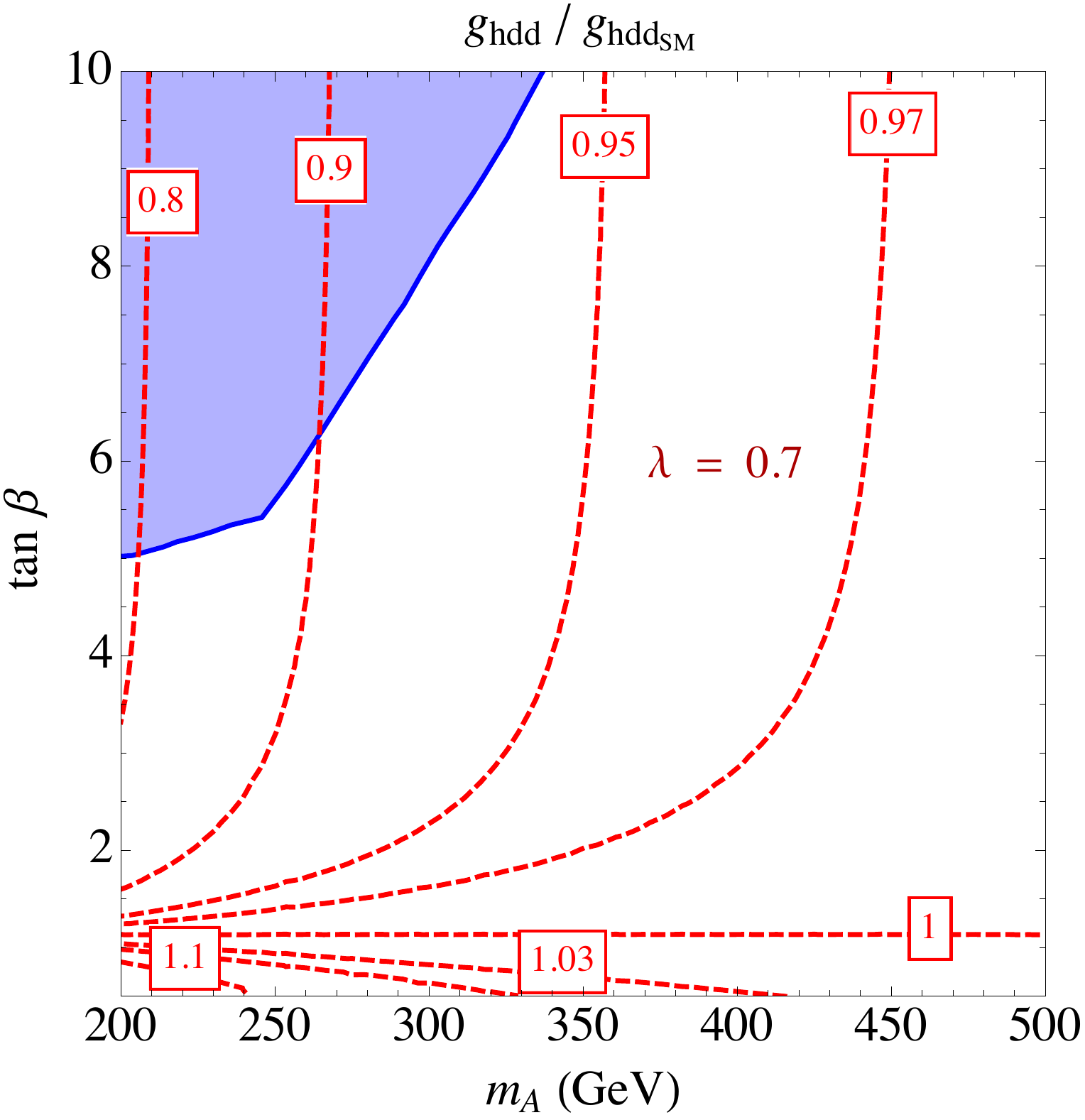}  \\
& \\
(iii) & (iv)\\
&\\
\includegraphics[width=0.48\textwidth]{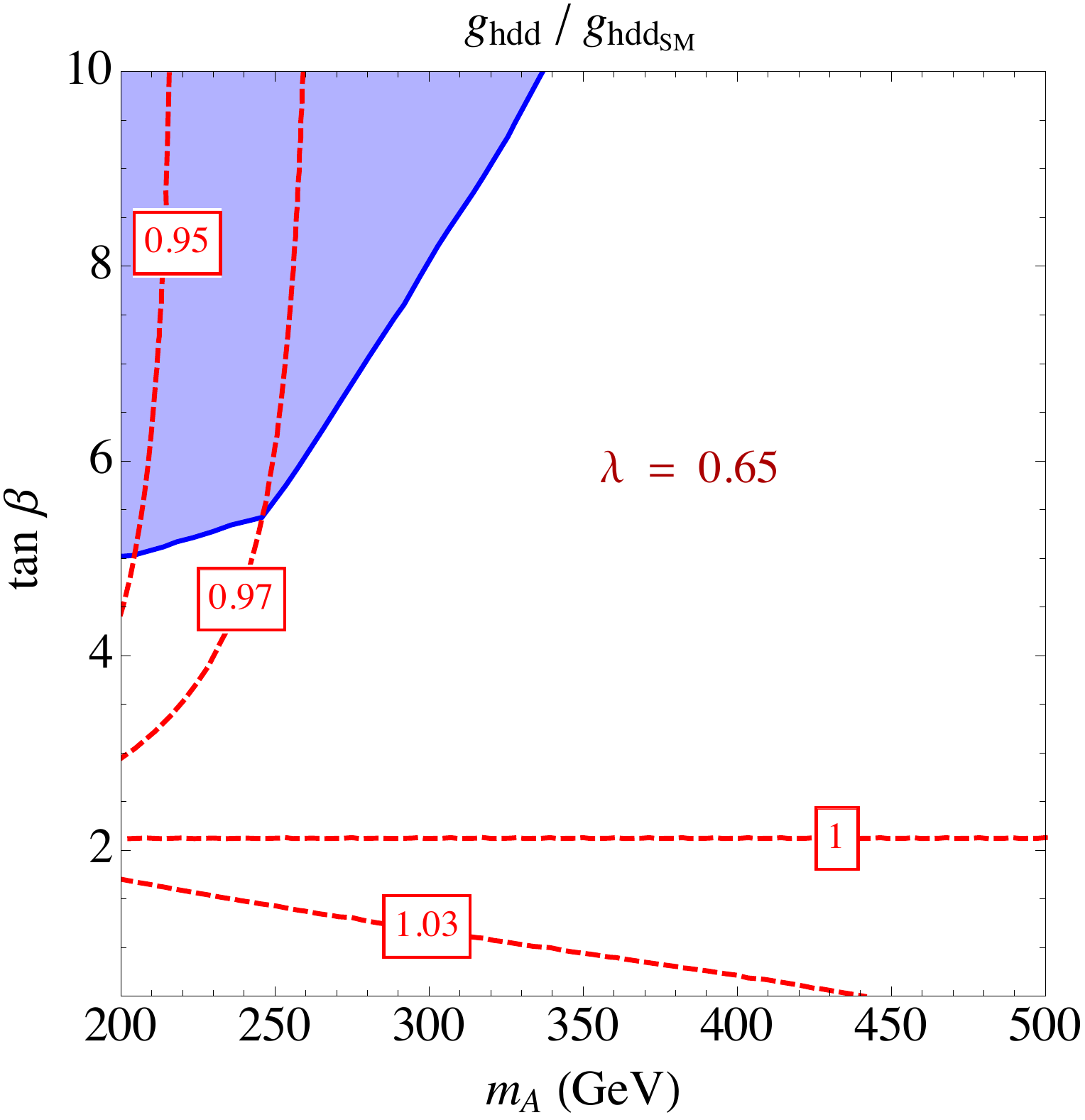}  &
\includegraphics[width=0.48\textwidth]{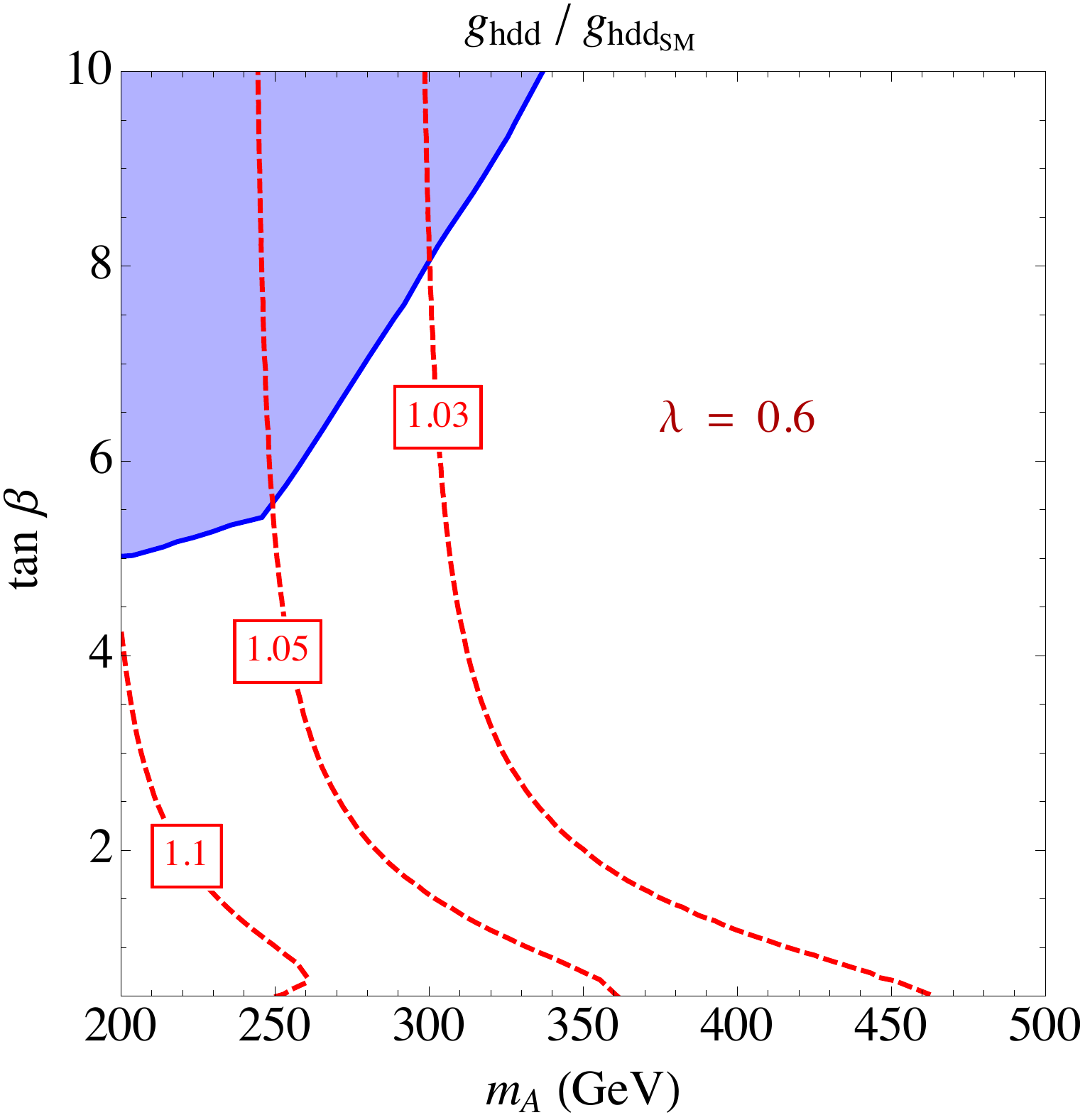}  
\end{tabular}
\end{center}
\caption{Blue shaded region denotes current LHC limits. The ratio of the Higgs coupling to down-type quarks to the SM limit is shown by the red dashed contours for various values of $\lambda$.}
\label{ghddNMSSM}
\end{figure}
%==============================================================================

Interestingly enough, for $c=1$ and  $t_\beta > 1$, Eq.~(\ref{eq:lambda}) shows that alignment can also  be obtained for $0.65 \simlt \lambda \simlt 0.75$, with smaller values
of $\lambda$ corresponding to the larger values of $t_\beta$ necessary to satisfy the alignment conditions.   Therefore, there is an overlap between the  widely studied NMSSM parameter space, $t_\beta\sim {\cal O}(1)$ and $\lambda \simeq 0.7$, and the ones necessary to  satisfy the  alignment conditions.  Small departures of $t_\beta$ from the alignment values lead to a variation of the bottom and tau couplings with respect to the SM values. Smaller values of $t_\beta$ lead to an enhancement of these couplings and larger values to a decrease in them.  This is shown in Fig.~\ref{ghddNMSSM} where the ratio of the down fermion coupling to the Higgs with respect to the SM values is presented for different values of $\lambda$ as a function of the heavy CP-odd Higgs mass, $m_A$. We have assumed that the singlet states are decoupled and do not affect the heavy Higgs bounds. 

In general, due to the presence of alignment, values of $0.75 \geq \lambda \geq 0.6$ lead to a drastic variation of $g_{hdd}$ with respect to the SM expectations as compared to  the small $\mu$ case in the MSSM~($\lambda = 0$) represented in Fig.~\ref{Treesalphabeta} and Eq.~(\ref{salphabeta}). The variations of these couplings depend very strongly on the precise value of $\lambda$.~\footnote{Because there is such strong dependance on the precise value of $\lambda$, the $t_\beta$ at alignment  could  be very sensitive to small loop corrections to $\tilde{\lambda}_3$, which enter into $\delta\tilde{\lambda}_3$ as in Eq.~(\ref{dellam3}), which have been neglected in this analysis.} 
As happened in the MSSM case, suppression is obtained for values of $t_\beta$ larger than the one leading to alignment, while enhancement is obtained otherwise. However, note that unlike in the MSSM, alignment can occur for  small values of $t_\beta \lesssim 5$. This implies that one can go to very low values of $m_A$ without significant variations of the Higgs couplings to fermions and gauge bosons (Fig.~\ref{ghddNMSSM} (iii)). In addition, as can be seen from  Figs.~\ref{ghddNMSSM} (i) and (ii), it is possible to have significant suppression of the down-type couplings in the wedge, unlike the MSSM case.

%%%%%%%%%%%%%%%%%%%%%%%%%%%%%
%%%%%%%%%%%%%%%%%%%%%%%%%%%%%
\section{Conclusions}
\label{sect:6}
%%%%%%%%%%%%%%%%%%%%%%%%%%%%%
%%%%%%%%%%%%%%%%%%%%%%%%%%%%%

Models with an extended Higgs sector appear in many theories beyond the SM. Among them, 2HDMs represent some of the simplest extensions of the SM Higgs sector and are realized in both supersymmetric and non-supersymmetric theories.   When masses of the non-standard scalar are much larger than the $Z$ mass, it is well-known that the mass eigenbasis, described by the mixing angle, $\alpha$, in the CP-even sector, aligns with the basis  characterized by the angle $\beta$, in which the Higgs VEV resides in only one of the doublets. This leads to the alignment limit where $\cos(\alpha-\beta) = 0$ and the lightest CP-even Higgs behaves as the SM one.  It has also been known for a long time that alignment  can be obtained without having to decouple the non-standard Higgs scalars \cite{Gunion:2002zf}. However, to the best of our knowledge a comprehensive study of the conditions for which alignment occurs in 2HDMs has not been carried out. In this article we present such an analysis. 

In general 2HDMs, absence of tree-level FCNC could be achieved by a discrete symmetry  forbidding the $\lambda_6$ and $\lambda_7$ couplings, in which case we show that alignment  occurs for specific values of the quartic couplings. More specifically, alignment only happens when $t_\beta$ is a common solution to two independent algebraic equations and is therefore not a natural occurrence in this scenario. Furthermore, if certain orderings in the quartic couplings are not satisfied, alignment would never occur; this is the case for the MSSM at tree-level.

More interestingly, if  $\lambda_6$ and $\lambda_7$ are non-zero but much smaller than the other couplings in the theory, perhaps as a result of a softly broken symmetry, we discover that alignment could exist for  generic values of the quartic couplings. In this case, one can show that alignment may occur at both small and large values of $t_\beta$. In particular, the large $t_\beta$ solution may be realized  in a wide region of parameter space.  Furthermore, this solution may be realized in the MSSM, where the coupling $\lambda_7$ is generated radiatively. A positive  $\lambda_7$ is required in this case. Moreover, we obtain that  small values of $(\mu/M_{\rm{SUSY}})$ are not compatible with alignment without decoupling. Therefore, small deviations from the SM expectations in the Higgs couplings to fermions and weak gauge bosons without any signal of non-standard Higgs bosons in direct searches,  would imply either large $m_A$ or large values of $(\mu/M_{\rm{SUSY}})$.

Other realizations of alignment without decoupling in supersymmetric theories can be obtained by going beyond the MSSM, which relaxes the strict relations among the tree-level quartic couplings of the MSSM Higgs potential. One simple possibility is to extend the MSSM Higgs sector by adding more scalars, such as an electroweak singlet (the NMSSM) or a triplet. When one takes the limit that the singlet/triplet scalars are heavy and integrated out, new contributions to the scalar quartic couplings are generated. More specifically, new contributions to $\tilde{\lambda}_3$ are generated, thereby allowing for alignment without decoupling even if $\lambda_6$ and $\lambda_7$ continue to be zero.

Last but not least, we consider the effectiveness of using precision measurements of Higgs couplings to fermions to probe the wedge region in the $m_A - \tan\beta$ plane. If in the future no relevant deviations of the Higgs-fermion couplings from their SM values are
observed, it could be naively inferred that large values of  $m_A$ are required, thereby disfavoring the wedge region. The possibility of alignment without decoupling opens up this region of parameter space, implying that new physics beyond the SM would have to be probed by other means. These would include, for example, measurements of loop-induced Higgs couplings, flavor physics, relic density and direct detection of dark matter, as well as direct searches for new physics beyond the 2HDMs. Moreover, alignment without decoupling also highlights the necessity of devising new search strategies to look for light non-standard Higgs bosons. We will return to these important topics in the near future.

 \begin {acknowledgements}
 I.L. would like to thank Tao Han and Markus Luty for discussions, and Howie Haber for an enlightening conversation on sign choices in 2HDM. This work was supported in part by the National Science Foundation under Grant No. PHYS-1066293 and the hospitality of the Aspen Center for Physics.
 Fermilab is operated by Fermi Research Alliance, LLC under Contract No. DE-AC02-07CH11359 with the U.S. Department of Energy. Work at ANL is supported in part by the U.S. Department of Energy under Contract No. DE-AC02-06CH11357. Work at Northwestern is supported in part by the U.S. Department of Energy under Contract No. DE-SC0010143. Work at KITP is supported by the National Science Foundation under Grant No. NSF PHY11-25915. M.C  was partially supported by a Simons Distinguished Visiting Scholar during her stay at the KITP. I.L.  was partially supported by the Simons Foundation under award No. 230683. N.R.S is supported by the DOE grant No. DE-SC0007859 and by the Michigan Center for Theoretical Physics.
 \end{acknowledgements}


\begin{thebibliography}{99}




%\cite{:2012gk,:2012gu}
\bibitem{:2012gk} 
  G.~Aad {\it et al.}  [ATLAS Collaboration],
  %``Observation of a new particle in the search for the Standard Model Higgs boson with the ATLAS detector at the LHC,''
  Phys.\ Lett.\ B {\bf 716}, 1 (2012)
  [arXiv:1207.7214 [hep-ex]].
  %%CITATION = ARXIV:1207.7214;%%

%\cite{:2012gu}
\bibitem{:2012gu} 
  S.~Chatrchyan {\it et al.}  [CMS Collaboration],
  %``Observation of a new boson at a mass of 125 GeV with the CMS experiment at the LHC,''
  Phys.\ Lett.\ B {\bf 716}, 30 (2012)
  [arXiv:1207.7235 [hep-ex]].
  %%CITATION = ARXIV:1207.7235;%%
  
%\cite{Low:2012rj}
\bibitem{Low:2012rj} 
  I.~Low, J.~Lykken and G.~Shaughnessy,
  %``Have We Observed the Higgs (Imposter)?,''
  arXiv:1207.1093 [hep-ph];
  %%CITATION = ARXIV:1207.1093;%%
  %\cite{Corbett:2012dm}
  T.~Corbett, O.~J.~P.~Eboli, J.~Gonzalez-Fraile and M.~C.~Gonzalez-Garcia,
  %``Constraining anomalous Higgs interactions,''
  Phys.\ Rev.\ D {\bf 86}, 075013 (2012)
  [arXiv:1207.1344 [hep-ph]];
  %%CITATION = ARXIV:1207.1344;%%
  P.~P.~Giardino, K.~Kannike, M.~Raidal and A.~Strumia,
  %``Is the resonance at 125 GeV the Higgs boson?,''
  arXiv:1207.1347 [hep-ph].
  %%CITATION = ARXIV:1207.1347;%%
  J.~Ellis and T.~You,
  %``Global Analysis of the Higgs Candidate with Mass ~ 125 GeV,''
  arXiv:1207.1693 [hep-ph];
  %%CITATION = ARXIV:1207.1693;%%
   J.~R.~Espinosa, C.~Grojean, M.~Muhlleitner and M.~Trott,
  %``First Glimpses at Higgs' face,''
  arXiv:1207.1717 [hep-ph];
  %%CITATION = ARXIV:1207.1717;%%
   M.~Montull and F.~Riva,
  %``Higgs discovery: the beginning or the end of natural EWSB?,''
  arXiv:1207.1716 [hep-ph];
  %%CITATION = ARXIV:1207.1716;%%
 D.~Carmi, A.~Falkowski, E.~Kuflik, T.~Volansky and J.~Zupan,
  %``Higgs After the Discovery: A Status Report,''
  arXiv:1207.1718 [hep-ph];
  %%CITATION = ARXIV:1207.1718;%%
  T.~Plehn and M.~Rauch,
  %``Higgs Couplings after the Discovery,''
  arXiv:1207.6108 [hep-ph];
  %%CITATION = ARXIV:1207.6108;%%


%\cite{Branco:2011iw}
\bibitem{Branco:2011iw} 
%\bibitem{Gunion:1989we} 
  For reviews see, for example, J.~F.~Gunion, H.~E.~Haber, G.~L.~Kane and S.~Dawson,
  %``The Higgs Hunter's Guide,''
  Front.\ Phys.\  {\bf 80}, 1 (2000);
  %%CITATION = FRPHA,80,1;%%
  %205 citations counted in INSPIRE as of 30 Aug 2013
%\cite{Carena:2002es}
%\bibitem{Carena:2002es} 
  M.~S.~Carena and H.~E.~Haber,
  %``Higgs boson theory and phenomenology,''
  Prog.\ Part.\ Nucl.\ Phys.\  {\bf 50}, 63 (2003)
  [hep-ph/0208209];
  %%CITATION = HEP-PH/0208209;%%
  %337 citations counted in INSPIRE as of 30 Aug 2013
    %
      G.~C.~Branco, P.~M.~Ferreira, L.~Lavoura, M.~N.~Rebelo, M.~Sher and J.~P.~Silva,
  %``Theory and phenomenology of two-Higgs-doublet models,''
  Phys.\ Rept.\  {\bf 516}, 1 (2012)
  [arXiv:1106.0034 [hep-ph]].
  %%CITATION = ARXIV:1106.0034;%%
  %180 citations counted in INSPIRE as of 02 Aug 2013
  
  %\cite{Craig:2012vn}
\bibitem{Craig:2012vn} 
 For examples of recent works since the discovery of the 125 GeV Higgs boson, see 
 N.~Craig and S.~Thomas,
  %``Exclusive Signals of an Extended Higgs Sector,''
  JHEP {\bf 1211}, 083 (2012)
  [arXiv:1207.4835 [hep-ph]];
  %%CITATION = ARXIV:1207.4835;%%
  %33 citations counted in INSPIRE as of 30 Aug 2013
%\cite{Alves:2012ez}
%\bibitem{Alves:2012ez} 
  D.~S.~M.~Alves, P.~J.~Fox and N.~J.~Weiner,
  %``Higgs Signals in a Type I 2HDM or with a Sister Higgs,''
  arXiv:1207.5499 [hep-ph];
  %\cite{Belanger:2012sd}
%\bibitem{Belanger:2012sd} 
  G.~Belanger, U.~Ellwanger, J.~F.~Gunion, Y.~Jiang and S.~Kraml,
  %``Two Higgs Bosons at the Tevatron and the LHC?,''
  arXiv:1208.4952 [hep-ph];
  %%CITATION = ARXIV:1208.4952;%%
  %20 citations counted in INSPIRE as of 30 Aug 2013
  %%CITATION = ARXIV:1207.5499;%%
  %31 citations counted in INSPIRE as of 30 Aug 2013
%\cite{Craig:2012pu}
%\bibitem{Craig:2012pu} 
  N.~Craig, J.~A.~Evans, R.~Gray, C.~Kilic, M.~Park, S.~Somalwar and S.~Thomas,
  %``Multi-Lepton Signals of Multiple Higgs Bosons,''
  JHEP {\bf 1302}, 033 (2013)
  [arXiv:1210.0559 [hep-ph]];
  %%CITATION = ARXIV:1210.0559;%%
  %10 citations counted in INSPIRE as of 30 Aug 2013
%\cite{Altmannshofer:2012ar}
%\bibitem{Altmannshofer:2012ar} 
  W.~Altmannshofer, S.~Gori and G.~D.~Kribs,
  %``A Minimal Flavor Violating 2HDM at the LHC,''
  Phys.\ Rev.\ D {\bf 86}, 115009 (2012)
  [arXiv:1210.2465 [hep-ph]];
  %%CITATION = ARXIV:1210.2465;%%
  %38 citations counted in INSPIRE as of 30 Aug 2013
%\cite{Bai:2012ex}
%\bibitem{Bai:2012ex} 
  Y.~Bai, V.~Barger, L.~L.~Everett and G.~Shaughnessy,
  %``The 2HDM-X and Large Hadron Collider Data,''
  Phys.\ Rev.\ D {\bf 87}, 115013 (2013)
  [arXiv:1210.4922 [hep-ph]];
  %%CITATION = ARXIV:1210.4922;%%
  %23 citations counted in INSPIRE as of 30 Aug 2013
%\cite{Drozd:2012vf}
%\bibitem{Drozd:2012vf} 
  A.~Drozd, B.~Grzadkowski, J.~F.~Gunion and Y.~Jiang,
  %``Two-Higgs-Doublet Models and Enhanced Rates for a 125 GeV Higgs,''
  JHEP {\bf 1305}, 072 (2013)
  [arXiv:1211.3580 [hep-ph]];
  %%CITATION = ARXIV:1211.3580;%%
  %34 citations counted in INSPIRE as of 30 Aug 2013
  %\cite{Chang:2012zf}
%\bibitem{Chang:2012zf} 
  J.~Chang, K.~Cheung, P.~-Y.~Tseng and T.~-C.~Yuan,
  %``Implications on the Heavy CP-even Higgs Boson from Current Higgs Data,''
  Phys.\ Rev.\ D {\bf 87}, no. 3, 035008 (2013)
  [arXiv:1211.3849 [hep-ph]];
  %%CITATION = ARXIV:1211.3849;%%
  %14 citations counted in INSPIRE as of 30 Aug 2013
%\cite{Chen:2013kt}
%\bibitem{Chen:2013kt} 
  C.~-Y.~Chen and S.~Dawson,
  %``Exploring Two Higgs Doublet Models Through Higgs Production,''
  Phys.\ Rev.\ D {\bf 87}, 055016 (2013)
  [arXiv:1301.0309 [hep-ph]];
  %%CITATION = ARXIV:1301.0309;%%
  %23 citations counted in INSPIRE as of 30 Aug 2013
   A.~Celis, V.~Ilisie and A.~Pich,
  %``LHC constraints on two-Higgs doublet models,''
  JHEP {\bf 1307}, 053 (2013)
  [arXiv:1302.4022 [hep-ph]];
  %%CITATION = ARXIV:1302.4022;%%
  %30 citations counted in INSPIRE as of 30 Aug 2013
  C.~-W.~Chiang and K.~Yagyu,
  %``Implications of Higgs boson search data on the two-Higgs doublet models with a softly broken $Z_2$ symmetry,''
  JHEP {\bf 1307}, 160 (2013)
  [arXiv:1303.0168 [hep-ph]];
  %%CITATION = ARXIV:1303.0168;%%
  %19 citations counted in INSPIRE as of 30 Aug 2013
   B.~'nGrinstein and P.~Uttayarat,
  %``Carving Out Parameter Space in Type-II Two Higgs Doublets Model,''
  JHEP {\bf 1306}, 094 (2013)
  [arXiv:1304.0028 [hep-ph]];
  %%CITATION = ARXIV:1304.0028;%%
  %21 citations counted in INSPIRE as of 30 Aug 2013
%\cite{Barroso:2013zxa}
%\bibitem{Barroso:2013zxa} 
  A.~Barroso, P.~M.~Ferreira, R.~Santos, M.~Sher and J.~‹o P.~Silva,
  %``2HDM at the LHC - the story so far,''
  arXiv:1304.5225 [hep-ph];
  %%CITATION = ARXIV:1304.5225;%%
  %12 citations counted in INSPIRE as of 30 Aug 2013
%\cite{Chen:2013rba}
%\bibitem{Chen:2013rba} 
  C.~-Y.~Chen, S.~Dawson and M.~Sher,
  %``Heavy Higgs Searches and Constraints on Two Higgs Doublet Models,''
  Phys.\ Rev.\ D {\bf 88}, 015018 (2013)
  [arXiv:1305.1624 [hep-ph]].
  %%CITATION = ARXIV:1305.1624;%%
  %16 citations counted in INSPIRE as of 30 Aug 2013


%\cite{Christensen:2012ei}
\bibitem{Christensen:2012ei} 
 J.~-M.~Gerard and M.~Herquet,
  %``A Twisted custodial symmetry in the two-Higgs-doublet model,''
  Phys.\ Rev.\ Lett.\  {\bf 98}, 251802 (2007)
  [hep-ph/0703051 [HEP-PH]];
  %%CITATION = HEP-PH/0703051;%%
   E.~Cervero and J.~-M.~Gerard,
  %``Minimal violation of flavour and custodial symmetries in a vectophobic Two-Higgs-Doublet-Model,''
  Phys.\ Lett.\ B {\bf 712}, 255 (2012)
  [arXiv:1202.1973 [hep-ph]];
  %%CITATION = ARXIV:1202.1973;%%
  N.~D.~Christensen, T.~Han and S.~Su,
  %``MSSM Higgs Bosons at The LHC,''
  Phys.\ Rev.\ D {\bf 85}, 115018 (2012)
  [arXiv:1203.3207 [hep-ph]];
  %%CITATION = ARXIV:1203.3207;%%
  %79 citations counted in INSPIRE as of 30 Aug 2013
%\cite{Han:2013mga}
%\bibitem{Han:2013mga} 
  T.~Han, T.~Li, S.~Su and L.~-T.~Wang,
  %``Non-Decoupling MSSM Higgs Sector and Light Superpartners,''
  arXiv:1306.3229 [hep-ph].
  %%CITATION = ARXIV:1306.3229;%%
  %1 citations counted in INSPIRE as of 30 Aug 2013

%\cite{Craig:2013hca}
\bibitem{Craig:2013hca} 
  N.~Craig, J.~Galloway and S.~Thomas,
  %``Searching for Signs of the Second Higgs Doublet,''
  arXiv:1305.2424 [hep-ph].
  %%CITATION = ARXIV:1305.2424;%%
  %9 citations counted in INSPIRE as of 31 Jul 2013
  

  
  %\cite{Gunion:2002zf}
\bibitem{Gunion:2002zf} 
  J.~F.~Gunion and H.~E.~Haber,
  %``The CP conserving two Higgs doublet model: The Approach to the decoupling limit,''
  Phys.\ Rev.\ D {\bf 67}, 075019 (2003)
  [hep-ph/0207010].
  %%CITATION = HEP-PH/0207010;%%
  %193 citations counted in INSPIRE as of 31 Jul 2013


 %\cite{Delgado:2013zfa}
\bibitem{Delgado:2013zfa} 
%\cite{Espinosa:1991wt}
%\bibitem{Espinosa:1991wt} 
  J.~R.~Espinosa and M.~Quiros,
  %``Higgs triplets in the supersymmetric standard model,''
  Nucl.\ Phys.\ B {\bf 384}, 113 (1992);
  %%CITATION = NUPHA,B384,113;%%
  %30 citations counted in INSPIRE as of 24 Sep 2013
  A.~Delgado, G.~Nardini and M.~Quiros,
  %``A Light Supersymmetric Higgs Sector Hidden by a Standard Model-like Higgs,''
  JHEP {\bf 1307}, 054 (2013)
  [arXiv:1303.0800 [hep-ph]].
  %%CITATION = ARXIV:1303.0800;%%
  %2 citations counted in INSPIRE as of 02 Aug 2013 

  
  
  %\cite{Haber:1993an}
\bibitem{Haber:1993an} 
  H.~E.~Haber and R.~Hempfling,
  %``The Renormalization group improved Higgs sector of the minimal supersymmetric model,''
  Phys.\ Rev.\ D {\bf 48}, 4280 (1993)
  [hep-ph/9307201].
  %%CITATION = HEP-PH/9307201;%%
  %270 citations counted in INSPIRE as of 04 Jul 2013
  

%\cite{Glashow:1976nt}
\bibitem{Glashow:1976nt} 
  S.~L.~Glashow and S.~Weinberg,
  %``Natural Conservation Laws for Neutral Currents,''
  Phys.\ Rev.\ D {\bf 15}, 1958 (1977).
  %%CITATION = PHRVA,D15,1958;%%
  %1190 citations counted in INSPIRE as of 02 Aug 2013
  


%\cite{Carena:2013qia}
\bibitem{Carena:2013qia} 
  M.~Carena, S.~Heinemeyer, O.~StŒl, C.~E.~M.~Wagner and G.~Weiglein,
  %``MSSM Higgs Boson Searches at the LHC: Benchmark Scenarios after the Discovery of a Higgs-like Particle,''
  arXiv:1302.7033 [hep-ph].
  %%CITATION = ARXIV:1302.7033;%%
  %23 citations counted in INSPIRE as of 04 Aug 2013
  

%\cite{Aaltonen:2012qt}
\bibitem{Aaltonen:2012qt} 
 ÊT.~Aaltonen {\it et al.} Ê[CDF and D0 Collaborations],
 Ê%``Evidence for a particle produced in association with weak bosons and decaying to a bottom-antibottom quark pair in Higgs boson searches at the Tevatron,''
 ÊPhys.\ Rev.\ Lett.\ Ê{\bf 109}, 071804 (2012)
 Ê[arXiv:1207.6436 [hep-ex]].
 Ê%%CITATION = ARXIV:1207.6436;%%
 Ê
 Ê\bibitem{LHCbottomtau} ÊG.~Aad {\it et al.} Ê[ATLAS Collaboration], ATLAS-CONF-2012-160, ATLAS-CONF-2012-161;
 Ê S.~Chatrchyan {\it et al.} Ê[CMS Collaboration], CMS-PAS-HIG-12-043, CMS-HIG-PAS-12-044.

%\cite{gagamu}
\bibitem{gagamu}
G.~Aad {\it et al.}  [ATLAS Collaboration],  ATLAS-CONF-2012-091;
S.~Chatrchyan {\it et al.} [CMS Collaboration], CMS-PAS-HIG-12-015.

%\cite{Okada:1990vk}
\bibitem{Okada:1990vk}
  Y.~Okada, M.~Yamaguchi and T.~Yanagida,
  %``Upper bound of the lightest Higgs boson mass in the minimal supersymmetric
  %standard model,''
  Prog.\ Theor.\ Phys.\  {\bf 85}, 1 (1991);
  %%CITATION = PTPKA,85,1;%%
%
%\cite{Ellis:1990nz}
%\bibitem{Ellis:1990nz}
  J.~R.~Ellis, G.~Ridolfi and F.~Zwirner,
  %``Radiative corrections to the masses of supersymmetric Higgs bosons,''
  Phys.\ Lett.\  B {\bf 257}, 83 (1991);
  %%CITATION = PHLTA,B257,83;%%
%
  %\cite{Haber:1990aw}
%\bibitem{Haber:1990aw}
  H.~E.~Haber and R.~Hempfling,
  %``Can the mass of the lightest Higgs boson of the minimal supersymmetric
  %model be larger than m(Z)?,''
  Phys.\ Rev.\ Lett.\  {\bf 66}, 1815 (1991);
  %%CITATION = PRLTA,66,1815;%%
%
%\bibitem{mhiggsRG1a}
  J.~A.~Casas, J.~R.~Espinosa, M.~Quiros and A.~Riotto,
  %``The Lightest Higgs boson mass in the minimal supersymmetric standard
  %model,''
  Nucl.\ Phys.\  B {\bf 436}, 3 (1995)
  [Erratum-ibid.\  B {\bf 439}, 466 (1995)]
  [arXiv:hep-ph/9407389];
  %%CITATION = NUPHA,B436,3;%%
%\bibitem{HHH} 
H.~Haber, R.~Hempfling and A.~Hoang,
              {\em Z. Phys.} {\bf C 75} (1997) 539,
              hep-ph/9609331;
              %%CITATION = HEP-PH 9609331; %%
%
    %\cite{Heinemeyer:1998yj}
%\bibitem{Heinemeyer:1998yj}
  S.~Heinemeyer, W.~Hollik and G.~Weiglein,
  %``FeynHiggs: A Program for the calculation of the masses of the neutral CP even Higgs bosons in the MSSM,''
  Comput.\ Phys.\ Commun.\  {\bf 124}, 76 (2000)
  [hep-ph/9812320];
  %%CITATION = HEP-PH/9812320;%%
  %\cite{Degrassi:2002fi}
  %\cite{Heinemeyer:1998np}
%\bibitem{Heinemeyer:1998np}
  S.~Heinemeyer, W.~Hollik and G.~Weiglein,
  %``The Masses of the neutral CP - even Higgs bosons in the MSSM: Accurate analysis at the two loop level,''
  Eur.\ Phys.\ J.\ C {\bf 9}, 343 (1999)
  [hep-ph/9812472];
  %%CITATION = HEP-PH/9812472;%%
  %\bibitem{Carena:2000dp}
  M.~S.~Carena, H.~E.~Haber, S.~Heinemeyer, W.~Hollik, C.~E.~M.~Wagner and G.~Weiglein,
  %``Reconciling the two loop diagrammatic and effective field theory computations of the mass of the lightest CP - even Higgs boson in the MSSM,''
  Nucl.\ Phys.\ B {\bf 580}, 29 (2000)
  [hep-ph/0001002];
  %%CITATION = HEP-PH/0001002;%%
%
%\cite{Martin:2002wn}
%\bibitem{Martin:2002wn}
  S.~P.~Martin,
  %``Complete two loop effective potential approximation to the lightest Higgs scalar boson mass in supersymmetry,''
  Phys.\ Rev.\ D {\bf 67}, 095012 (2003)
  [hep-ph/0211366];
  %%CITATION = HEP-PH/0211366;%%
%
%\cite{Okada:1990vk}
%\bibitem{Degrassi:2002fi}
  G.~Degrassi, S.~Heinemeyer, W.~Hollik, P.~Slavich and G.~Weiglein,
  %``Towards high precision predictions for the MSSM Higgs sector,''
  Eur.\ Phys.\ J.\ C {\bf 28}, 133 (2003)
  [hep-ph/0212020].
  %%CITATION = HEP-PH/0212020;%%
  %\cite{Frank:2006yh}
    %\cite{Aaltonen:2011gs}

%\cite{Okada:1990vk}
\bibitem{mhiggsRG1}
 M.~Carena, J.~Espinosa, M.~Quir\'os and C.~Wagner,
                    {\em Phys. Lett.} {\bf B 355} (1995) 209,
                    hep-ph/9504316;
                    %%CITATION = HEP-PH 9504316;%%
                    M.~Carena, M.~Quir\'os and C.~Wagner,
                    {\em Nucl. Phys.} {\bf B 461} (1996) 407,
                    hep-ph/9508343.
                    %%CITATION = HEP-PH 9508343;%%

%\cite{Carena:2011aa}
\bibitem{Carena:2011aa} 
  M.~Carena, S.~Gori, N.~R.~Shah and C.~E.~M.~Wagner,
  %``A 125 GeV SM-like Higgs in the MSSM and the $\gamma \gamma$ rate,''
  JHEP {\bf 1203}, 014 (2012)
  [arXiv:1112.3336 [hep-ph]].
  %%CITATION = ARXIV:1112.3336;%%
  %225 citations counted in INSPIRE as of 31 Aug 2013

\bibitem{deltamb2}   D.~M.~Pierce, J.~A.~Bagger, K.~T.~Matchev and R.~-j.~Zhang,
  %``Precision corrections in the minimal supersymmetric standard model,''
  Nucl.\ Phys.\ B {\bf 491}, 3 (1997)
  [hep-ph/9606211];
  %%CITATION = HEP-PH/9606211;%%

\bibitem{deltamb} 
L.~J.~Hall, R.~Rattazzi and U.~Sarid,
  %``The Top quark mass in supersymmetric SO(10) unification,''
  Phys.\ Rev.\ D {\bf 50}, 7048 (1994)
  [hep-ph/9306309];
  %%CITATION = HEP-PH/9306309;%%
  %732 citations counted in INSPIRE as of 31 Aug 2013
R.~Hempfling,
  %``Yukawa coupling unification with supersymmetric threshold corrections,''
  Phys.\ Rev.\ D {\bf 49}, 6168 (1994);
  %%CITATION = PHRVA,D49,6168;%%
  %357 citations counted in INSPIRE as of 31 Aug 2013
%
%\bibitem{deltamb1} 
M.~S.~Carena, M.~Olechowski, S.~Pokorski and C.~E.~M.~Wagner,
  %``Electroweak symmetry breaking and bottom - top Yukawa unification,''
  Nucl.\ Phys.\ B {\bf 426}, 269 (1994)
  [hep-ph/9402253].
  %%CITATION = HEP-PH/9402253;%%
  %650 citations counted in INSPIRE as of 31 Aug 2013
  




%\cite{Carena:2002es}
\bibitem{Carena:2002es} 
 M.~S.~Carena, S.~Mrenna and C.~E.~M.~Wagner,
  %``MSSM Higgs boson phenomenology at the Tevatron collider,''
  Phys.\ Rev.\ D {\bf 60}, 075010 (1999)
  [hep-ph/9808312];  
  %%CITATION = HEP-PH/9808312;%%
   M.~S.~Carena, S.~Mrenna and C.~E.~M.~Wagner,
  %``The Complementarity of LEP, the Tevatron and the CERN LHC in the search for a light MSSM Higgs boson,''
  Phys.\ Rev.\ D {\bf 62}, 055008 (2000)
  [hep-ph/9907422].
  %%CITATION = HEP-PH/9907422;%%
  %140 citations counted in INSPIRE as of 31 Aug 2013
   H.~E.~Haber, M.~J.~Herrero, H.~E.~Logan, S.~Penaranda, S.~Rigolin and D.~Temes,
  %``SUSY QCD corrections to the MSSM h0 $b \bar{b}$ vertex in the decoupling limit,''
  Phys.\ Rev.\ D {\bf 63}, 055004 (2001)
  [hep-ph/0007006];
  %%CITATION = HEP-PH/0007006;%%
   M.~S.~Carena, H.~E.~Haber, H.~E.~Logan and S.~Mrenna,
  %``Distinguishing a MSSM Higgs boson from the SM Higgs boson at a linear collider,''
  Phys.\ Rev.\ D {\bf 65}, 055005 (2002)
  [Erratum-ibid.\ D {\bf 65}, 099902 (2002)]
  [hep-ph/0106116];
  %%CITATION = HEP-PH/0106116;%%
  %82 citations counted in INSPIRE as of 31 Aug 2013
  %%CITATION = HEP-PH/0007006;%%
  %86 citations counted in INSPIRE as of 31 Aug 2013
  M.~S.~Carena and H.~E.~Haber,
  %``Higgs boson theory and phenomenology,''
  Prog.\ Part.\ Nucl.\ Phys.\  {\bf 50}, 63 (2003)
  [hep-ph/0208209]; 
    J.~Guasch, W.~Hollik and S.~Penaranda,
  %``Distinguishing Higgs models in H ---> b anti-b / H ---> tau+ tau-,''
  Phys.\ Lett.\ B {\bf 515}, 367 (2001)
  [hep-ph/0106027].
  %%CITATION = HEP-PH/0106027;%%
  %%CITATION = HEP-PH/0208209;%%


\bibitem{Patrick} 
M.~Carena, P.~Draper, T.~Liu and C.~Wagner,
  %``The 7 TeV LHC Reach for MSSM Higgs Bosons,''
  Phys.\ Rev.\ D {\bf 84}, 095010 (2011)
  [arXiv:1107.4354 [hep-ph]].
  %%CITATION = ARXIV:1107.4354;%%
  %19 citations counted in INSPIRE as of 08 Oct 2013

 \bibitem{LHCtautauMSSM} Atlas Collaboration, ATLAS-CONF-2012-094; CMS Collaboration, CMS-PAS-HIG-12-050.  




%\cite{Ellwanger:2009dp}
%\cite{Maniatis:2009re}
\bibitem{Ellwanger:2009dp} 
  M.~Maniatis,
  %``The Next-to-Minimal Supersymmetric extension of the Standard Model reviewed,''
  Int.\ J.\ Mod.\ Phys.\ A {\bf 25}, 3505 (2010)
  [arXiv:0906.0777 [hep-ph]];
  %%CITATION = ARXIV:0906.0777;%%
  %128 citations counted in INSPIRE as of 25 Sep 2013
  U.~Ellwanger, C.~Hugonie and A.~M.~Teixeira,
  %``The Next-to-Minimal Supersymmetric Standard Model,''
  Phys.\ Rept.\  {\bf 496}, 1 (2010)
  [arXiv:0910.1785 [hep-ph]].
  
  
  %\cite{Espinosa:1991gr}
\bibitem{Espinosa:1991gr} 
  J.~R.~Espinosa and M.~Quiros,
  %``On Higgs boson masses in nonminimal supersymmetric standard models,''
  Phys.\ Lett.\ B {\bf 279}, 92 (1992);
  %%CITATION = PHLTA,B279,92;%%
  %187 citations counted in INSPIRE as of 24 Sep 2013
  %%CITATION = ARXIV:0910.1785;%%
  %315 citations counted in INSPIRE as of 23 Aug 2013
%\cite{King:2012is}
%\bibitem{King:2012is} 
%\cite{Barger:2006dh}
%\bibitem{Barger:2006dh} 
%\cite{Miller:2003ay}
%\bibitem{Miller:2003ay} 
  D.~J.~Miller, 2, R.~Nevzorov and P.~M.~Zerwas,
  %``The Higgs sector of the next-to-minimal supersymmetric standard model,''
  Nucl.\ Phys.\ B {\bf 681}, 3 (2004)
  [hep-ph/0304049];
  %%CITATION = HEP-PH/0304049;%%
  %152 citations counted in INSPIRE as of 24 Sep 2013
  V.~Barger, P.~Langacker, H.~-S.~Lee and G.~Shaughnessy,
  %``Higgs Sector in Extensions of the MSSM,''
  Phys.\ Rev.\ D {\bf 73}, 115010 (2006)
  [hep-ph/0603247];
  %%CITATION = HEP-PH/0603247;%%
  %140 citations counted in INSPIRE as of 24 Sep 2013
  %\cite{Dermisek:2008uu}
%\bibitem{Dermisek:2008uu} 
  R.~Dermisek and J.~F.~Gunion,
  %``Many Light Higgs Bosons in the NMSSM,''
  Phys.\ Rev.\ D {\bf 79}, 055014 (2009)
  [arXiv:0811.3537 [hep-ph]];
 %%CITATION = ARXIV:0811.3537;%%
  %37 citations counted in INSPIRE as of 25 Sep 2013
   L.~J.~Hall, D.~Pinner and J.~T.~Ruderman,
  %``A Natural SUSY Higgs Near 126 GeV,''
  JHEP {\bf 1204}, 131 (2012)
  [arXiv:1112.2703 [hep-ph]];
  %%CITATION = ARXIV:1112.2703;%%
  %240 citations counted in INSPIRE as of 08 Oct 2013
   U.~Ellwanger,
  %``A Higgs boson near 125 GeV with enhanced di-photon signal in the NMSSM,''
  JHEP {\bf 1203}, 044 (2012)
  [arXiv:1112.3548 [hep-ph]];
  %%CITATION = ARXIV:1112.3548;%%
  %149 citations counted in INSPIRE as of 08 Oct 2013
  S.~F.~King, M.~Muhlleitner and R.~Nevzorov,
  %``NMSSM Higgs Benchmarks Near 125 GeV,''
  Nucl.\ Phys.\ B {\bf 860}, 207 (2012)
  [arXiv:1201.2671 [hep-ph]];
  %%CITATION = ARXIV:1201.2671;%%
  %102 citations counted in INSPIRE as of 24 Sep 2013
  %\cite{Vasquez:2012hn}
%\bibitem{Vasquez:2012hn} 
  D.~A.~Vasquez, G.~Belanger, C.~Boehm, J.~Da Silva, P.~Richardson and C.~Wymant,
  %``The 125 GeV Higgs in the NMSSM in light of LHC results and astrophysics constraints,''
  Phys.\ Rev.\ D {\bf 86}, 035023 (2012)
  [arXiv:1203.3446 [hep-ph]];
  %%CITATION = ARXIV:1203.3446;%%
  %61 citations counted in INSPIRE as of 24 Sep 2013
%\bibitem{Cao:2012yn} 
  J.~Cao, Z.~Heng, J.~M.~Yang and J.~Zhu,
  %``Status of low energy SUSY models confronted with the LHC 125 GeV Higgs data,''
  JHEP {\bf 1210}, 079 (2012)
  [arXiv:1207.3698 [hep-ph]].
  %%CITATION = ARXIV:1207.3698;%%
  %64 citations counted in INSPIRE as of 24 Sep 2013

%\cite{Lu:2013cta}
\bibitem{Lu:2013cta} 
  X.~Lu, H.~Murayama, J.~T.~Ruderman and K.~Tobioka,
  %``A Natural Higgs Mass in Supersymmetry from Non-Decoupling Effects,''
  arXiv:1308.0792 [hep-ph].
  %%CITATION = ARXIV:1308.0792;%%
  %2 citations counted in INSPIRE as of 24 Sep 2013




  %\cite{Nomura:2005rk}
\bibitem{Nomura:2005rk} 
  Y.~Nomura, D.~Poland and B.~Tweedie,
  %``mu B - driven electroweak symmetry breaking,''
  Phys.\ Lett.\ B {\bf 633}, 573 (2006)
  [hep-ph/0509244].
  %%CITATION = HEP-PH/0509244;%%
  %16 citations counted in INSPIRE as of 23 Aug 2013


\end{thebibliography}
\end{document}